\documentclass[12pt]{article}
\usepackage[latin9]{inputenc}
\usepackage{geometry}
\usepackage[active]{srcltx}
\usepackage{float}
\usepackage{amsmath}
\usepackage{amsthm}
\usepackage{amssymb}
\usepackage{graphicx}
\usepackage{rotfloat}
\usepackage{setspace}
\usepackage{esint}
\usepackage[authoryear]{natbib}
\usepackage[font=normalsize,labelfont=bf]{caption}
\usepackage{comment}
\usepackage{xr-hyper}
\usepackage{hyperref}
\usepackage{nicefrac}
\usepackage{mathtools}
\usepackage[caption=false]{subfig}
\usepackage{algorithm}
\usepackage{algpseudocode}
\makeatletter
\usepackage{bbm} 
\usepackage{longtable}
\expandafter\def\csname bm\endcsname{\boldmath}
\usepackage{booktabs} 
\usepackage[mathlines]{lineno}

\floatstyle{ruled}
\newfloat{algorithm}{tbp}{loa}
\providecommand{\algorithmname}{Algorithm}
\floatname{algorithm}{\protect\algorithmname}

\usepackage{latexsym}
\usepackage{amsthm}\usepackage{amsfonts}\usepackage{graphicx}\usepackage{epsfig}
\expandafter\def\csname bm\endcsname{\boldmath}

\newcommand*{\patchAmsMathEnvironmentForLineno}[1]{%
      \expandafter\let\csname old#1\expandafter\endcsname\csname #1\endcsname
      \expandafter\let\csname oldend#1\expandafter\endcsname\csname end#1\endcsname
      \renewenvironment{#1}%
         {\linenomath\csname old#1\endcsname}%
         {\csname oldend#1\endcsname\endlinenomath}}%
    \newcommand*{\patchBothAmsMathEnvironmentsForLineno}[1]{%
      \patchAmsMathEnvironmentForLineno{#1}%
      \patchAmsMathEnvironmentForLineno{#1*}}%
    \AtBeginDocument{%
    \patchBothAmsMathEnvironmentsForLineno{equation}%
    \patchBothAmsMathEnvironmentsForLineno{align}%
    \patchBothAmsMathEnvironmentsForLineno{flalign}%
    \patchBothAmsMathEnvironmentsForLineno{alignat}%
    \patchBothAmsMathEnvironmentsForLineno{gather}%
    \patchBothAmsMathEnvironmentsForLineno{multline}%
    }

\usepackage{setspace}
\def\dispmuskip{\thinmuskip= 3mu plus 0mu minus 2mu \medmuskip=  4mu plus 2mu minus 2mu \thickmuskip=5mu plus 5mu minus 2mu}
\def\textmuskip{\thinmuskip= 0mu                    \medmuskip=  1mu plus 1mu minus 1mu \thickmuskip=2mu plus 3mu minus 1mu}
\def\beq{\dispmuskip\begin{equation}}    \def\eeq{\end{equation}\textmuskip}
\def\beqn{\dispmuskip\begin{displaymath}}\def\eeqn{\end{displaymath}\textmuskip}
\def\bea{\dispmuskip\begin{eqnarray}}    \def\eea{\end{eqnarray}\textmuskip}
\def\bean{\dispmuskip\begin{eqnarray*}}  \def\eean{\end{eqnarray*}\textmuskip}

\usepackage{bm}  

\newcommand{\bs}[1]{\bm{#1}}           
\newcommand{\thetad}{\bs{\theta}_d}    

\def\E{{\mathbb E}}                         

\def\N{{\cal N}}

\def\Y{{\cal Y}}

\def\vech{\text{\rm vech}}

\usepackage[mathlines]{lineno} 

\makeatletter
\newcommand*{\addFileDependency}[1]{
  \typeout{(#1)}
  \@addtofilelist{#1}
  \IfFileExists{#1}{}{\typeout{No file #1.}}
}
\makeatother

\usepackage{subfig}
\usepackage[export]{adjustbox}

\newcommand{\bb}[1]{\boldsymbol{#1}}

\newcommand{\thetacop}{\bs{\theta}_{\textrm{cop}}}

\newcommand{\diagEta}{\bb{D}}

\newcommand{\ymar}{\bs{y}_{(d)}}

\newcommand{\Gammamar}{\bs{\gamma}_{\textrm{mar}}}
\newcommand{\Lambdamar}{\bs{\lambda}_{\textrm{mar}}}
\usepackage{comment}

\let\oldref\ref
\renewcommand{\ref}[1]{(\oldref{#1})}

\usepackage{amsthm}
\theoremstyle{definition}
\usepackage{authblk}

\usepackage{bbm}
\usepackage{xr}
\usepackage{multirow}
\usepackage{threeparttable}

\usepackage{tikz}
\usetikzlibrary{positioning, decorations.pathreplacing, shapes.misc, fit}
\usetikzlibrary{arrows.meta}

\usepackage{placeins} 

\usetikzlibrary{calc}

\begin{document}

\title{\bf Neural Inference Functions for Margins for Time Series Copula Models}	
\author[a]{Daniel Fynn}
\author[a]{David Gunawan}
\author[b,a]{Andrew Zammit-Mangion}
\affil[a]{School of Mathematics and Physics, University of Wollongong, Wollongong, New South Wales, Australia}
\affil[b]{School of Mathematics and Statistics, University of New South Wales, Sydney, New South Wales, Australia}
\renewcommand\Authands{ and }
\maketitle

\maketitle

\begin{abstract}
Copula models are widely employed in multivariate time series analysis because they permit flexible modelling of marginal distributions independently  of the dependence structure, which is fully characterised by the copula function. However, Bayesian inference with these models becomes computationally demanding as the number of variables in the time series increases. Motivated by the classical inference functions for margins (IFM) approach, we propose a new neural-network based inference framework for estimating parameters in copula models, termed the neural inference functions for margins (N-IFM). N-IFM enables rapid parameter estimation for new data, fast sequential prediction, and efficient model comparison via time-series validation. We assess the performance of N-IFM using both simulated and real datasets and compare it to Hamiltonian Monte Carlo, demonstrating substantial computational gains with comparable inferential accuracy.

\end{abstract}

\section{Introduction \label{sec:Introduction-1}}

Multivariate time series models are important inferential tools, as they provide a powerful framework for analysing volatility and dependence across multiple series. In financial applications, such models are used to study volatility clustering, co-movement among asset returns, portfolio risk, and systemic risk arising from complex dependence across assets \citep{Patton2009Feb, Kastner2019May, deng2025large}. 

Copula-based time series models are particularly useful in this setting because they allow the marginal distributions of the individual series to be modelled separately from the dependence structure linking them \citep{Patton2012, Smith2023}. A copula can be used to model serial dependence within each univariate series \citep{ BladtMcNeil2022}, contemporaneous cross-sectional dependence across series after accounting for marginal dynamics \citep{Acar2019, NasriRemillard2019}, or serial and cross-sectional dependence jointly within a unified multivariate time-series model \citep{smith2015copula, LoaizaMaya2018, Nagler2022, Zhao2022}. In this paper, we use the copula to model contemporaneous cross-sectional dependence among the series, conditional on the marginal time-series dynamics. Specifically, serial dependence and conditional heteroskedasticity for each series are captured through the generalised autoregressive conditional heteroskedasticity (GARCH) model \citep{bollerslev1986generalized}, while Gaussian \citep{song2000} and Student's-\(t\) \citep{demarta2005t} copulas are used to capture dependence across the transformed series at each time point.



Exact Bayesian inference for high-dimensional copula models using Markov chain Monte Carlo (MCMC) methods is computationally expensive \citep{smith2012modelling, smith2015copula, gunawan2020mixed}.
Recently, variational Bayes (VB) inference has been used to estimate high-dimensional time-series copula models, reducing computational cost and improving scalability \citep{loaiza2019variational,nguyen2020variational,deng2025large}. However, 
financial and economics practitioners are interested in updating parameter and volatility estimates in real time, in order to regularly produce one-step or multi-step ahead predictive distributions. Sequential updating using MCMC or VB methods is computationally intensive, as each new observation {typically} requires rerunning full MCMC and VB algorithms. 




Neural amortised inference, a type of simulation-based inference method, has recently emerged as a powerful approach for Bayesian inference; see \citet{zammit2024neural} for a recent review. 
Rather than performing computationally intensive inference from scratch for each new dataset, these methods train neural networks to learn reusable mappings from data to posterior distributions of model parameters \citep{radev2020bayesflow, zammit2024neural}.
Once trained, these neural networks amortise their training cost by providing fast, near-instantaneous inference for new observations, making them particularly well suited to scenarios that require repeated inference, such as sequential parameter updating or time-series validation for model selection. Often, flexible deep learning architectures \citep{Goodfellow2016} are employed to model the mapping between observations and the posterior distribution of the model parameters.



In this paper, we propose an efficient neural amortised inference approach to fit high-dimensional time-series copula models. Motivated by the classical inference functions for margins (IFM) approach \citep{joe1997multivariate}, which first estimates the parameters of the marginal distributions and subsequently estimates the copula parameters conditional on the marginal parameter estimates, we propose neural inference functions for margins (N-IFM), which leverages neural networks to perform both stages of inference in an amortised manner. Likelihood-free neural approaches have also recently been considered for copula models; for example, \citet{Andre2025Dec} develop a
likelihood-free neural Bayes estimator for bivariate copula models. In contrast, our two-stage construction is designed for high-dimensional time-series copula models. 
The two-stage construction is not a requirement of simulation-based inference, but is adopted here to improve scalability. As noted by \citet{Joe2005Jun} and \citet{Arachchige2025Jan}, numerical computations for multivariate models become increasingly challenging as the dimension grows, motivating the use of two-stage estimation procedures. {Such two-stage estimation is well established in Bayesian copula modelling, where the marginal distributions are estimated first and the copula parameters are then inferred conditional on the fitted margins \citep{LoaizaMaya2018, SmithKlein2021, Smith2023}.} This yields a modular neural estimator that separates marginal time-series learning from cross-sectional dependence learning.

The primary computational motivation of the proposed method is to facilitate repeated {parameter inference and prediction} for multivariate time series observed over fixed-length windows. The substantial computational cost of training the neural networks is incurred only once for a chosen model specification, after which the trained neural networks can be reused to obtain fast posterior approximations for many new datasets of the same input dimension. 
This setting commonly arises in financial applications where forecasts and risk measures are updated using rolling windows. 
The method is therefore especially well suited to rolling-window prediction and time-series validation, but not directly applicable to fixed-origin expanding-window schemes. 
We empirically evaluate the performance of N-IFM on simulated and real datasets, benchmarking it against
Hamiltonian Monte Carlo (HMC; \citealp{neal2011mcmc}) and showing substantial computational
gains with comparable inferential accuracy.



The rest of this paper is organised as follows. Section \ref{sec:TimeSeriesCopulaModels} presents background material on copula models, while Section \ref{sec:NeuralVariationalInference} presents N-IFM. Section \ref{sec:predictiveperformance} outlines the criteria used to assess predictive performance. Section \ref{sec:SimulationStudy} describes simulation studies conducted to evaluate the accuracy of N-IFM, and Section \ref{sec:RealData}
applies N-IFM to a real-world dataset. Section \ref{sec:Conclusion} discusses our results
and findings. The paper also has an online supplement with additional technical details and examples.

\section{Background \label{sec:TimeSeriesCopulaModels}}
Section~\ref{sec:CopulaModel} provides a brief overview of copula models. 
Section~\ref{sec:implicitcopula} introduces implicit copula models, and 
Section~\ref{sec:GaussianCopula} describes the Gaussian and $t$ copula specifications. 
Section~\ref{sec:factorcopula} details the factor-based parameterisation of the correlation structure for the Gaussian and $t$ copulas. 
Finally, Section~\ref{sec:marginalmodel} presents the  generalised autoregressive conditional heteroskedasticity (GARCH) model that we use to model the marginal time series.

\subsection{Copula models\label{sec:CopulaModel}}

Copulas offer a general framework for modelling multivariate distributions in which the marginal distributions and the dependence structure across variables are modelled separately. By Sklar's theorem \citep{Sklar1959}, the joint cumulative distribution function of a multivariate continuous random variable \( \bs{Y} \equiv (Y_1, \ldots, Y_D)^\top \) is given by 
\begin{equation}
F_{\bs{Y}}(\bs{y}) = C(F_{Y_1}(y_1), \ldots, F_{Y_D}(y_D)), \label{eq:copula_cdf}
\end{equation}
\noindent where \( \bs{y} \equiv (y_1, \ldots, y_D)^\top \) denotes a realisation of the random vector \( \bs{Y} \), \( F_{Y_d}(y_d) \) is the marginal cumulative distribution function (CDF) of the \( d \)-th component \( Y_d \), and \( C: [0,1]^D \rightarrow [0,1] \) is the copula function, which captures the dependence structure among the components of \( \bs{Y} \). 

To obtain the joint probability density function (PDF) \( p_{\bs{Y}}(\bs{y}) \), we differentiate \eqref{eq:copula_cdf} with respect to all components of \( \bs{Y} \):

\begin{equation}
p_{\bs{Y}}(\bs{y}) = \frac{\partial^D}{\partial y_1 \cdots \partial y_D} F_{\bs{Y}}(\bs{y}) = c(F_{Y_1}(y_1), \ldots, F_{Y_D}(y_D)) \prod_{d=1}^{D} p_{Y_d}(y_d), \label{eq:copula_pdf}
\end{equation}
where \( p_{Y_d}(y_d) = \frac{\partial}{\partial y_d} F_{Y_d}(y_d) \) is the marginal density of \( Y_d \)  and \( c(\bs{u}) = \frac{\partial^D}{\partial u_1 \cdots \partial u_D} C(\bs{u}) \) is the {copula density}, where \( \bs{u} \equiv (u_1, \ldots, u_D)^\top \in [0,1]^D \) with \( u_d = F_{Y_d}(y_d) \).

\subsection{Implicit copula models \label{sec:implicitcopula}}
\citet{mcneil2015quantitative} use the term implicit copula to refer to the copula that is implicitly defined by the multivariate distribution of a continuous random vector \(\bs{Z} \equiv (Z_1, \ldots, Z_D)^\top\). This copula is obtained by inverting Sklar's theorem, through what is called the inversion method \citep{nelsen2006introduction}. If \(\bs{Z}\) has a joint distribution function \(F_{\bs{Z}}\) with marginal distributions \(F_{Z_1}, \ldots, F_{Z_D}\), then the implicit copula function is given by
\begin{equation}
C(\bs{u}) = F_{\bs{Z}}\left( F_{Z_1}^{-1}(u_1), \ldots, F_{Z_D}^{-1}(u_D) \right)\label{eq:InversionCopula}.
\end{equation}
Then, by the change of variables formula, we obtain the copula density
\begin{equation}
c(\bs{u})
= \frac{\partial^D}{\partial u_1 \cdots \partial u_D} C(\bs{u}) 
= \frac{p_{\bs{Z}}(\bs{z})}{\prod_{d=1}^D p_{Z_d}(z_d)},
\end{equation}
where $\bs{z}\equiv(z_1,\ldots,z_D)^\top$ with $z_d=F_{Z_d}^{-1}(u_d)$ for $d=1,\ldots,D$; further details are given in Section \ref{sec:copula_deriv} of the online supplement.
Here, $p_{\bs{Z}}(\bs{z})$ denotes the joint density of $\bs{Z}$,
${
p_{\bs{Z}}(\bs{z})=\frac{\partial^D}{\partial z_1\cdots \partial z_D}F_{\bs{Z}}(\bs{z}),}
$
and $p_{Z_d}(z_d)$ denotes the marginal density of $Z_d$,
${
p_{Z_d}(z_d)=\frac{\partial}{\partial z_d}F_{Z_d}(z_d), \  d=1,\ldots,D.}
$
In this paper we consider two commonly used implicit copula models, the Gaussian copula \citep{song2000} and the $t$ copula \citep{demarta2005t}. Since their margins are constrained to be uniform, copula models are invariant to location and scale. Therefore, when defining the joint distribution $F_{\bs Z}$, the location parameters are 
irrelevant, and the scale matrices can be restricted to correlation matrices \citep{nelsen2006introduction}. 



\subsection{Gaussian and $t$ copulas \label{sec:GaussianCopula}}
This section briefly describes the widely-used Gaussian copula of \citet{song2000} and $t$ copula of \cite{demarta2005t}. Both these copulas are constructed using the implicit method of Section \ref{sec:implicitcopula}.

Let $\Phi(\cdot)$ and $\phi(\cdot)$ be the CDF and PDF, respectively, of a univariate standard Gaussian random variable $\bs Z$, and let \(\Phi_D(\cdot; \bs{\mu}, \bs{\Omega})\) and \(\phi_D(\cdot; \bs{\mu}, \bs{\Omega})\) be the CDF and PDF of a $D$-dimensional multivariate normal distribution with mean vector \(\bs{\mu}\) and correlation matrix \(\bs{\Omega}\), respectively. The Gaussian copula is derived from a multivariate normal distribution \(\bs{Z} \sim \mathcal{N}_D(\bs{0}, \bs{\Omega})\), where \(\bs{\Omega}\) is a \(D \times D\) correlation matrix. 
From \eqref{eq:InversionCopula}, the Gaussian copula function is defined as
\begin{equation}
C_{\textrm{G}}(\bs{u}; \bs{\Omega}) = \Phi_D\left( \Phi^{-1}(u_1), \ldots, \Phi^{-1}(u_D); \bs{0}, \bs{\Omega} \right), \label{eqn:appendix_SRE_gaussia_copula}
\end{equation}


\noindent where \(\Phi^{-1}(\cdot)\) is the inverse of the standard normal CDF, applied component-wise to the vector $\bs{u} \equiv (u_1, \ldots, u_D)^{\top}$. 
The {Gaussian copula density} is given by:
\[
c_{\textrm{G}}(\bs{u}; \bs{\Omega}) = \frac{\phi_D(\bs{z}; \bs{0}, \bs{\Omega})}{\prod_{d=1}^D \phi(z_d)},
\]
where \(\bs{z} \equiv (\Phi^{-1}(u_1), \ldots, \Phi^{-1}(u_D))^\top\). Despite its popularity, the Gaussian copula does not allow for tail dependence: the tendency of variables to exhibit dependence during extreme co-movements.

The $t$ copula is constructed by modelling the vector \(\bs{Z}\) as a multivariate {$t$}-distributed with degrees of freedom \(\nu > 2\). 
A key advantage of the {$t$} copula over the Gaussian copula lies in its ability to model {tail dependence}. The {$t$} copula also allows for heavy tails, providing a more realistic description of dependence for financial data.

Let $T_D(\cdot;\bs{0},\bs{\Omega},\nu)$ and 
$t_D(\cdot;\bs{0},\bs{\Omega},\nu)$ denote the CDF and PDF, respectively, 
of the $D$-dimensional multivariate $t$-distribution with $\nu > 2$ degrees of freedom, a zero mean vector, 
and a positive definite scale matrix $\bs{\Omega}$.  For $0 \leq u_1, ..., u_D \leq 1$, the $D$-dimensional $t$ copula ($C_{T}$) and its joint copula density ($c_{t}$) are given by
\begin{align}
C_{T}(\bs{u}; \bs{\Omega}, \nu) &= T_D(z_1, ..., z_D; \bs{0},\bs{\Omega}, \nu)\label{eqn:appendix_SRE_t_copula},\\
c_{t}(\bs{u}; \bs{\Omega}, \nu) &= \frac{t_D(z_1, ..., z_D; \bs{0},\bs{\Omega}, \nu)}{\prod_{d=1}^D t_\nu(z_d)},\label{eqn:appendix_SRE_t_copula_density}
\end{align}
where $t_\nu(\cdot)$ is the PDF of a univariate $t$-distribution with $\nu > 2$ degrees of freedom and $z_d=T^{-1}_{\nu}(u_d)$ for $d=1,...,D$, where $T_\nu^{-1}(\cdot)$ is the quantile function of standard univariate $t$-distribution with $\nu$ degrees of freedom.

\subsection{Factor correlation matrix \label{sec:factorcopula}}
When the dimensionality \( D \) is large, estimating an unrestricted correlation matrix \( {\bs\Omega} \) for Gaussian and $t$ copulas becomes challenging due to the large number of parameters and the constraints required for positive definiteness. To address this issue, we follow the approach of \citet{murray2013} and employ a factor structure for the correlation matrix for  \( \bs{Z} \). Specifically, we factorise the correlation matrix as
\begin{equation}
\label{eq:Omega}
\bar{\bs\Omega} = \bs{R}_1 \bs{R} \bs{R}_1
= \bs{R}_1 (\bs{G}\bs{G}^\top + \bs{I}_D)\bs{R}_1,
\end{equation}
where $\bs{I}_D$ denotes the $D \times D$ identity matrix, $\bs{R}_1=\mathrm{diag}(\bs{R})^{-1/2}$ scales $\bs{R}$ to have unit diagonal entries thereby converting it into a valid correlation matrix, and $\bs{G}$ is a $D \times k$ factor loading matrix with $k < D$. 
For identifiability, $\bs{G}$ is constrained so that $G_{ij}=0$ for $j>i$ and $G_{ii}>0$ for $i = 1, \dots, k$.
To facilitate satisfying the positivity constraints when making inference, we reparameterise the loading matrix $\bs{G}$ as $\widetilde{\bs{G}}$ where 
\begin{equation*}
\widetilde G_{ij} =
\begin{cases}
\log(G_{ij}), & \text{if } i = j \leq k,\\[2mm]
G_{ij}, & \text{if } i \neq j.
\end{cases}
\end{equation*}



\subsection{Marginal time series model\label{sec:marginalmodel}}

The GARCH model is one of the most commonly used for modelling volatility clustering \citep{bollerslev1986generalized}. In most applications, the GARCH(1,1) model is sufficient for capturing the volatility dynamics of financial data, and it is hence the model we use for each (marginal) time series. 
Let $y_{d,t}$ denote an observation from the $d$-th time series at time $t$, for $d = 1,\ldots,D$ and $t = 1,\ldots,T$, where $T$ denotes the length of the time window.
The GARCH(1,1) model with Gaussian errors is given by
    \begin{align}
    y_{d,t} &= \sigma_{d,t} \, \epsilon_{d,t} && \text{for } t = 1, \dots, T, \label{eqn:garch_y}\\
    \sigma_{d,t}^2 &= \gamma_d + \alpha_{1,d} \, y_{d,t-1}^2 + \alpha_{2,d} \, \sigma_{d,t-1}^2 && \text{for } t = 2, \dots, T, \label{eqn:garch_var}\\
    \sigma_{d,1}^2 &= \frac{\gamma_d}{1 - \alpha_{1,d} - \alpha_{2,d}}, \label{eqn:garch_init}
    \end{align}
where $\gamma_d > 0$, $\alpha_{1,d} \geq 0$, $\alpha_{2,d} \geq 0$, $\alpha_{1,d} + \alpha_{2,d} < 1$, and $\epsilon_{d,t} \sim\N(0,1)$ for $d=1,...,D$. 
Instead of making inference on the constrained $\alpha_{1,d}, \alpha_{2,d}$ and $\gamma_{d}$, we instead make inference on the transformed parameters $\varphi_{1,d},\varphi_{2,d}$ and $ \varphi_{3,d}$, where the transformations are detailed in Section \ref{supp:datagen} of the online supplement.
In addition to Gaussian errors, 
we also consider error terms that follow a {standardised} Student's-$t$ distribution with $\widetilde{\nu}_d$ degrees of freedom, for $d = 1, \dots, D$.

\section{Neural Posterior Inference 
\label{sec:NeuralVariationalInference}}
Section \ref{sec:classinfmargins} discusses the classical inference functions for margins (IFM). Section \ref{sec:N-IFM} discusses the proposed neural inference functions for margins (N-IFM). 
\subsection{Classical inference functions for margins \label{sec:classinfmargins}}

This section discusses the classical IFM approach proposed by \citet{joe1996estimation}. The IFM method applies to models in which the univariate marginal distributions can be modelled separately from the dependence structure, as is the case when the dependence is modelled using a copula discussed in Section \ref{sec:CopulaModel}.
Here, we consider the case where the margins are generated through a time series.
Let $\boldsymbol{\theta}_d$ denote the vector of (possibly transformed) marginal parameters for the $d$-th series, $d = 1,\ldots,D$, 
and let $\boldsymbol{\theta}_{\mathrm{cop}}$ denote the vector of (possibly transformed) copula parameters. The full parameter vector is then given by
\[
\boldsymbol{\theta}
\equiv
\bigl(
\boldsymbol{\theta}_1^{\top},
\ldots,
\boldsymbol{\theta}_D^{\top},
\boldsymbol{\theta}_{\mathrm{cop}}^{\top}
\bigr)^{\top}.
\]
Consider observations
$\bs{y}_{t}\equiv\left({y}_{1,t},...,{y}_{D,t}\right)^{\top}$, for $t=1,...,T$  from the copula model with parameters $\bs\theta$, and let $\bs{y}\equiv(\bs{y}^{\top}_1,...,\bs{y}^{\top}_T)^{\top} $.
Based on the joint density of $\boldsymbol{Y}$ given in \eqref{eq:copula_pdf} and the observed data $\boldsymbol{y}_t$, $t = 1, \ldots, T$, the corresponding full log-likelihood is given by
\begin{equation}
\ell(\bs{\theta}; \bs{y}) = \sum_{t=1}^T \ell_{C,t}(\thetacop; F_{Y_{1,t}}(y_{1,t}; \bs{\theta}_1), \ldots, F_{Y_{D,t}}(y_{D,t}; \bs{\theta}_D)) + \sum_{d=1}^D \sum_{t=1}^T \ell_{d,t}(\bs{\theta}_d; y_{d,t}), \label{eq:loglikcop}
\end{equation} 
where \( \ell_{C,t}(\thetacop; F_{Y_{1,t}}(y_{1,t}; \bs{\theta}_1), \ldots, F_{Y_{D,t}}(y_{D,t}; \bs{\theta}_D)) = \log c(F_{Y_{1,t}}(y_{1,t}; \bs{\theta}_1), \ldots, F_{Y_{D,t}}(y_{D,t}; \bs{\theta}_D); \thetacop) \) and \( \ell_{d,t}(\bs{\theta}_d; y_{d,t}) = \log p_{Y_{d,t}}(y_{d,t}; \bs{\theta}_d) \), for $d=1,...,D$. The full maximum likelihood (ML) estimator  \( \widehat{\bs{\theta}}_{\textrm{ML}} \) is then defined by
\[
\hat{\bs{\theta}}_{\textrm{ML}} = \arg\max_{\bs{\theta} \in \Theta} \ell(\bs{\theta}; \bs{y}),
\]
where $\Theta$ is some suitable parameter space.
This optimisation is typically performed using numerical optimisation methods and can be computationally prohibitive, particularly in high-dimensional settings, which motivates the need for IFM. 
Consider the \( D \) log-likelihood functions for the univariate marginal distributions:
\[
\ell_d(\bs{\theta}_d;\{y_{d,t}\}_{t=1}^T) = 
\sum_{t=1}^T \ell_{d,t}(\thetad; y_{d,t}) =
\sum_{t=1}^T \log p_{Y_{d,t}}(y_{d,t}; \bs{\theta}_d), \quad d = 1, \ldots, D,
\] 
and the log-likelihood function for the copula:
\[
\ell_C(\thetacop; F_{Y_{1,1}}(y_{1,1}; \bs{\theta}_1), \ldots, F_{Y_{D,T}}(y_{D,T}; \bs{\theta}_D)) = \sum_{t=1}^T \log c(F_{Y_{1,t}}(y_{1,t}; {\bs{\theta}}_1), \ldots, F_{Y_{D,t}}(y_{D,t}; {\bs{\theta}}_D); \thetacop).
\] 
IFM proceeds as follows:
\begin{itemize}
    \item[(a)] Maximise the univariate log-likelihoods \( \ell_d \), for \( d = 1, \ldots, D \),  separately, to obtain the ML estimates \( \widehat{\bs{\theta}}_1, \ldots, \widehat{\bs{\theta}}_D \).
    \item[(b)] Maximise the log-likelihood function \( \ell_C \) with respect to the copula parameter vector \( \thetacop \) to obtain the ML estimate \( \widehat{\bs\theta}_{\textrm{cop}} \) given the marginal parameter estimates \( \widehat{\bs{\theta}}_1, \ldots, \widehat{\bs{\theta}}_D \) and the copula ``data'' $\bs{u} \equiv (\bs{u}_{1}^\top,...,\bs{u}_{T}^\top)^{\top}$, where $\bs{u}_{t}\equiv(u_{1,t},...,u_{D,t})^{\top}$, and where $u_{d,t} = F_{Y_{d,t}}(y_{d,t};\widehat{\bs{\theta}}_d)$.
\end{itemize}
The two-step estimation procedure is computationally simpler than jointly estimating all parameters \( \bs{\theta} \) by directly maximising the full log-likelihood in \eqref{eq:loglikcop}. 
By decoupling the estimation of marginal and dependence parameters, IFM reduces computational complexity while still providing consistent estimators under standard regularity conditions \citep{joe1996estimation}.
In a Bayesian version of the IFM one equips $\bs \theta_1,\dots,\bs \theta_D$ with a prior distribution, and then uses a Bayesian functional  of the posterior distribution (e.g., the posterior mean) of $\bs \theta_1,\dots,\bs \theta_D$ as plug-in for finding the posterior distribution of $\bs \theta_{\textrm{cop}}$. 

\subsection{Neural inference functions for margins\label{sec:N-IFM}}

For ease of exposition, from here on we omit variable subscripts and write conditional and marginal
densities simply as $p( \cdot  \mid  \cdot)$; for example, $p(\boldsymbol{y})$ denotes the density of $\boldsymbol{Y}$.

In this section we propose a two-stage approach, which we call N-IFM, to obtain an approximation to the posterior distribution of marginal and copula parameters motivated by the classical IFM discussed in Section \ref{sec:classinfmargins}. The approach leverages the representation capacity of neural networks to efficiently compute accurate approximations of posterior distributions from data \citep{zammit2024neural}.
{
{We begin by defining the target posterior through an IFM-based factorisation that separates the model parameters into marginal and copula components. Following the IFM approach of \citet{joe1996estimation}, we consider the two-stage posterior
\begin{equation}
\label{eq:fullposterior}
{p}\!\left(
\bs{\theta}_1,\ldots,\bs{\theta}_D,
\thetacop
\mid \bs{y}
\right)
:=
p\!\left(
\thetacop \mid \bs{y}, \hat{\bs\theta}_1, \dots, \hat{\bs\theta}_D
\right)
\prod_{d=1}^{D}
p\!\left(
\bs{\theta}_d \mid \bs{y}_{(d)}
\right), 
\end{equation} 
where $\bs y_{(d)} \equiv (y_{d,1}, \dots, y_{d,T})^\top$ is the $d$-th time series data, and $\hat{\bs\theta}_d
$ is the plug-in estimate for the marginal parameters, derived from the posterior distribution over  $\boldsymbol{\theta}_d$ for $d=1,...,D$. As in classical IFM, it is used as a computationally convenient alternative to the full posterior. This factorisation separates inference for the marginal parameters from inference for the 
dependence structure.}}



{To enable scalable computation, we use an approximation that mirrors the same factorisation in \eqref{eq:fullposterior}:
\begin{equation}
q\!\left(
\bs{\theta}_1,\ldots,\bs{\theta}_D,
\thetacop; {\bs{\lambda}}
\right)
=
q_{\textrm{cop}}\!\left(
\thetacop; {\bs{\lambda}_{\textrm{cop}}}
\right)
\prod_{d=1}^{D}
q_d\!\left(
\bs{\theta}_d; {\bs{\lambda}_d}
\right),
\label{eq:approx-ifm}
\end{equation}
where \( \bs{\lambda} = \left( \bs{\lambda}_1^{\top}, \ldots, \bs{\lambda}_D^{\top}, \bs{\lambda}_{\textrm{cop}}^{\top} \right)^{\top} \) denotes the vector of parameters of the approximate posterior distributions, \( \bs{\lambda}_d \) are the  parameters associated with the approximate posterior distribution \( q_d(\bs{\theta}_d; {\bs{\lambda}_d} )\) of the parameters \( \bs{\theta}_d \) in the \( d \)-th marginal model for $d=1,...,D$, and \( \bs{\lambda}_{\textrm{cop}} \) are the parameters governing the approximation \( q_{\textrm{cop}}(\thetacop; {\bs{\lambda}_{\textrm{cop}}}) \) of the copula parameters \( \thetacop  \).} 

N-IFM is carried out in two stages. 
In the first stage, we train a set of neural networks to approximate the posterior 
distributions of the marginal model parameters $\bs{\theta}_{d}$ for 
$d = 1,\ldots,D$. 
In the second stage, we train a separate neural network to approximate the posterior 
distribution of the copula parameters $\thetacop$, using 
the copula ``data'' $\bs{u}$ as the inputs and estimates of the marginal model parameters estimated from the first stage.

The proposed approach is designed for data with a fixed number of series \(D\) and a fixed time-series
window length \(T\). The neural network for the marginal model maps a
univariate time series of length \(T\) to an approximate posterior distribution for the marginal
parameters. Hence, once trained for a given marginal specification
and window length \(T\), the same neural network can be applied independently to any
number of marginal time series. This makes the neural network for the marginal model reusable when additional financial assets
are included, provided the marginal model and \(T\) remain unchanged. For the copula
stage, however, the neural network maps the copula data from \(D\) series to an approximate posterior distribution for the copula parameters.
Consequently, the copula network is specific to the chosen value of \(D\), as well
as to the copula family and factor dimension. Handling a variable dimension \(D\) would require a more sophisticated neural network architecture.

Sections \ref{sec:firststage} and \ref{sec:secondstage} discuss the neural approximation for the posterior distributions of marginal and copula parameters, respectively. Section \ref{sec:N-IFMalgorithm} gives the N-IFM algorithm.





\subsubsection{Neural approximation for marginal models\label{sec:firststage}}
This section discusses the neural approximation for the posterior distribution of the marginal model parameters. Consider the prior density $p(\thetad)$, and the likelihood for the $d$-th marginal model $p(\bs{y}_{(d)}\mid\thetad)$ where $\bs Y_{(d)} \equiv (Y_{d,1}, \dots, Y_{d,T})^\top $. Write the posterior density of the $d$-th marginal model as $p(\thetad\mid\bs{y}_{(d)})$, for $ d=1,...,D$. 
For a given $\bs y_{(d)}$, $\bs \lambda_d$ can be optimised by minimising the (forward)
Kullback-Leibler divergence between 
$p(\bs\theta_{d}\mid\bs{y}_{(d)})$ and $q_{d}\left(\bs\theta_{d};\bs\lambda_{d}\right)$, 
for $d=1,...,D$, as follows: 
\begin{equation}
\bs{\lambda}^*_{d} = \arg\min_{\bs\lambda_{d}} \, \textrm{KL}\left\{ p(\thetad \mid \bs{y}_{(d)}) \,\|\, q_d(\thetad; \bs\lambda_{d}) \right\},
\label{eq:ForwardKL}
\end{equation} 
for $d=1,...,D$. 
The optimisation problem in \eqref{eq:ForwardKL} is typically computationally expensive to solve for a single dataset \( \bs{y}_{(d)} \). 
When this inference procedure must be repeated for many different datasets (e.g., in validation and real-time forecasting), the computational burden becomes prohibitive. As a result, such approaches are often impractical in applications that demand fast, repeated inference. 

In neural amortised inference, we construct neural network models that map the observed data to the parameters defining the approximate posterior distribution of the model parameters. We replace  $q_d(\thetad; \bs\lambda_{d})$ with $q_d(\thetad; \bs\lambda_{d}(\bs{y}_{(d)};\bs\gamma_{d}))$, where $\bs\gamma_d$ are the neural network parameters. 
The problem of estimating the posterior distribution $p(\bs\theta_{d}\mid\bs{y}_{(d)})$
can be written as an optimisation problem of finding the optimal neural-network parameters, $\bs\gamma_{d}^{*}$, using the expected forward Kullback-Leibler divergence, as follows \citep{zammit2024neural}:
\begin{eqnarray}
\bs{\gamma}_{d}^{*} & = & \underset{\bs\gamma_{d}}{\textrm{arg min}}\quad
{\E_{\bs{Y}_{(d)}}} \left[\textrm{KL}\left(p(\bs\theta_{d}\mid\bs{y}_{(d)})||q_{d}\left(\bs\theta_{d};\bs\lambda_{d}(\bs{y}_{(d)};\bs\gamma_{d})\right)\right)\right]\\
 & = & \underset{\bs\gamma_{d}}{\textrm{arg min}}\int_{\Theta}\int_{\Y_{(d)}}\left[-\log q_{d}\left(\bs\theta_{d};\bs\lambda_{d}(\bs{y}_{(d)};\bs\gamma_{d})\right) \right] p_{{Y}_{d}}(\bs{y}_{(d)} \mid\bs\theta_{d})  p\left(\bs\theta_{d}\right)\textrm{d}\bs{y}_{(d)} \textrm{d}\bs\theta_d,\label{eq:LossFunction1}
\end{eqnarray}
where $\mathcal{Y}_{(d)}$ is the sample space of $\bs y_{(d)}$.
We approximate \eqref{eq:LossFunction1} with a Monte Carlo approximation
using $N$ sets of simulated parameters and corresponding datasets from our prior distribution and marginal models: 
$\left\{ \left\{ \bs\theta^{\left(n\right)}_{d},\bs{y}^{\left(n\right)}_{(d)}\right\} :n=1,...,N\right\} $.
The optimisation problem is therefore approximated as 
\begin{equation} \label{eqn:marg_opt}
\bs\gamma_{d}^{*}\approx \underset{\bs{\gamma}_{d}}{\textrm{arg min}}\frac{1}{N}\sum_{n=1}^{N}-\log q_{d}\left(\bs\theta^{(n)}_{d};\bs\lambda_{d}(\bs{y}^{(n)}_{(d)};\bs\gamma_{d})\right),
\end{equation}
where $\bs\theta^{\left(n\right)}_{d}\sim p\left(\bs\theta_{d}\right)$ and $\bs{y}_{(d)}^{\left(n\right)}\sim p(\bs{y}_{(d)}\mid\bs\theta^{(n)}_{d})$   for $d=1,...,D$.   

As posterior approximation for the (possibly transformed) marginal parameters, we use a Gaussian distribution of the form 
\begin{equation} \label{eqn:garch_variationalform}
q_{d}\left(\bs\theta_{d};\bs{\lambda}_{d}(\bs{y}_{(d)};\bs{\gamma}_{d})\right)=\N\left(\bs\mu_{d}\left(\bs{y}_{(d)};{\bs\gamma_{d}}\right), \bs\Sigma_{d}\left(\bs{y}_{(d)};{\bs\gamma_{d}}\right)\right),
\end{equation}
where $\bs\mu_{d}\left(\bs{y}_{(d)};{\bs\gamma_{d}}\right)$ is an $m_{d} \times 1$ mean vector and $\bs\Sigma_{d}\left(\bs{y}_{(d)};{\bs\gamma_{d}}\right)$
is an $m_{d}\times m_{d}$ covariance matrix of the approximate posterior, where $m_{d}$ is the number of parameters in the $d$-th marginal model. We reparameterise the covariance matrix 
\begin{equation} \label{eqn:cholesky_decom}
\bs\Sigma_{d}\left(\bs{y}_{(d)};{\bs\gamma_{d}}\right)=\bs{L}_{d}\left(\bs{y}_{(d)};{\bs\gamma_{d}}\right)\bs{L}_{d}\left(\bs{y}_{(d)};{\bs\gamma_{d}}\right)^{\top},
\end{equation}
where $\bs{L}_{d}\left(\bs{y}_{(d)};{\bs\gamma_{d}}\right)$ is an $m_{d}\times m_{d}$ lower triangular
Cholesky factor with positive diagonal elements. Once the neural network has been trained, given a new dataset, the trained neural network outputs the mean and covariance matrix of the approximate posterior distribution for the parameters of the $d$-th marginal model. In this case, ${\bs\lambda_d(\cdot; \bs\gamma_d) \equiv \left( \bs\mu_d(\cdot; \bs\gamma_d)^\top, \vech(\bs L_d(\cdot; \bs\gamma_d))^\top\right)}^\top$. 


In our applications, all marginal models are a GARCH(1,1) model (described in Section \ref{sec:marginalmodel}), so it is sufficient to train a single neural network ${\bs \lambda_{\textrm{mar}}(\cdot; \bs \gamma_{\textrm{mar}}) \equiv (\bs \mu_{\textrm{mar}}(\cdot; \bs \gamma_{\textrm{mar}})^\top, \vech(\bs L_{\textrm{mar}}(\cdot; \bs \gamma_{\textrm{mar}}))^\top)^\top}$, for all marginal time series, where \( \Gammamar\) are the neural network parameters.
We let $\boldsymbol{\theta}_{d} \equiv (\varphi_{1,d}, \varphi_{2,d}, \varphi_{3,d})^\top$ denote the vector of transformed GARCH(1,1)  parameters (see Section~\ref{supp:datagen} of the online supplement) and $\ymar$ denote the training observations generated from the corresponding GARCH(1,1) model, for $d = 1, \dots,D$.  

Algorithm \ref{alg:firststage} gives the steps needed to find optimal neural network parameters $\Gammamar^{*}$, which map the observations from each marginal time series $\bs{y}_{(d)}$ to the mean ${\bs\mu_{d}(\bs{y}_{(d)};\Gammamar^{*})}$ and Cholesky factor $\bs L_{d}(\bs{y}_{(d)};\Gammamar^{*})$ to obtain the approximate posterior distribution $q_{d}\left(\bs\theta_{d};\bs\lambda_{d}(\bs{y}_{(d)};\Gammamar^{*})\right)$,  for $d=1,\ldots,D$. The details of 
the algorithm to generate observations from the GARCH(1,1) model, and
the convolutional neural network (CNN) model used to map between observations $\bs{y}_{(d)}$ and $\bs \lambda_d \equiv \bs{\lambda}_{d}(\bs{y}_{(d)};\bs{\gamma}^{*}_{\textrm{mar}})$ 
are given in Sections {\ref{supp:datagen} and \ref{supp:cnn} of the online supplement, respectively.}


\begin{algorithm}[t!]
\caption{Marginal Posterior Approximation}
\label{alg:firststage}
\begin{algorithmic}[1]
\Require Prior density $p(\bs\theta_{\textrm{mar}}) \equiv p(\bs \theta_1) = \cdots = p(\bs \theta_D)$, likelihood ${p(\bs{y}_{\textrm{mar}} \mid \bs\theta_{\textrm{mar}}) \equiv  {p(\bs y_{(1)} \mid \bs \theta_1)}= \cdots = p(\bs y_{(D)} \mid \bs \theta_D)}$. 

\Statex

\State Draw $N$ Monte Carlo samples $\{ (\bs\theta_{\textrm{mar}}^{(n)}, \bs{y}_{\textrm{mar}}^{(n)}) \}_{n=1}^{N}$ with
\[
{\bs\theta_{\textrm{mar}}^{(n)} \sim p(\bs\theta_{\textrm{mar}}) }, \quad \bs{y}_{\textrm{mar}}^{(n)} \sim p(\bs{y}_{\textrm{mar}} \mid \bs\theta_{\textrm{mar}}^{(n)}).
\] 
\State Optimise $\Gammamar$ through
\[
\Gammamar^{*} = \arg\min_{\Gammamar} \, \frac{1}{N} \sum_{n=1}^{N} -\log q_{\textrm{mar}}\!\left(\bs\theta_{\textrm{mar}}^{(n)}; \Lambdamar(\bs{y}_{\textrm{mar}}^{(n)};\Gammamar)\right),
\]
where 
\[
q_{\textrm{mar}}(\bs\theta_{\textrm{mar}}; \Lambdamar(\bs{y}_{\textrm{mar}}^{(n)};\Gammamar)) = \N\!\bigl(\bs\mu_{\textrm{mar}}(\bs{y}_{\textrm{mar}}^{(n)};\Gammamar), \, \bs\Sigma_{\textrm{mar}}(\bs{y}_{\textrm{mar}}^{(n)};\Gammamar)\bigr),
\]
with covariance matrix
\[
\bs\Sigma_{\textrm{mar}}(\bs{y}_{\textrm{mar}}^{(n)};\Gammamar) = \bs{L}_{\textrm{mar}}(\bs{y}_{\textrm{mar}}^{(n)};\Gammamar) \bs{L}_{\textrm{mar}}(\bs{y}_{\textrm{mar}}^{(n)};\Gammamar)^{\top}.
\]

\Statex
\State \textbf{Output:} Trained neural network with parameters $\Gammamar^{*}$ to map the observations from each marginal series $\bs{y}_{(d)}$ to the mean ${\bs\mu_{d}(\bs{y}_{(d)};\Gammamar^{*})}$ and covariance matrix $\bs \Sigma_d(\bs{y}_{(d)};\bs \gamma_{\textrm{mar}}^{*})$ to obtain the approximate posterior distribution $q_{d}\left(\bs\theta_{d};\bs\lambda_{d}(\bs{y}_{(d)};\Gammamar^{*})\right)$, for $d=1,\ldots,D$.
\end{algorithmic}
\end{algorithm}

\begin{figure}[t!]
  \centering
  \tikzset{
    >={Latex[length=2.2mm]},
    line/.style={line width=0.9pt},
    nodeC/.style={circle, draw, minimum size=11mm, inner sep=0pt, font=\large},
    nodeR/.style={rectangle, rounded corners=2pt, draw, minimum height=11mm, minimum width=18mm, inner sep=2pt, font=\bfseries},
    bracebelow/.style={decorate, decoration={brace, amplitude=7pt, mirror}},
    labelsmall/.style={font=\small},
    annot/.style={font=\scriptsize, inner sep=1pt, fill=white, text opacity=1},
    dataClr/.style ={draw=black},
    netClr/.style  ={draw=black, fill=black!3},
    varClr/.style  ={draw=black},
    parClr/.style  ={draw=black},
    arrowClr/.style={draw=black},
    sampleClr/.style={draw=black}
  }
  \begin{tikzpicture}[node distance=2.6cm, baseline={(current bounding box.center)}]
    \node[nodeC, dataClr]   (data)   {$\bs{y_{(d)}}$};
    \node[nodeR, right=of data, netClr] (nn) {Neural Network};
    \node[nodeC, right=of nn, varClr]   (lambda) {$\bs{\lambda}_d$};
    \node[nodeC, right=of lambda, parClr] (theta)  {$\bs{\theta}_d$};
    \draw[line, arrowClr, ->] (data) -- (nn);
    \draw[line, arrowClr, ->] (nn) -- (lambda);
    \draw[line, sampleClr, dashed, ->] (lambda) -- node[above, annot]{sample} (theta);
    \draw[bracebelow] ([yshift=-6pt]data.south west) -- ([yshift=-6pt]data.south east)
      node[midway, below=10pt, labelsmall]{Data};

    \draw[bracebelow] ([yshift=-6pt]nn.south west) -- ([yshift=-6pt]nn.south east)
      node[midway, below=10pt, labelsmall, align=center]{Inference};

    \draw[bracebelow] ([yshift=-6pt]lambda.south west) -- ([yshift=-6pt]lambda.south east)
      node[midway, below=10pt, labelsmall, align=center]{Approx.\ Posterior Dist.\\ Parameters};

    \draw[bracebelow] ([yshift=-6pt]theta.south west) -- ([yshift=-6pt]theta.south east)
      node[midway, below=10pt, labelsmall, align=center]{Model\\Parameters};
  \end{tikzpicture}
  \caption{Inference framework for the marginal models in N-IFM.}
  \label{fig:inference_neural_bayes_marginals}
\end{figure}
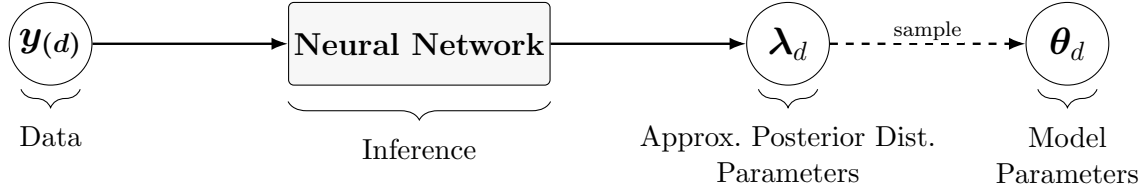

Figure~\ref{fig:inference_neural_bayes_marginals} summarises the 
inference framework for the marginal models. The observed marginal data 
$\bs{y}_{(d)}$ are first passed through a neural network, which outputs $\bs\lambda_{d}(\bs{y}_{(d)};\Gammamar^{*})$, the parameters of the approximate posterior 
distribution for the marginal model parameters. Conditional on 
$\bs\lambda_{d}(\bs{y}_{(d)};\Gammamar^{*})$, samples of the marginal parameters 
$\bs{\theta}_d$ are then drawn from this approximate posterior distribution. 


\subsubsection{Neural approximation for copula models\label{sec:secondstage}}




This section discusses the neural approximation for the posterior distribution of the copula parameters. Let $\thetacop$ be the vector of copula parameters of dimension $m_{\textrm{cop}}$, and
consider the prior density $p(\thetacop)$
and the joint copula density $c(\bs{u};\thetacop)$. For the Gaussian copula, ${\thetacop \equiv (\textrm{vech}(\bs {\widetilde G}))}$, while for the $t$ copula, $\thetacop \equiv (\textrm{vech}(\bs{\widetilde G})^{\top}, \nu)^{\top}$. The posterior density
is $p(\thetacop\mid\bs{u})$.
We aim to construct a flexible neural network architecture to map the copula ``data'' $\bs{u}$ to the parameters of the approximate posterior distribution of the copula parameters $q_{\textrm{cop}}(\thetacop;\bs{\lambda}_{\textrm{cop}}(\bs{u};{\bs\gamma_{\textrm{cop}}}))$, where $\bs\gamma_{\textrm{cop}}$ are the neural network parameters. Similar to the marginal model in Section \ref{sec:firststage}, the problem of estimating the posterior distribution 
can be written as an optimisation problem of finding the optimal neural-network parameters, $\bs\gamma_{\textrm{cop}}^{*}$, using the expected forward Kullback-Leibler divergence, as follows:
\begin{eqnarray}
\bs{\gamma}_{\textrm{cop}}^{*} & = & \underset{\bs\gamma_{\textrm{cop}}}{\textrm{arg min}}\quad \E_{U}\left[\textrm{KL}\left(p(\thetacop \mid \bs{u})\mid \mid q_{\textrm{cop}}\left(\thetacop;\bs{\lambda}_{\textrm{cop}}(\bs{u};{\bs\gamma_{\textrm{cop}}})\right)\right)\right]\\
 & = & \underset{\bs\gamma_{\textrm{cop}}}{\textrm{arg min}}\int_{\Theta}
 {\int_{[0,1]^{DT}}} \left[ -\log q_{\textrm{cop}}\left(\thetacop;\bs{\lambda}_{\textrm{cop}}(\bs{u};{\bs\gamma_{\textrm{cop}}})\right) \right] 
 \nonumber \\
 & & \quad\quad\quad\quad\quad\quad\quad\quad 
 {
 c(\bs{u}}\mid \thetacop)p\left(\thetacop\right)\textrm{d}\bs{u} \textrm{d}\thetacop, \label{eq:LossFunction}
\end{eqnarray}
{where $U$ is the sample space of $\bs u$}.
We approximate \eqref{eq:LossFunction} with a Monte Carlo approximation
using $N$ sets of simulated parameters and corresponding ``data'' simulated with standard uniform margins,
$\left\{ \left\{ \thetacop^{(n)},\bs{u}^{\left(n\right)}\right\} :n=1,...,N\right\} $.
The optimisation problem then becomes 
\begin{equation} \label{eqn:cop_opt}
\bs\gamma_{\textrm{cop}}^{*}=\underset{\bs{\gamma}_{\textrm{cop}}}{\textrm{arg min}}\frac{1}{N}\sum_{n=1}^{N}-\log q_{\textrm{cop}}\left(\thetacop^{(n)};\bs\lambda_{\textrm{cop}}(\bs{u}^{(n)};\bs\gamma_{\textrm{cop}})\right),
\end{equation}
where $\thetacop^{(n)} \sim p(\thetacop)$ and $\bs{u}^{\left(n\right)}\sim c(\bs{u}\mid \thetacop^{(n)})$. The algorithm to generate copula ``data'' for the Gaussian and $t$ copulas is given in Algorithm~\ref{alg:copula_sim} in Section \ref{supp:datagen} of the online supplement. 

We now discuss the posterior approximation for the copula parameters. 
We adopt a {mean-field} Gaussian approximation of the form
\begin{equation}\label{eqn:cop_variationalform_mf}
q_{\textrm{cop}}\!\left(\thetacop;\bs{\lambda}_{\textrm{cop}}(\bs{u};{\bs\gamma_{\textrm{cop}}})\right)
=\prod_{\ell=1}^{m_{\textrm{cop}}} 
\N\!\Big(
{\theta}_{\textrm{cop},\ell}\,\Big|\,
\mu_{\textrm{cop},\ell}\!\left(\bs{u};\bs{\gamma}_{\textrm{cop}}\right),\,
\sigma_{\textrm{cop}, \ell}^{2}\!\left(\bs{u};\bs{\gamma}_{\textrm{cop}}\right)
\Big),
\end{equation}
where $\bs \lambda _{\textrm{cop }}(\cdot; \bs \gamma_{\textrm{cop}}) \equiv \left(   \mu_{\textrm{cop},1}(\cdot; \bs \gamma_{\textrm{cop}}) , \dots, \mu_{\textrm{cop},m_{\textrm{cop}}}(\cdot; \bs \gamma_{\textrm{cop}}), \sigma^2_{\textrm{cop},1}(\cdot; \bs \gamma_{\textrm{cop}}), \dots,  \sigma^2_{\textrm{cop},m_{\textrm{cop}}}(\cdot; \bs \gamma_{\textrm{cop}})  \right)^{\top}, $ $\bs{u}\equiv(\bs{u}^{\top}_{1},\ldots,\bs{u}^{\top}_{T})^{\top}$ are the copula ``data'', $m_{\textrm{cop}}$ is the number of copula parameters, 
and $\bs{\gamma}_{\textrm{cop}}$ parameterises the neural network that outputs the 
marginal means and variances of the univariate Gaussian approximation for each copula parameter. 
The mean-field assumption enforces posterior independence among the copula parameters. 
{In Section~\ref{supp:gauscop_factor_q} of the online supplement, we also consider a Gaussian approximation with a 
{rank-one factor covariance}
for the Gaussian copula parameters, thereby extending~\eqref{eqn:cop_variationalform_mf}; however, the results in Section~\ref{supp:gauscop_factor_q} show that this extension does not materially improve the posterior approximation for the models we consider here.}
Algorithm \ref{alg:secondstage} gives the steps needed to find optimal neural network parameters $\bs{\gamma}_{\textrm{cop}}^{*}$, which map the copula ``data'', $\bs u$, to the mean ${\mu_{\textrm{cop}, \ell}(\bs{u};\bs{\gamma}_{\textrm{cop}}^{*})}$ and variances $\sigma^2_{\textrm{cop}, \ell}(\bs{u};\bs{\gamma}_{\textrm{cop}}^{*})$ for $\ell=1,\ldots,m_{\textrm{cop}}$, to obtain the approximate posterior distribution $q_{\textrm{cop}}(\bs\theta_{\textrm{cop}};\bs\lambda_{\textrm{cop}}(\bs{u};\bs{\gamma}_{\textrm{cop}}^{*}))$.

\begin{algorithm}[t]
\caption{Copula Posterior Approximation}
\label{alg:secondstage}
\begin{algorithmic}[1]
\Require Prior density $p(\thetacop)$, likelihood $c(\bs{u} \mid \thetacop)$.

\Statex

\State Draw $N$ Monte Carlo samples 
\[
\{ (\thetacop^{(n)}, \bs{u}^{(n)}) \}_{n=1}^{N}
\]
with
\[
\thetacop^{(n)} \sim p(\thetacop), 
\quad 
\bs{u}^{(n)} \sim c(\bs{u} \mid \thetacop^{(n)}) \quad \text{(see lines 1--12 of Algorithm~\ref{alg:copula_sim} of the online supplement}).
\] 
\State Optimise neural network parameters $\bs\gamma_{\textrm{cop}}$ through
\[
\bs\gamma_{\textrm{cop}}^{*} 
= \arg\min_{\bs\gamma_{\textrm{cop}}} 
\frac{1}{N} \sum_{n=1}^{N} 
- \log q_{\textrm{cop}}\!\left(
\thetacop^{(n)}; 
\bs\lambda_{\textrm{cop}}(\bs{u}^{(n)};\bs\gamma_{\textrm{cop}})
\right).
\]


\Statex
\State \textbf{Output:} Trained neural network, with parameters 
$\bs\gamma_{\textrm{cop}}^{*}$, to map copula ``data'' $\bs{u}$ 
to approximate posterior means $\mu_{\textrm{cop},\ell}(\bs{u};\bs\gamma^{*}_{\textrm{cop}})$ 
and variances $\sigma_{\textrm{cop}, \ell}^{2}(\bs{u};\bs\gamma^{*}_{\textrm{cop}})$ 
for $\ell=1,\ldots,m_{\textrm{cop}}$.
\end{algorithmic}
\end{algorithm}

Since $\bs u_1, \dots, \bs u_T$ are mutually independent, we employ the Deep Sets neural network architecture of \citet{zaheer_deepsets} to learn a flexible mapping from sets of copula ``data'' to copula parameters; see Section \ref{supp:deepsets} of the online supplement and also \cite{sainsbury2024likelihood} for further details. 

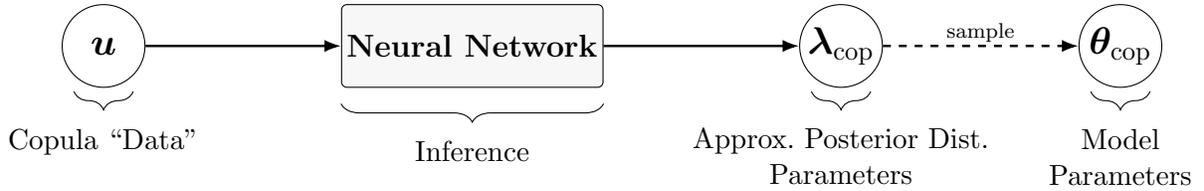
\begin{figure}[t]
  \centering
  \tikzset{
    >={Latex[length=2.2mm]},
    line/.style={line width=0.9pt},
    nodeC/.style={circle, draw, minimum size=11mm, inner sep=0pt, font=\large},
    nodeR/.style={rectangle, rounded corners=2pt, draw, minimum height=11mm, minimum width=18mm, inner sep=2pt, font=\bfseries},
    bracebelow/.style={decorate, decoration={brace, amplitude=7pt, mirror}},
    labelsmall/.style={font=\small},
    annot/.style={font=\scriptsize, inner sep=1pt, fill=white, text opacity=1},
    dataClr/.style ={draw=black},
    netClr/.style  ={draw=black, fill=black!3},
    varClr/.style  ={draw=black},
    parClr/.style  ={draw=black},
    arrowClr/.style={draw=black},
    sampleClr/.style={draw=black}
  }
  \begin{tikzpicture}[node distance=2.6cm, baseline={(current bounding box.center)}]
    \node[nodeC, dataClr]   (data)   {$\bs{u}$};
    \node[nodeR, right=of data, netClr] (nn) {Neural Network};
    \node[nodeC, right=of nn, varClr]   (lambda) {$\bs{\lambda_{\textrm{cop}}}$};
    \node[nodeC, right=of lambda, parClr] (theta)  {$\thetacop$};
    \draw[line, arrowClr, ->] (data) -- (nn);
    \draw[line, arrowClr, ->] (nn) -- (lambda);
    \draw[line, sampleClr, dashed, ->] (lambda) -- node[above, annot]{sample} (theta);
    \draw[bracebelow] ([yshift=-6pt]data.south west) -- ([yshift=-6pt]data.south east)
      node[midway, below=10pt, labelsmall]{Copula ``Data''};

    \draw[bracebelow] ([yshift=-6pt]nn.south west) -- ([yshift=-6pt]nn.south east)
      node[midway, below=10pt, labelsmall, align=center]{Inference};

    \draw[bracebelow] ([yshift=-6pt]lambda.south west) -- ([yshift=-6pt]lambda.south east)
      node[midway, below=10pt, labelsmall, align=center]{Approx.\ Posterior Dist.\\ Parameters};

    \draw[bracebelow] ([yshift=-6pt]theta.south west) -- ([yshift=-6pt]theta.south east)
      node[midway, below=10pt, labelsmall, align=center]{Model\\Parameters};
  \end{tikzpicture}
  \caption{Inference framework for the copula model in N-IFM.}
  \label{fig:inference_neural_bayes_copula}
\end{figure}

Figure~\ref{fig:inference_neural_bayes_copula} illustrates the neural inference framework for the copula model. The copula ``data'' $\bs{u}$ are 
first fed into a neural network. This network outputs $\bs\lambda_{\textrm{cop}}(\bs{u};\bs\gamma^{*}_{\textrm{cop}})$, the 
parameters of the approximate posterior distribution for the copula parameters. 
Conditional on $\bs\lambda_{\textrm{cop}}(\bs{u};\bs\gamma^{*}_{\textrm{cop}})$, we then draw samples of the 
copula parameters $\thetacop$ from this approximate 
posterior distribution.

\subsubsection{Neural inference functions for margins (N-IFM) algorithm\label{sec:N-IFMalgorithm}}

Given the trained neural network parameters $\Gammamar^{*}$ for the marginal models and the observed data $\boldsymbol{y}_{(d)}$ as input, the ``marginal'' neural-network outputs the parameters of the approximate posterior distribution of the marginal model parameters, denoted by $\boldsymbol{\lambda}_{d}(\boldsymbol{y}_{(d)}; \Gammamar^{*})$, for $d = 1, \ldots, D$. From each approximate posterior $q_d(\boldsymbol{\theta}_d; \boldsymbol{\lambda}_{d}(\boldsymbol{y}_{(d)};\Gammamar^{*}))$, we then draw $J = 1000$ samples and compute the posterior mean estimates $\widehat{\boldsymbol{\theta}}_d$ for each marginal model.  Using these posterior mean estimates, we then construct the copula ``data'' $\boldsymbol{u}$, which serve as the input to the neural network for the copula model. Finally, given the trained copula-network parameters $\bs{\gamma}_{\textrm{cop}}^{*}$ and the copula ``data'' as input, 
the network outputs the parameters of the approximate posterior distributions of the copula parameters, 
$\bs{\lambda}_{\textrm{cop}}(\bs{u}; \bs{\gamma}^{*}_{\textrm{cop}})$ to obtain the approximate posterior $q_{\textrm{cop}}\!\left(
\thetacop;
\bs\lambda_{\textrm{cop}}(\bs{u}; \bs\gamma^{*}_{\textrm{cop}})
\right)$.  

We write the full approximate posterior distribution based on N-IFM as:
\begin{equation}
q(\bs\theta; \bs{\lambda}(\bs{y};\bs{\gamma}^{*}))
=
q_{\textrm{cop}}\!\left(
\thetacop;
\bs\lambda_{\textrm{cop}}(\bs{u}; \bs\gamma^{*}_{\textrm{cop}})
\right)
\prod_{d=1}^{D}
q_d(\boldsymbol{\theta}_d; \boldsymbol{\lambda}_{d}(\boldsymbol{y}_{(d)};\Gammamar^{*})),
\label{eq:approx-ifm-amortised}
\end{equation}
where $\bs{\gamma}^* \equiv
(\bs{\gamma}_{\mathrm{mar}}^{*\top},
 \bs{\gamma}_{\mathrm{cop}}^{*\top})^{\top}$ is the trained neural network parameters and the copula ``data'' $\bs{u} \equiv (\bs{u}_{1}^\top,...,\bs{u}_{T}^\top)^{\top}$ where $\bs{u}_{t}\equiv(u_{1,t},...,u_{D,t})^{\top}$, and $u_{d,t} = F_{Y_{d,t}}(y_{d,t};\widehat{\bs{\theta}}_d)$. Algorithm \ref{alg:neural_inference_margins} describes the proposed neural inference functions for margins. 


{A fully joint amortised estimator could be trained by simulating datasets from the joint prior predictive distribution and learning a direct mapping from the multivariate time series to the posterior of all marginal and copula parameters. We do not adopt this strategy because, even in relatively low-dimensional settings, the network would need to learn a very complex mapping: it would need to simultaneously learn the marginal time-series dynamics, the transformation to copula scale, and the cross-sectional dependence structure. As the number of series increases, the dimension of the joint parameter vector also grows, further increasing the complexity of the learning problem and the neural architecture required.}

{The proposed N-IFM approach is instead a modular approximation motivated by the classical IFM estimator. By decomposing inference into marginal and dependence components, the marginal neural network only needs to learn the posterior distribution of the parameters of a univariate time-series model, while the copula neural network only needs to learn the posterior distribution of the dependence parameters from the transformed copula data. 
It also provides additional flexibility: the neural network for the marginal model trained for a given marginal specification can be reused across all series. 
} 

{This modularity is especially useful in the repeated-inference settings considered in this paper, such as rolling-window prediction and time-series validation, where the same trained neural estimators are applied many times for different combinations of marginal and copula models. Although the resulting approximation targets an IFM-style posterior rather than the exact fully joint posterior, it preserves the main inferential separation used in classical copula estimation and greatly reduces training and computational cost. Marginal uncertainty can still be propagated into predictive inference through draws from the marginal posterior approximations, and the empirical results below show that the resulting posterior and predictive distributions are close to those obtained from HMC and HMC-IFM, while being much faster to compute.}

\begin{algorithm}[t]
\caption{Neural Inference Functions for Margins}
\label{alg:neural_inference_margins}
\begin{algorithmic}[1]
\Require Trained marginal neural network parameters $\Gammamar^{*}$, trained copula neural network parameters $\bs\gamma_{\textrm{cop}}^{*}$, data $\{\bs y_{(d)}\}_{d=1}^D$.

    \For{$d = 1$ to $D$}
        \State{{Evaluate the trained marginal CNN to obtain $\bs\lambda_{d}(\bs y_{(d)};\Gammamar^{*})$}.
        }

        \State {Compute posterior means of marginal parameters:}
        \[
        \widehat{\bs\theta}_d = \E_{q_d(\bs\theta_d;\bs\lambda_{d}(\bs y_{(d)};\Gammamar^{*}))}[\bs\theta_d] ,\quad 
        \quad
        \]
        \State Compute copula ``data'' $\bs u$ through:
        \[
        u_{d,t} = F_{Y_{d,t}}(y_{d,t};\widehat{\bs\theta}_d),\quad  
        \text{for~~} t=1,...,T.
        \]
    \EndFor
    \State {Compute copula posterior parameters using Algorithm \ref{alg:secondstage} with Deep Sets architecture: $\bs\lambda_{\textrm{cop}}(\bs u;\bs{\gamma}^{*}_{\textrm{cop}})$.}
    
    \State Output: $\{   \bs\lambda_{d}(\bs y_{(d)};\Gammamar^{*})   \}_{d=1}^D$ and $\bs\lambda_{\textrm{cop}}(\bs u;\bs{\gamma}^{*}_{\textrm{cop}})$.
\end{algorithmic}
\end{algorithm}

\section{Measure of predictive performance \label{sec:predictiveperformance}}
This section outlines the performance criteria used to evaluate and compare the predictive accuracy of the competing models. Section \ref{sec:preddensity} discusses the approximate predictive density. Section \ref{sec:timeseriescv} discusses the time-series validation.

\subsection{Approximate predictive density \label{sec:preddensity}}
Let $\bs{Y}_{T+h}$ denote the unobserved value of the variable $\bs{Y}$ at time $T+h$, where $h>0$.
The $h$-step-ahead  predictive density of $\bs{Y}_{T+h}=\bs{y}_{T+h}$ is
\begin{equation}
\label{eq:pred_true}
p_{{}}(\bs{y}_{T+h}\mid \bs{y})
=\int_{
\Theta} p_{{}}(\bs{y}_{T+h}\mid {\bs y,} \bs\theta)
\, p(\bs\theta\mid \bs{y})
\,  \textrm{d}\bs\theta,
\end{equation} 
where 
$\Theta$ is the parameter space of 
$\bs \theta$ and 
$p(\boldsymbol{\theta}\mid\boldsymbol{y})$ denotes the posterior distribution given the data up to time $T$ in {\eqref{eq:fullposterior}}.
A  consistent Monte Carlo estimate of
\eqref{eq:pred_true} is
\begin{equation}
\label{eq:pred_mc_true}
\widehat{p}_{}(\bs{y}_{T+h}\mid \bs{y})
=\frac{1}{J}\sum_{j=1}^J p_{ }\!\left(\bs{y}_{T+h}\mid \bs{ {\bs y,}\theta}^{(j)}\right), \quad \bs \theta^{(j)} \sim p(\bs \theta \mid \bs y).
\end{equation} 
When evaluated at the observed value $\bs{y}_{T+h}=\bs{y}_{T+h}^{\textrm{obs}}$, \eqref{eq:pred_mc_true} yields the $h$-step-ahead
predictive density $ \widehat{p}_{}(\bs{y}_{T+h}^{\textrm{obs}}\mid \bs{y})$.
One may also
simulate $\bs{y}_{T+h}^{(j)} \sim p_{}(\,\cdot\,\mid {\bs y,} \bs\theta^{(j)})$, for $j=1,...,J$ and form a kernel
density estimate of $p_{}(\bs{y}_{T+h}\mid \bs{y})$.


For high-dimensional time-series copula models, obtaining the exact posterior
$p(\bs\theta\mid \bs{y})$ can be computationally expensive. 
By applying \eqref{eq:approx-ifm-amortised}, we obtain the corresponding approximate $h$-step-ahead predictive density
\begin{equation}
\label{eq:pred_vb_def}
g(\bs{y}_{T+h}\mid \bs{y})
=\int_{
\Theta} p_{}(\bs{y}_{T+h}\mid {\bs y,}\bs\theta)
\, q(\bs\theta; \bs{\lambda}(\bs{y};\bs{\gamma}^{*}))
\, \textrm{d}\bs\theta.
\end{equation}
We approximate  
\eqref{eq:pred_vb_def} by
\begin{equation}
\label{eq:pred_mc_vb}
\widehat{g}(\bs{y}_{T+h}\mid \bs{y})
=\frac{1}{J}\sum_{j=1}^J
p_{}\!\left(\bs{y}_{T+h}\mid {\bs y,}\bs\theta^{(j)}\right), \quad \bs \theta^{(j)} \sim q(\bs \theta ; \bs \lambda(\bs y;\bs \gamma^*)).
\end{equation}
Evaluating \eqref{eq:pred_mc_vb} at the observed value $\bs y_{T+h} =\bs{y}_{T+h}^{\textrm{obs}}$ yields the approximate predictive density 
$\widehat{g}(\bs{y}_{T+h}^{\textrm{obs}}\mid \bs{y})$.
Simulations of $\bs y_{T+h}^{(j)} \sim p(\cdot \mid {\bs y,}\bs\theta^{(j)})$, for $j = 1,\dots,J$, can also be used to construct a kernel density estimate of $g(\bs{y}_{T+h}\mid \bs{y})$.


\subsection{Time series validation \label{sec:timeseriescv}}
We assess out-of-sample performance using time-series validation with a fixed-width rolling window of length $T$. The neural networks we use in N-IFM
are trained to map datasets of a fixed size to posterior approximations. A rolling-window
scheme keeps the input dimension fixed as new observations arrive, allowing the same
pre-trained networks to be reused at each forecast update. 

Assume that we have a time series of length $T+K$. 
{For $i=0,\dots,K-h$, we fit the model on the window $\bs y_{1+i:T+i}\equiv(\bs y_{1+i}^\top,\dots,\bs y_{T+i}^\top)^\top$ and produce an $h$-step-ahead forecast for $\bs y_{T+h+i}$. Equations \eqref{eq:pred_true}--\eqref{eq:pred_mc_vb} are for the window ending at time $T$; for window $i$, $\bs y$ is replaced by $\bs y_{1+i:T+i}$. Given $J$ posterior draws $\{\bs\theta^{(i,j)}\}_{j=1}^{J}$ from the approximate posterior based on $\bs y_{1+i:T+i}$, we compute the $h$-step-ahead predictive density $p(\bs y^{\mathrm{obs}}_{T+h+i}\mid \bs y_{1+i:T+i},\bs\theta^{(i,j)})$.}
The $h$-step-ahead log-predictive density score (LPDS) is the sum of the log of the $h$-step posterior predictive density score over the $K-h+1$ rolls:
\begin{equation}
\label{eq:LPDS}
    \textrm{LPDS}
    \;=\;
    \sum_{i=0}^{K-h}
    \log\!\left\{
        \frac{1}{J}\sum_{j=1}^{J}
        p_{}\!\left(\bs y^{\textrm{obs}}_{T+h+i}\,\middle|\, {\bs y_{1+i:T+i}, }\bs\theta^{(i,j)}\right)
    \right\}.
\end{equation}
In all experiments we set $J=1000$. 
Larger LPDS values are indicative of better predictive performance \citep{gelman2014understanding}.

\section{Simulation Study \label{sec:SimulationStudy}}

This section details a simulation study designed to assess the proposed N-IFM method for data simulated from a
Gaussian copula model. A second simulation study, based on a {$t$-copula}, is provided
in Section \ref{sec:sim_stcop} of the {online supplement}.
We compare N-IFM to Hamiltonian Monte Carlo (HMC), specifically
the No-U-Turn Sampler (NUTS) of \citet{hoffman2014no}, as implemented in \texttt{Stan}
\citep{carpenter2017stan}. The posterior distributions obtained from HMC are treated as ground truth.
In addition, we consider {HMC-IFM}, which mirrors the two-step
structure of N-IFM. HMC-IFM first fits each marginal time series separately to obtain posterior draws and the
corresponding posterior mean estimates of the marginal parameters; conditional on these marginal
estimates, it then estimates the copula parameters.
We assess agreement across methods by comparing posterior densities of both marginal and copula
parameters. We also conduct the time-series validation procedure described in
Section~\ref{sec:timeseriescv} using the N-IFM, HMC and HMC-IFM methods, to highlight the computational
advantages of N-IFM in repeated, time-consuming predictive exercises.



For all examples, we run the HMC {and HMC-IFM algorithms on two cores} with 2 chains of $4000$ iterations each, discarding the first $2000$ iterations of each chain as burn-in, unless stated otherwise. These settings were chosen to target a minimum effective sample size (ESS) of approximately $1000$ for all parameters. 
Training and validation of N-IFM was carried out on a high-end server, with an
NVIDIA Tesla A100 80GB graphics processing unit.

Section~\ref{sec:Priors} describes the prior distributions specified for the marginal and copula parameters. 
Section~\ref{sec:simdatagen} outlines the procedure used to generate the training data, while 
Section~\ref{sec:trainingneuralnetwork} discusses the training of the neural network models. 
Section~\ref{sec:sim_gauscop} presents a simulation study based on a Gaussian copula with GARCH(1,1) 
marginals with Gaussian errors. {Section~\ref{sec:calib_gauscop} presents a calibration study for the Gaussian copula with GARCH(1,1) 
marginals with Gaussian errors, assessing the frequentist properties of the N-IFM posteriors for one to four factors.}

\subsection{Simulating parameters from priors}
\label{sec:Priors}

Training the N-IFM models requires generating large collections of synthetic datasets from the GARCH(1,1) and copula models. This is accomplished by first sampling model parameters from their prior distributions and then simulating data conditional on these parameters. 

Certain model parameters are subject to validity constraints enforced
through the choice of prior distributions. For the GARCH(1,1) parameters,
the prior is specified subject to the constraint $\alpha_{1,d} + \alpha_{2,d} < 1$ for $d = 1, \dots, D$, 
to ensure stationarity.
Similarly, in the $t$ copula model and in the GARCH(1,1) model with Student's-$t$ errors, the degrees-of-freedom parameters are assigned gamma priors subject to the constraints  $\nu >2$ and $\widetilde{\nu}_d > 2$ for $d = 1, \dots, D$, to ensure finite variance. 
Table~\ref{tab:parameters} in Section~\ref{supp:datagen} of the online supplement summarises the prior distributions assigned to all model parameters, and provides a detailed discussion of the associated prior choices.
Since neural-network outputs are unconstrained, we apply suitable transformations to map the parameters to the real line; these transformed parameters are then used to define ${\bs \theta_{\text{mar}}}$ and  $\bs \theta_{\text{cop}}$. The specific transformations used are described in Section~\ref{supp:datagen} of the online supplement.
The resulting collection of parameter draws forms the basis for generating synthetic datasets used to train the neural networks in both stages of N-IFM, which we discuss next.

\subsection{Generation of simulated data} \label{sec:simdatagen}

For the marginal time series, we consider two alternative specifications: a
GARCH$(1,1)$ model with Gaussian errors and a GARCH$(1,1)$ model with Student's-$t$
errors. 
In Section~\ref{sec:trainingneuralnetwork} we discuss our procedure for training the neural networks in iterative steps, where each iterative step is analogous to the usual training epoch, but where the data are re-simulated at the beginning of each step (also known as `simulation-on-the-fly'; \cite{Chan2018}).
To generate the training datasets from GARCH(1,1) with Gaussian errors, we first generate
$N = 30,000$ independent parameter draws per 
{iterative step}
from the prior distributions described in
Section~\ref{sec:Priors}. Specifically, for $n = 1, \ldots, N$, we sample ${\bs{ \theta}_{\text{mar}}^{(n)} \equiv ({\varphi}_{1,\textrm{mar}}^{(n)}, {\varphi}_{2,\textrm{mar}}^{(n)}, {\varphi}_{3, \textrm{mar}}^{(n)})^\top}$, construct the GARCH parameters
$\alpha_{1, \textrm{mar}}^{(n)}$, $\alpha_{2, \textrm{mar}}^{(n)}$ and $\gamma_{\textrm{mar}}^{(n)}$, and use these to simulate a synthetic time
series $\bs{y}_{\textrm{mar}}^{(n)}$, 
which is used to train the  neural-networks for the marginal models. 
A similar process is used for GARCH(1,1) with Student's-$t$ errors, but an additional degrees-of-freedom parameter 
$\widetilde \nu_{\textrm{mar}}$
is required to simulate  $\bs{y}_{\textrm{mar}}^{(n)}$. The number of 
{iterative steps}
we use, and hence the total number of draws we need, is determined during training, as detailed in Section \ref{sec:trainingneuralnetwork}.

For the Gaussian copula model, we simulate $N=30,000$ draws per 
{iterative step}
for each non-zero element in factor loading matrix $\widetilde{\bs{G}}^{(n)}$ for $n = 1, \ldots, N$. Each factor loading matrix $\widetilde{\bs{G}}^{(n)}$ is transformed into $\bs G^{(n)}$ and then into a correlation matrix
$\bar{\bs{\Omega}}^{(n)}$ according to \eqref{eq:Omega}. For each $\bar{\bs{\Omega}}^{(n)}$, we generate $1000$ independent realisation $\bs{z}^{(n)}$  from multivariate normal distribution with a zero mean and a covariance matrix $\bar{\bs{\Omega}}^{(n)}$.
The variable $\bs{z}^{(n)}$ is then transformed to copula ``data'' $\bs{u}^{(n)}$ using the appropriate marginal CDF; see Section \ref{supp:datagen} of the online supplement for further details.
The $\bs{u}^{(n)}$ are used to train
the neural networks for the copula models. 
Note that for the copula network we do not need data from the marginal models, only uniform variables constructed via $\bs Z$ with the correct dependence.

\subsection{Training the neural networks\label{sec:trainingneuralnetwork}}

{We train a separate network for each of the two marginal specifications and for each combination of copula family (Gaussian, $t$) and factor dimension (one to four).}
For the convolutional neural network used to estimate the GARCH(1,1) model with Gaussian errors, the inputs consist of the simulated datasets $\bs{y}_{\textrm{mar}}^{(n)}$ for $n=1,..,N$ generated in Section~\ref{sec:simdatagen}. The target outputs are the transformed GARCH(1,1) parameters:
{$\bs \theta_{\textrm{mar}}^{(n)} \equiv ({\varphi}_{1,\textrm{mar}}^{(n)}, {\varphi}_{2,\textrm{mar}}^{(n)}, {\varphi}_{3, \textrm{mar}}^{(n)})^\top$}, which are obtained by transforming
${\alpha}_{1,\textrm{mar}}^{(n)}$, ${\alpha}_{2,\textrm{mar}}^{(n)}$ and ${\gamma}_{\textrm{mar}}^{(n)}$ as described in
Section~\ref{supp:datagen} of the online supplement. For the GARCH(1,1) model with Student's-$t$ errors, the outputs of the neural network include the transformed degrees-of-freedom parameter as described in Section~\ref{supp:datagen} of the online supplement.

{At each iterative step, the newly simulated datasets are split into training and validation sets, with 90\% of the datasets used for training and the remaining 10\% used to compute the validation loss. 
}
{An example of the training and validation loss curves is shown in Figure \ref{suppfig:gauscopd20f1_loss} in Section~\ref{sec:nifm-training} of the online supplement.} During training, the data are
processed in mini-batches of size $32$. Optimisation is performed using the Adam stochastic gradient
descent algorithm \citep{Kingma2015}, with gradients of the objective function computed via backpropagation {and a learning rate of $9 \times 10^{-5}$}. The optimisation proceeds until either the validation loss does not decrease for 100 consecutive 
{iterative steps}
or a maximum of 4000 
{iterative steps}
is reached, 
{where one iterative step involves training using 30,000 samples with a batch size of 32}. 
{Because a fresh set of datasets is simulated each iterative step and then discarded, no data are reused, and the number of iterative steps reflects the volume of simulated data seen in training.}

An analogous training procedure is employed for the neural network models to estimate the Gaussian and $t$ copulas. In this case, 
the training inputs consist of the copula ``data''
$\bs{u}^{(n)}$ and their associated parameters {$\textrm{vech}(\widetilde{\bs{G}})^{(n)}$} for $n=1,\ldots,N$ and degrees-of-freedom parameter $\nu$ for the $t$ copula. The transformations required for 
the copula parameters are discussed in detail in Section \ref{supp:datagen} of the online supplement. 

{
Tables \ref{supptab:garch_arch} and \ref{supptab:copula_arch} present the neural-network architectures used for the GARCH(1,1) marginal and copula models, respectively. We selected these architectures by comparing validation losses across several candidate configurations. The networks are similar in size to those used by \citet{sainsbury2024likelihood}, although their study considers models with fewer parameters to be inferred.
}



\begin{table}[t]
\caption{Neural-network training time and training-sample count for the marginal and copula N-IFM models, and the amortisation payback against HMC-IFM and HMC fits. \emph{Payback} is the number of data points HMC {or HMC-IFM} fits whose combined run time equals the one-off cost of training the N-IFM networks (marginal $+$ copula) and inference, for Gaussian copula with GARCH(1,1) marginals with Gaussian errors (Gaussian copula with GARCH marginals with Student's-$t$-errors in parentheses). 
}
    \centering
    \begin{tabular}{lcccc}
    \hline \hline
        Models & \shortstack{Computing\\Time (s)} & \shortstack{Training\\ Samples} & \shortstack{Payback\\HMC-IFM\\Gauss.\ ($t$)} & \shortstack{Payback\\HMC\\Gauss.\ ($t$)} \\
        \hline
        GARCH(1,1) w/ Gaussian errors & 3,500 & 10,320,000 & - & - \\
        GARCH(1,1) w/ Student's-$t$ errors & 15,000 & 10,020,000 & - & - \\
        Gaussian copula one-factor & 8,500 & 28,470,000 & 20 (43) & 10 (3) \\
        Gaussian copula two-factors & 17,000 & 35,250,000 & 17 (26) & 11 (3) \\
        Gaussian copula three-factors & 24,000 & 40,800,000 & 16 (22) & 12 (4) \\
        Gaussian copula four-factors$^*$ & 33,000 & 56,700,000 & 2 (3) & 2 (1) \\
        $t$ copula one-factor & 52,000 & 35,010,000 & 8 (12) & 4 (3) \\
        $t$ copula two-factors & 48,500 & 30,720,000 & 4 (5) & 2 (3) \\
        $t$ copula three-factors & 86,500 & 50,940,000 & 4 (5) & 4 (3) \\
        $t$ copula four-factors$^*$ & 97,500 & 44,910,000 & 1 (1) & 1 (1) \\
        \hline
    \end{tabular}
    \vspace{4pt}
\noindent\parbox{\linewidth}{\footnotesize $^*$The four-factor HMC and HMC-IFM models require longer Markov chains (20,000 iterations) due to poor mixing and a low effective sample size.}
    \label{tab:amortisedtime}
\end{table}


{Training the N-IFM networks is a one-off cost; the trained networks are then reused for all subsequent
inference at virtually no additional cost. Table~\ref{tab:amortisedtime} reports this training cost, which ranges from approximately \(1\) hour for the GARCH\((1,1)\) models with Gaussian errors to about \(27\) hours for the four-factor \(t\) copula model on our hardware. The \(t\) copula and Student's-\(t\) GARCH\((1,1)\) marginal models each take more time to train than their Gaussian counterparts,  because 
their training datasets are more involved to generate; this simulation cost could be dramatically reduced with parallelisation.}

{The payback columns of Table~\ref{tab:amortisedtime} report the number of data points that the corresponding benchmark method, HMC-IFM or HMC, could fit sequentially during the time required to train the N-IFM networks. For example, the four-factor $t$ copula with GARCH(1,1) marginals with Gaussian errors has a one-off training cost of \(97{,}500 + 3{,}500 = 101{,}000\) s (copula plus marginal network), which {matches the combined run time of about $1$} HMC-IFM or $1$ HMC fits. Values in parentheses correspond to the same copula with GARCH(1,1) marginals with Student's-$t$
errors.
The N-IFM computation times reported in the other tables are post-training inference times. They represent the cost of applying the trained networks to new rolling-window datasets and do not include the upfront training cost.}


\subsection{Simulation study: Gaussian copula} \label{sec:sim_gauscop}

We conduct a simulation study to compare the accuracy of the N-IFM method to that of the HMC and HMC-IFM methods. 
The simulated data are generated from a one-factor Gaussian copula model with GARCH(1,1) 
marginals with Gaussian errors, comprising $D = 20$ series of length $T = 1000$, following Algorithm \ref{alg:copula_sim} in Section \ref{supp:datagen} of the online supplement. 

We obtain the parameters of the approximate posterior distributions of marginal parameters $\bs\lambda_d$ for $d=1,...,D$ and copula parameters 
$\bs\lambda_{\textrm{cop}}$ using Algorithm~\ref{alg:neural_inference_margins} in Section \ref{sec:N-IFMalgorithm}. Then, we generate posterior samples of the marginal parameters $\bs\theta_d^{(j)}$ for $d=1,...,D$ and the copula parameters $\thetacop^{(j)}$, for $j = 1, \dots, 1000$.

\begin{figure}[t]
    \centering
    \includegraphics[width=\linewidth]{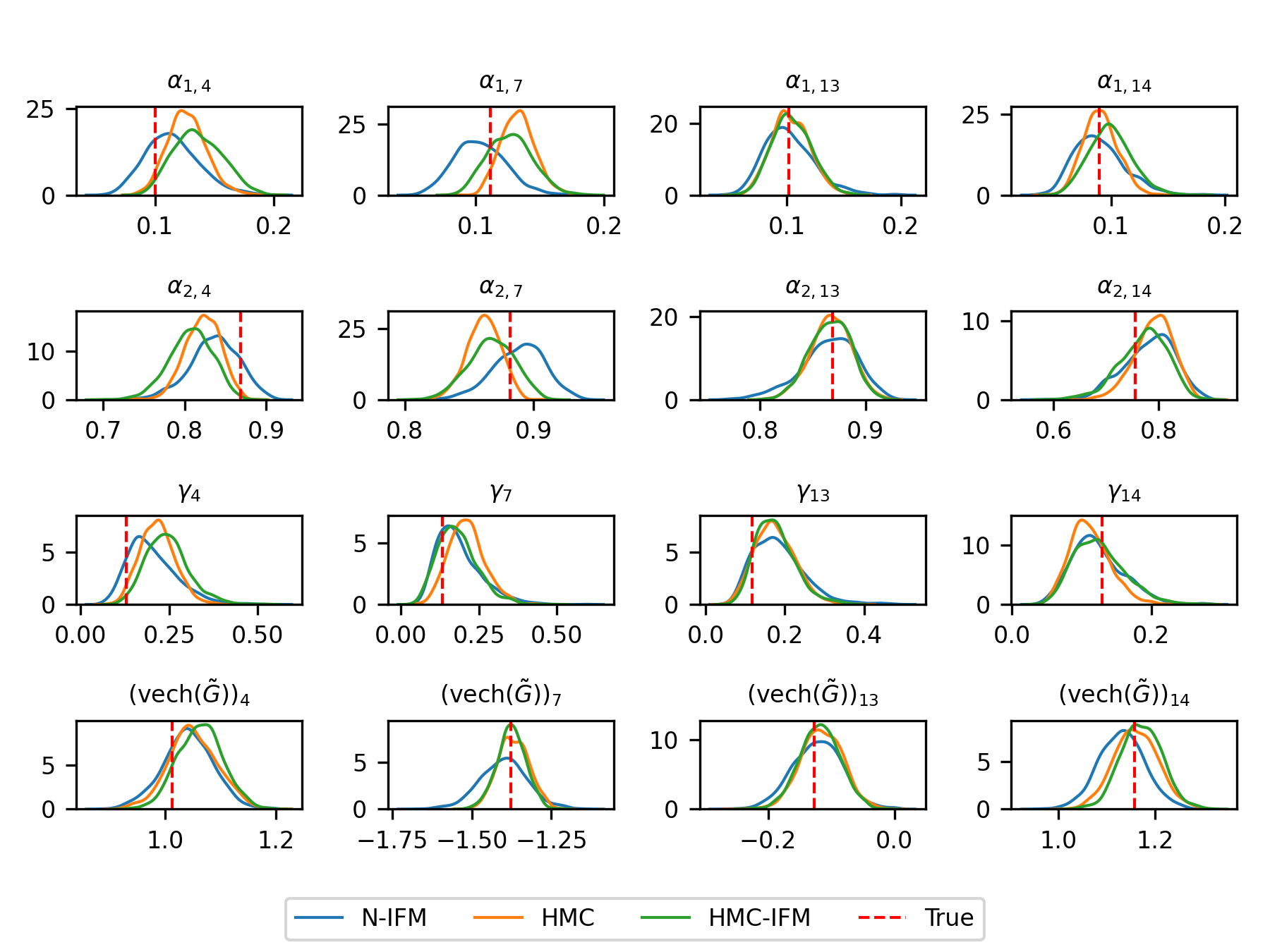}
    \caption{Posterior distributions of selected GARCH(1,1) parameters and Gaussian copula parameters estimated using the N-IFM, HMC and HMC-IFM methods for simulated data generated from a one-factor Gaussian copula model with GARCH(1,1) marginals with Gaussian errors, comprising $D = 20$ series of length $T = 1000$. Results from N-IFM, HMC and HMC-IFM are compared with the true parameter values, indicated by the vertical dashed lines.}
    \label{fig:gauscop_posterior_comparison}
\end{figure} 


Figure~\ref{fig:gauscop_posterior_comparison} 
shows posterior distributions of selected GARCH(1,1) parameters and Gaussian copula parameters estimated using the N-IFM, HMC and HMC-IFM methods for 
a one-factor Gaussian copula model with GARCH(1,1) marginals with Gaussian errors, comprising $D = 20$ series of length $T = 1000$. The figure shows the posterior distributions of model parameters obtained using the N-IFM method closely matching the HMC and HMC-IFM posteriors, with only minor discrepancies.

Figures~\ref{suppfig:alpha1_gauscop_f1} to \ref{suppfig:gtilde_gauscop_f1} in
Section~\ref{supp:hmc_vs_hmcifm} of the online supplement present the posterior distributions
of all model parameters obtained using N-IFM, HMC, and HMC-IFM. Across all parameters, the
posterior distributions produced by N-IFM and HMC-IFM closely align with those obtained from HMC. This strong agreement indicates that {for this experiment} the two-stage N-IFM and HMC-IFM approaches provide 
accurate approximations to the full joint posterior distribution of the model parameters. Table~\ref{tab:hmc_gaucsop_simf1} reports the effective sample sizes (ESS) and the potential scale reduction factors ($\widehat{R}$) for all model parameters estimated using HMC. The large ESS values and $\widehat{R}$ close to~1 indicate good mixing across chains and provide evidence of satisfactory Markov chain convergence.

\begin{figure}[t]
    \centering
    \includegraphics[width=\linewidth]{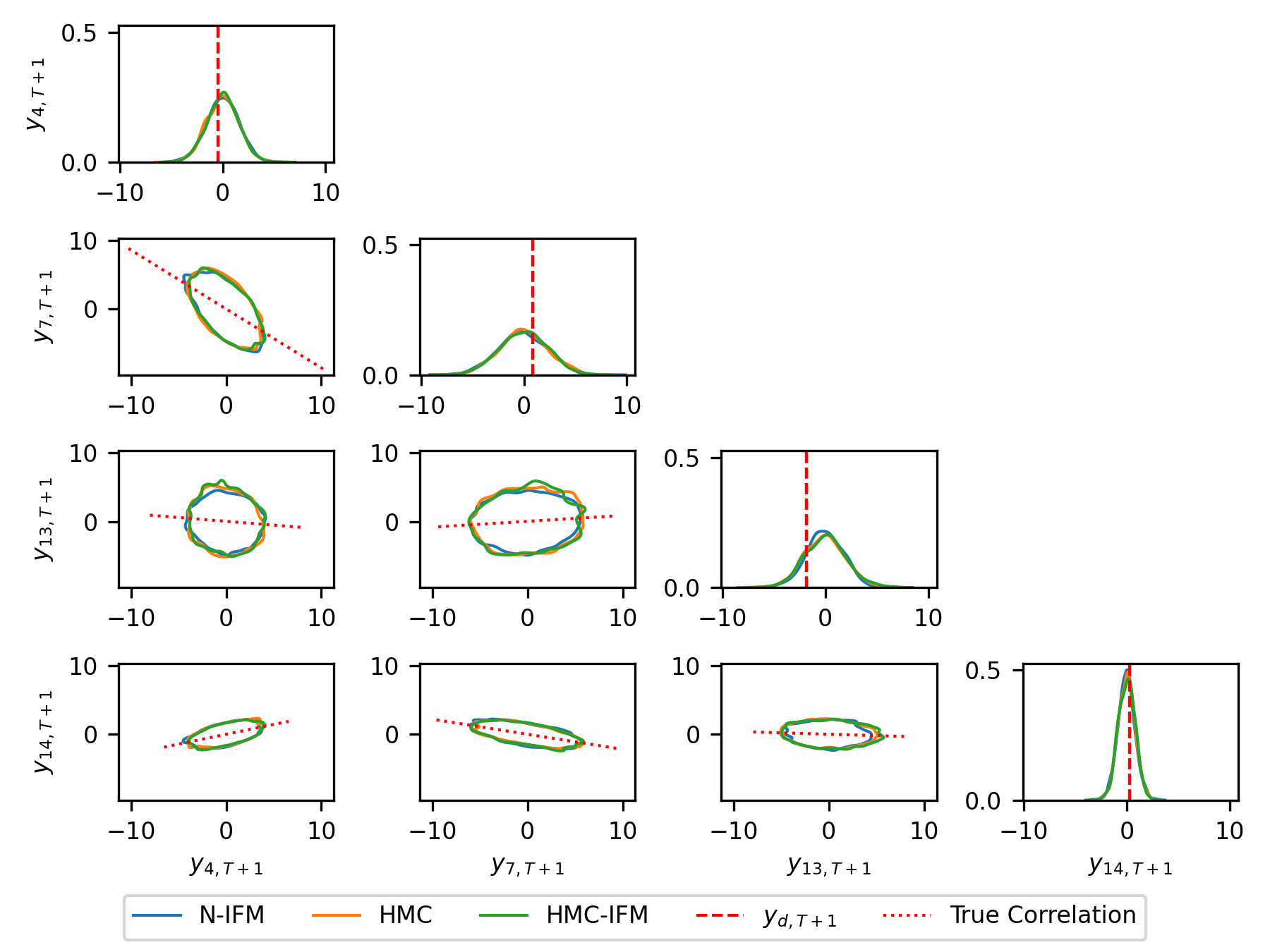}
    \caption{Selected one-step-ahead posterior predictive densities obtained from N-IFM, HMC and HMC-IFM for a one-factor Gaussian copula model
    with GARCH(1,1) marginals with Gaussian errors, comprising $D = 20$ series of length $T = 1000$. Diagonal panels show the marginal predictive distributions, while
    off-diagonal panels display the corresponding bivariate predictive distributions, {shown as 95\% highest-density contours obtained from kernel density estimates of the predictive draws}. In the diagonal panels, the vertical lines indicate the observed $y_{d,T+1}$, while in the off-diagonal panels, the dotted lines 
    {have slope $\rho_{ij}\,\sigma_i/\sigma_j$, where $\rho_{ij}$ is the true correlation between series $i$ and $j$ and $\sigma_i$, $\sigma_j$ are the standard deviations of the predictive draws.}
    }
    \label{fig:gauscop_bipredplot}
\end{figure}

Figure~\ref{fig:gauscop_bipredplot} shows some of the univariate and bivariate one-step-ahead posterior predictive densities obtained from N-IFM, HMC and HMC-IFM for a one-factor Gaussian copula model
with GARCH(1,1) marginals with Gaussian errors. The predictive densities obtained from N-IFM, HMC and HMC-IFM are very similar. The lower triangular
panels display the corresponding bivariate predictive distributions, which exhibit strong
agreement across the three methods. 
{In particular, the distributions obtained from N-IFM, HMC and HMC-IFM are very similar to one another and tilt in the direction of the true correlation.}
Table \ref{tab:timings} in Section \ref{sec:computingtime} of the online supplement shows the computing time of N-IFM, HMC and HMC-IFM for all simulation studies and real data applications. 
{In terms of computational efficiency, N-IFM is approximately $1,500$ times faster than HMC and $900$ times faster than HMC-IFM post-training.}
{Section~\ref{supp:gauscop_factor_q} of the online supplement shows that enriching the mean field approximation in ~\eqref{eqn:cop_variationalform_mf} with a rank-one factor covariance structure for the Gaussian copula parameters leaves the posterior for $\textrm{vech}(\widetilde{\boldsymbol{G}})$ essentially unchanged 
{for the one-factor Gaussian copula model.}
}

We now investigate the performance of the N-IFM and HMC-IFM methods using the one-step-ahead LPDS
and computing time as evaluation criteria. 
We generate 1100 observations from a three-factor Gaussian copula model with GARCH(1,1)
marginals with Gaussian errors, {following Algorithm \ref{alg:copula_sim} in Section \ref{supp:datagen} of the online supplement}. 
The one-step-ahead LPDS is computed using a
rolling-window scheme with window length $T = 1000$. The approximate posterior
distributions are first obtained using the first $1000$ observations, and the log of the one-step-ahead predictive density is
computed at time $t = 1001$. The window is then advanced by one time step, the approximate
posterior distributions are obtained using observations $t = 2, \ldots, 1001$, and the
log of the one-step-ahead predictive density is evaluated at time $t = 1002$. This procedure is repeated until the final log of one-step-ahead predictive density  is
computed at time $t = 1100$. Then, the one step ahead LPDS is computed using \eqref{eq:LPDS} for models with one to four factor Gaussian copula models with GARCH(1,1) marginals with Gaussian errors.

Table~\ref{tab:factorselection_sim} shows the one-step-ahead LPDS and computation time for one to four factor Gaussian copula models with GARCH(1,1) marginals with Gaussian errors for the $100$ one-step-ahead predictions. 
The table shows that both N-IFM and HMC-IFM correctly identify the true data-generating process, namely the three-factor model. Moreover, the two methods produce identical rankings across all competing models. In terms of computational efficiency, N-IFM is much faster than HMC-IFM when performing model selection.
From the simulation study, we conclude that {N-IFM is substantially faster than both HMC and HMC-IFM}, while producing posterior and predictive distributions that closely agree with those
obtained from HMC-IFM that themselves are generally similar to those from HMC.

{The N-IFM computational times in Table~\ref{tab:factorselection_sim} exclude the one-off cost of simulating synthetic
training datasets and training the neural networks; these upfront costs are reported in
Table~\ref{tab:amortisedtime}. 
The comparison in Table~\ref{tab:factorselection_sim} therefore reflects the
post-training cost of applying N-IFM repeatedly in the rolling-window prediction exercise.}

    \begin{table}[t]
    \centering
    \caption{Comparison of N-IFM and HMC-IFM for one to four factor models in terms of the one-step-ahead LPDS and computational time (in seconds) excluding training time (see Table \ref{tab:amortisedtime}). The simulated data are generated from a three-factor Gaussian copula model with GARCH(1,1) marginals with Gaussian errors. Values shown in bold indicate the best-performing model according to the LPDS.
    {N-IFM values and computing times are reported as the mean (with the standard deviation in parentheses) obtained from 30 repeated LPDS computations.}}
    \begin{tabular}{cccccc}
    \hline \hline
    Methods & Criteria & 1-factor & 2-factor & 3-factor & 4-factor\\
    \hline \noalign{\vskip 4pt}
    N-IFM & LPDS & -3045.67 & -2666.32 & \textbf{-2412.77} & -2467.73\\
     & & (0.19) & (0.26) & \textbf{(1.16)} & (2.76)\\[4pt]
     & Time (s) & 103.90 & 105.03 & 106.73 & 109.29\\
     & & (1.97) & (1.64) & (1.79) & (1.60)\\[4pt]
    HMC-IFM & LPDS & -2989.01 & -2654.90 & \textbf{-2370.20} & -2371.01\\
     & Time (s) & 58,600 & 84,500 & 184,000 & 1,540,000$^*$\\
    \hline
    \end{tabular}
    
    \vspace{4pt}
    \noindent\parbox{\linewidth}{\footnotesize $^*$The four-factor HMC-IFM model requires longer Markov chains (20,000 iterations) due to poor mixing and a low effective sample size.}
    \label{tab:factorselection_sim}
    \end{table}


\subsection{{Calibration study:} Gaussian copula}\label{sec:calib_gauscop}

{The simulation study in Section~\ref{sec:sim_gauscop} compares N-IFM, HMC and HMC-IFM on a single simulated dataset.  
We now assess frequentist calibration of N-IFM. For each of one to four factors, we generate $1000$ datasets from a Gaussian copula model with GARCH(1,1) marginals and Gaussian errors, where each dataset consists of $20$ time series of length $T=1000$ with the true parameters held fixed. 
For each dataset, we draw $1000$ samples from the N-IFM posterior and evaluate the bias and RMSE of the posterior mean, together with the empirical coverage probability and mean width of the nominal $95\%$ credible interval.
For one representative dataset, we fit all three methods, N-IFM, HMC-IFM, and HMC, and compare their posterior distributions using the univariate Wasserstein-1 distance \citep{villani2009optimal}. We report the distances for each pair of methods: N-IFM versus HMC-IFM, N-IFM versus HMC, and HMC versus HMC-IFM. Definitions and the supporting figures and tables are in Section~\ref{sec:supp_gauscop} of the online supplement. The results for the $t$ copula with GARCH(1,1) marginals with Student's-$t$ errors are given in Section~\ref{sec:calib_stcop} of the online supplement.}

\begin{figure}[t]
    \centering
    \includegraphics[width=\linewidth]{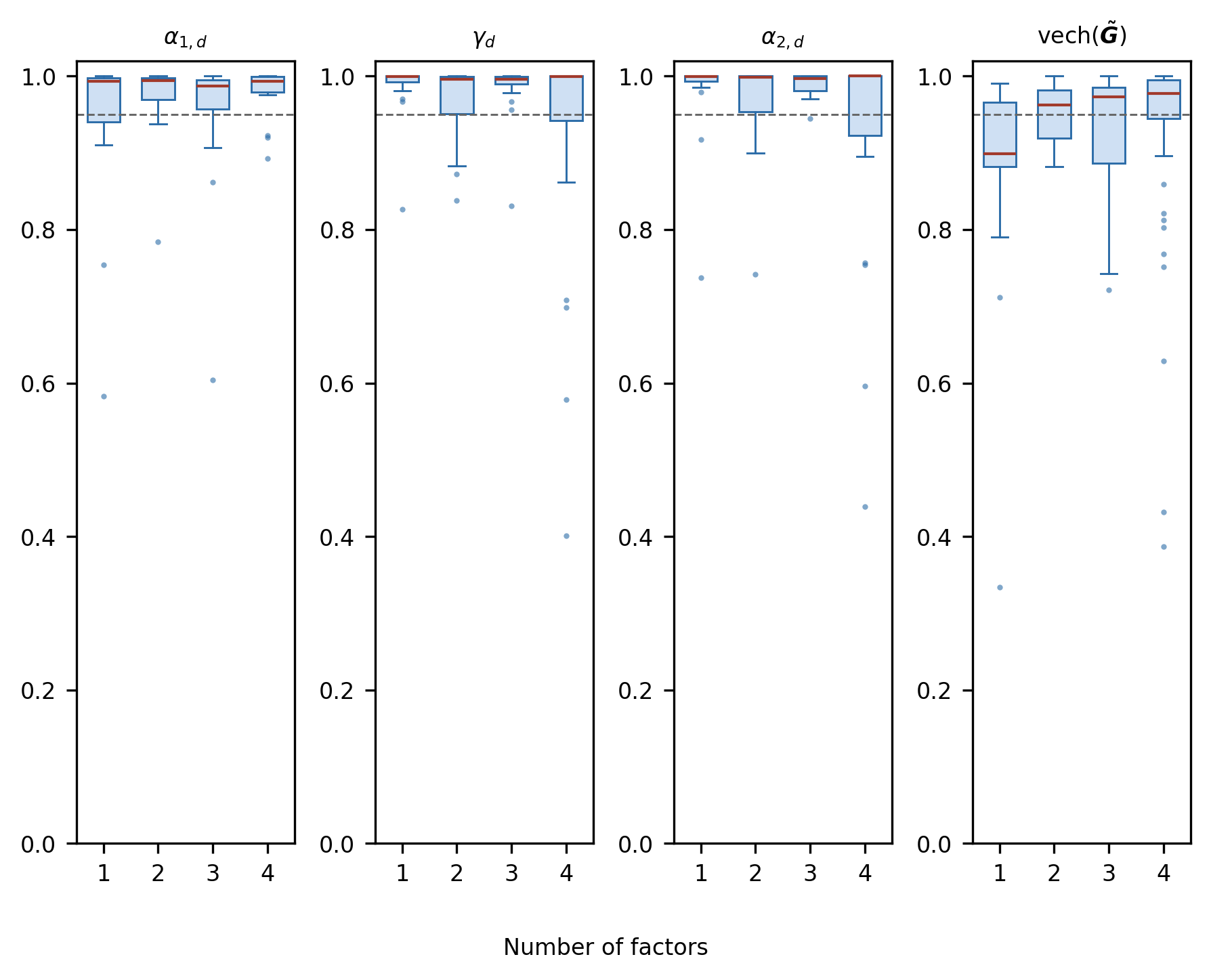}
    \caption{Empirical coverage probabilities of the nominal $95\%$ credible
    intervals produced by N-IFM for the Gaussian copula model with GARCH(1,1)
    marginals and Gaussian errors. For each parameter group and number of 
    factors, the boxplot summarises the parameter-specific coverage probabilities.
    The dashed horizontal line indicates the nominal coverage level of $0.95$.}
    \label{fig:calib_cov_gauscop}
\end{figure}

{The GARCH(1,1) marginal parameters, $\alpha_{1,d}$, $\alpha_{2,d}$, and
$\gamma_d$, for $d=1,\ldots,D$, are recovered accurately across all 
dimensions considered. In particular, the estimates exhibit no systematic
bias (Figure~\ref{fig:calib_bias_gauscop}), the RMSEs
{are similar across the four settings}
(Table~\ref{tab:calib_rmse_gauscop}), and the
empirical coverage probabilities are 
{at or above the nominal level}
(Figure~\ref{fig:calib_cov_gauscop}; the outliers are discussed in the online supplement). The
corresponding credible-interval widths 
{are also similar across the four settings}
(Table~\ref{tab:calib_width_gauscop}). Table~\ref{tab:calib_wass_gauscop} shows close agreement across the three posterior approximations for the GARCH(1,1) marginal parameters, with mean univariate Wasserstein-1 distances no greater than $0.04$ for all numbers of factors considered.} 


{The copula loadings, $\textrm{vech}(\widetilde{\bs{G}})$, exhibit a different
pattern. Although the bias remains close to zero and the empirical coverage
remains near the nominal level for all numbers of factors considered, the
precision of the N-IFM posterior deteriorates as the number of factors
increases. In particular, the mean width of the $95\%$ credible intervals for
$\textrm{vech}(\widetilde{\bs{G}})$ increases from $0.36$ for one factor to
$1.05$ for four factors (Table~\ref{tab:calib_width_gauscop}), while the RMSE
approximately doubles over the same range
(Table~\ref{tab:calib_rmse_gauscop} in the online supplement). The univariate
Wasserstein-1 distances provide further insight into this deterioration. For
$\textrm{vech}(\widetilde{\bs{G}})$, the HMC and HMC-IFM posteriors remain
close, with Wasserstein-1 distances no greater than $0.03$, whereas the N-IFM
posterior increasingly deviates from both HMC-based posteriors as the number
of factors increases. Specifically, the distances involving N-IFM increase
from approximately $0.07$ for one factor to approximately $0.26$ for four
factors (Table~\ref{tab:calib_wass_gauscop} in the online supplement). These
results indicate that the deterioration in posterior precision and
approximation accuracy for the copula loadings becomes more pronounced for
N-IFM as the number of factors increases.}
{N-IFM is therefore well calibrated for the GARCH(1,1) marginal parameters at every number of factors, while its posterior for the copula loadings becomes increasingly less accurate as the number of copula parameters grows. }

\begin{table}[t] \caption{Mean width of the nominal $95\%$ N-IFM credible intervals for the marginal and copula parameters in the Gaussian copula model with GARCH(1,1) marginals and Gaussian errors, reported for models with one to four copula factors. Here, $\bs\alpha_{1,1:20}=(\alpha_{1,1},\ldots,\alpha_{1,20})$, $\bs\gamma_{1:20}=(\gamma_1,\ldots,\gamma_{20})$, and $\bs\alpha_{2,1:20}=(\alpha_{2,1},\ldots,\alpha_{2,20})$ denote the corresponding GARCH(1,1) parameters across the $20$ marginal series. The notation $\textrm{ave}(\cdot)$ denotes averaging over the parameters within the corresponding parameter group. Each interval width is first averaged over the $1000$ simulation replications.} \centering \begin{tabular}{lcccc} \hline\hline Parameter & 1-factor & 2-factor & 3-factor & 4-factor \\ \hline $\textrm{ave}(\bs\alpha_{1,1:20})$ & 0.091 & 0.090 & 0.089 & 0.090 \\ $\textrm{ave}(\bs\gamma_{1:20})$ & 0.176 & 0.183 & 0.192 & 0.162 \\ $\textrm{ave}(\bs\alpha_{2,1:20})$ & 0.175 & 0.172 & 0.155 & 0.184 \\ $\textrm{ave}\!\left\{\textrm{vech}(\widetilde{\bs{G}})\right\}$ & 0.359 & 0.476 & 0.519 & 1.054 \\ \hline \end{tabular} \label{tab:calib_width_gauscop} \end{table}

{Further simulation studies reported in Section~\ref{sec:sim_stcop} of the online supplement also show that N-IFM performs well for the $t$ copula, 
{yielding posterior distributions that are close to those obtained under HMC-IFM; in this case, both differ slightly from those of full HMC.}
Section~\ref{sec:calib_stcop} discusses the calibration study for the $t$ copula with GARCH(1,1) marginals with Student's-$t$ errors.
Section~\ref{sec:stochterrors} presents an additional model selection simulation study. Taken together, these findings demonstrate that N-IFM provides a computationally efficient and reliable framework for both parameter inference and model selection.}


\section{Real data application \label{sec:RealData}}

We apply the proposed N-IFM method to a panel of $D = 20$ daily industry portfolio returns from 2 January 2015 to 16 May 2019. The dataset consists of $1100$ de-meaned, average value-weighted daily returns obtained from the Center for Research in Security Prices (CRSP) database\footnote{Tuck School of Business at Dartmouth, \url{https://mba.tuck.dartmouth.edu/pages/faculty/ken.french/index.html}}.
Further details on the data are provided in Section~\ref{sec:realdata_extrainfo} of the online supplement, including the list of industries, reported in Table~\ref{tab:siccodes}. 

Figure~\ref{fig:realdata_all} in Section~\ref{sec:realdata_extrainfo} of the online supplement shows {daily returns for the 20 industry portfolios over the first 1,000-day estimation window.} The agricultural industry is labelled as $\boldsymbol y_1$. The $\boldsymbol{y}_1$ series has a minimum (-6.42) and maximum (7.66) return and shows the presence of some extreme movements and occasional large shocks. Several other series  exhibit extreme movements and pronounced shocks, most notably $\boldsymbol{y}_{7}$ and $\boldsymbol{y}_{8}$. The series $\boldsymbol{y}_{1}$ and $\boldsymbol{y}_{8}$ show substantial excess kurtosis (8.23 and 33.16 respectively, compared with 0 under a Gaussian distribution), indicating heavy tails and a higher frequency of extreme observations. In addition, series $\boldsymbol{y}_{1}$ illustrates clear volatility clustering, highlighting the heteroskedastic nature of the data. These features support the use of GARCH-type volatility models with Student's-$t$ errors rather than Gaussian errors.

Figure {\ref{fig:realdata_all_hist}} in Section \ref{sec:realdata_extrainfo} of the online supplement shows the histograms of all series, with some appearing Gaussian and others resembling a Student's-$t$ distribution. The histogram for the agricultural sector confirms that a Student's-$t$ distribution is appropriate.

We consider {zero} to four factors of the following models: (1) The Gaussian copula with GARCH(1,1) marginals with Gaussian errors; (2) The Gaussian copula with GARCH(1,1) marginals with Student's-$t$ errors; (3) The $t$ copula with GARCH(1,1) marginals with Gaussian errors; (4) The $t$ copula with GARCH(1,1) marginals with Student's-$t$ errors. The zero-factor model is equivalent to assuming independent GARCH(1,1) processes for each series. The best model is selected using the time series validation with the highest LPDS value discussed in Section \ref{sec:timeseriescv}.

\begin{table}[t]
\caption{Comparison of the LPDS values obtained using the N-IFM method for {zero} to four factors of the four models for the industry dataset: (1) The Gaussian copula with GARCH marginals with Gaussian errors; (2) The Gaussian copula with GARCH marginals with Student's-$t$ errors; (3) The $t$ copula with GARCH marginals with Gaussian errors; (4) The $t$ copula with GARCH marginals with Student's-$t$ errors. {The zero-factor model is equivalent to assuming independent GARCH(1,1) processes for each series.} Values shown in bold indicate the best-performing model according to the one-step-ahead LPDS. {LPDS values are reported as the mean (with the standard deviation) obtained by repeating the LPDS computation $30$ times on the industry dataset.}
    }
    \centering
    \begin{tabular}{cccccc}
    \hline \hline
Model &   {Zero-factor}& 1-factor&  2-factor&  3-factor& 4-factor\\
       \hline \noalign{\vskip 4pt}
        (1) & -3410.83 & -2818.08 & -2819.99 & -2842.13 & -2803.64 \\
         & (0.16) & (0.60) & (0.70) & (1.51) & (2.28) \\[4pt]
        (2) & -3335.66 & -2708.59 & -2700.08 & -2714.10 & -2675.12 \\
         & (0.12) & (0.22) & (0.24) & (0.70) & (1.60) \\[4pt]
        (3) & -3410.83 & -2776.82 & -2765.29 & -2762.43 & -2734.20 \\
         & (0.16) & (0.35) & (0.28) & (0.67) & (0.86) \\[4pt]
        (4) & -3335.66 & -2671.22 & -2663.25 & -2655.15 & \textbf{-2631.06} \\
         & (0.12) & (0.21) & (0.25) & (0.33) & \textbf{(0.62)} \\[4pt]
         \hline
    \end{tabular}
    \label{tab:factorselection_real}
\end{table}

Table~\ref{tab:factorselection_real} reports the LPDS for each model based on 100 one-step-ahead predictions. The zero-factor model delivers substantially lower LPDS values than the copula models with one to four factors, indicating markedly poorer predictive performance and highlighting the importance of modelling cross-series dependence via one or more factors. The highest LPDS, shown in bold, corresponds to the four-factor {$t$ copula} model with GARCH(1,1) marginals with Student's-$t$ errors. Therefore, further analysis
in this section is based on the four-factor {$t$ copula} model with GARCH(1,1) marginals with Student's-$t$ errors.

{We assess the accuracy of the posterior distributions of the model parameters estimated using N-IFM by comparing them with those obtained from the {HMC and HMC-IFM methods} on the real dataset. 
Figures~\ref{suppfig:alpha1_stcop_f4} to \ref{suppfig:copnu_stcop_f4} in Section~\ref{supp:realdata} of the online supplement present the posterior distributions of all model parameters estimated using {N-IFM, HMC and HMC-IFM}. 
Overall, the three methods produce substantial overlap in the posterior distributions. However, noticeable discrepancies appear for some marginal model parameters. {In particular, the N-IFM approximation for some $\alpha_{1,d}$ parameters, and to a lesser extent for $\alpha_{2,d}$ and $\gamma_d$, differs from the corresponding HMC, while still having strong agreement with HMC-IFM; this corroborates findings from the simulation studies.} In contrast, the posterior distributions of {$\textrm{vech}(\widetilde{\boldsymbol{G}})$} obtained using N-IFM, except for a few cases, closely agree with those from HMC and HMC-IFM. 
Table~\ref{tab:hmc_stcop_realf4} in the online supplement reports the ESS and the $\widehat{R}$ for all model parameters estimated using HMC. The large ESS values and $\widehat{R}$ close to~1 indicate good mixing across chains and provide evidence of satisfactory Markov chain convergence. }

{
Figure~\ref{fig:stcop_posterior_density_real} 
shows selected univariate and bivariate one-step-ahead posterior predictive densities obtained from N-IFM, HMC and HMC-IFM for the industry dataset. 
The predictive densities obtained from N-IFM, HMC and HMC-IFM are very similar. The lower triangular
panels display the corresponding bivariate predictive distributions, which exhibit strong
agreement across the three methods.
In terms of computational efficiency, N-IFM is approximately {$300,000$ times faster than HMC-IFM  and $430,000$} times faster than HMC post-training as shown in Table \ref{tab:timings} in Section \ref{sec:computingtime} of the online supplement, while delivering posterior estimates of comparable accuracy.

\begin{figure}[t]
    \centering
    \includegraphics[width=\linewidth]{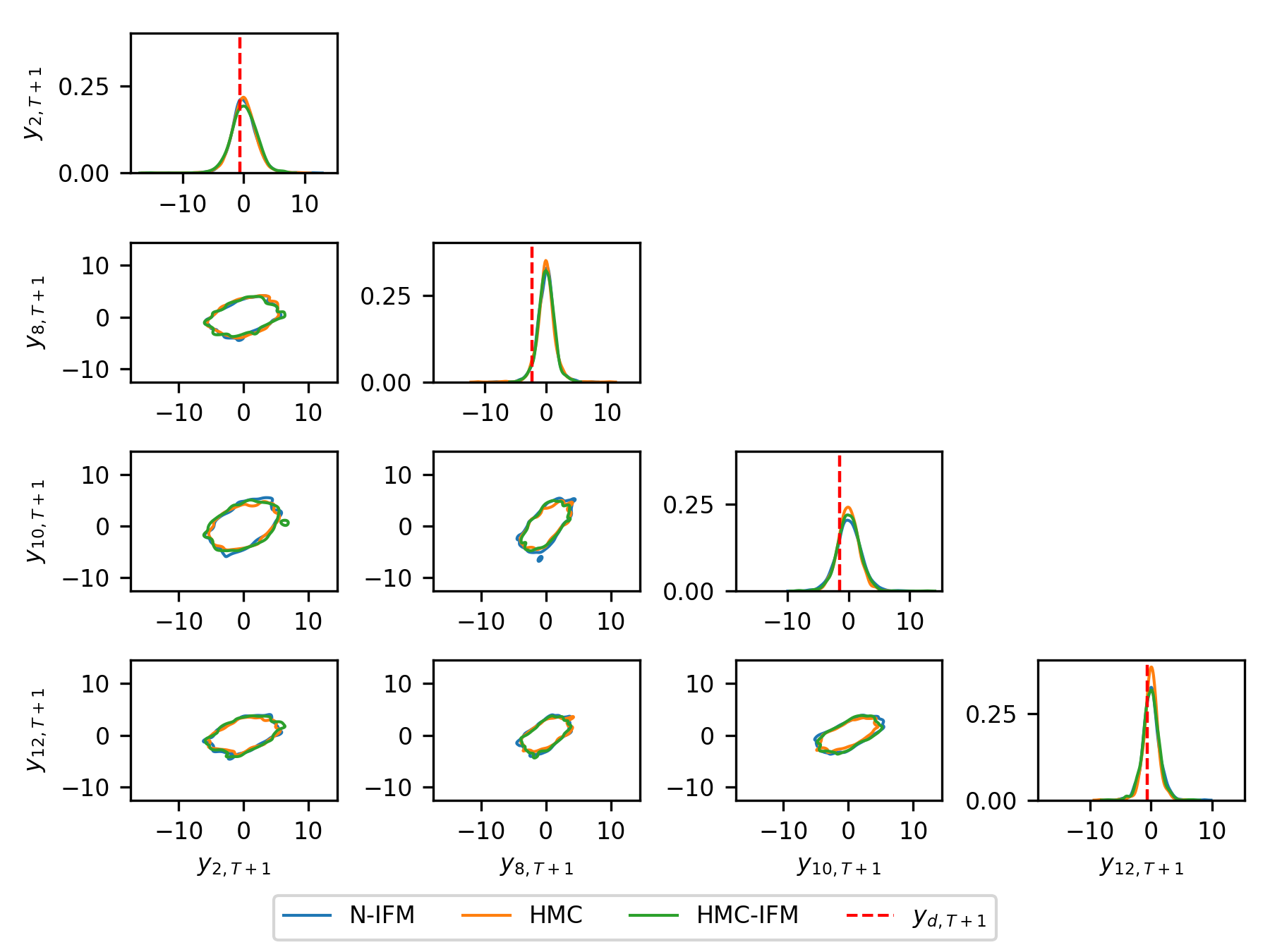}
    \caption{Selected one-step-ahead posterior predictive densities obtained from N-IFM, HMC and HMC-IFM for a four-factor $t$ copula model with GARCH(1,1) marginals with Student's-$t$ errors, comprising $D = 20$ series of length $T = 1000$ for the industry dataset. Diagonal panels show the marginal predictive distributions, while off-diagonal panels display the corresponding bivariate predictive distributions, 
    {shown as 95\% highest-density contours obtained from kernel density estimates of the predictive draws.} 
    In the diagonal panels, the vertical lines indicate the observed $y_{d,T+1}$.}
    \label{fig:stcop_posterior_density_real}
\end{figure}

Figure~\ref{fig:stcop_heatmap_pred} (left panel) shows the one-step-ahead predicted correlations from N-IFM for the 20 industry portfolio series. The correlations vary across industry pairs, from below $0.2$ to about $0.8$, and none are negative. The strongest dependence is between Fabricated Metal Products and Furniture and
Fixtures, with a correlation of about $0.8$. The weakest dependence involves Tobacco Products, whose correlations with Mining and Primary Metal Industries are below $0.2$.

The right panel of Figure~\ref{fig:stcop_heatmap_pred}  shows the 90\% prediction intervals across the 100 rolling-origin one-step-ahead forecasts for the four-factor $t$ copula model with GARCH(1,1) marginals with Student's-$t$ errors. The 90\% prediction interval expands and contracts over time, reflecting volatility clustering, and only a small proportion of observations fall outside the interval. The 90\% prediction interval of $\bs{y}_{12}$ is relatively wide, spanning approximately $[-4, 4]$. After around 20 time steps, the interval narrows to approximately $[-2, 2]$ and remains stable thereafter, closely tracking the observed values.

\begin{figure}[t]
    \centering
\includegraphics[width=\linewidth]{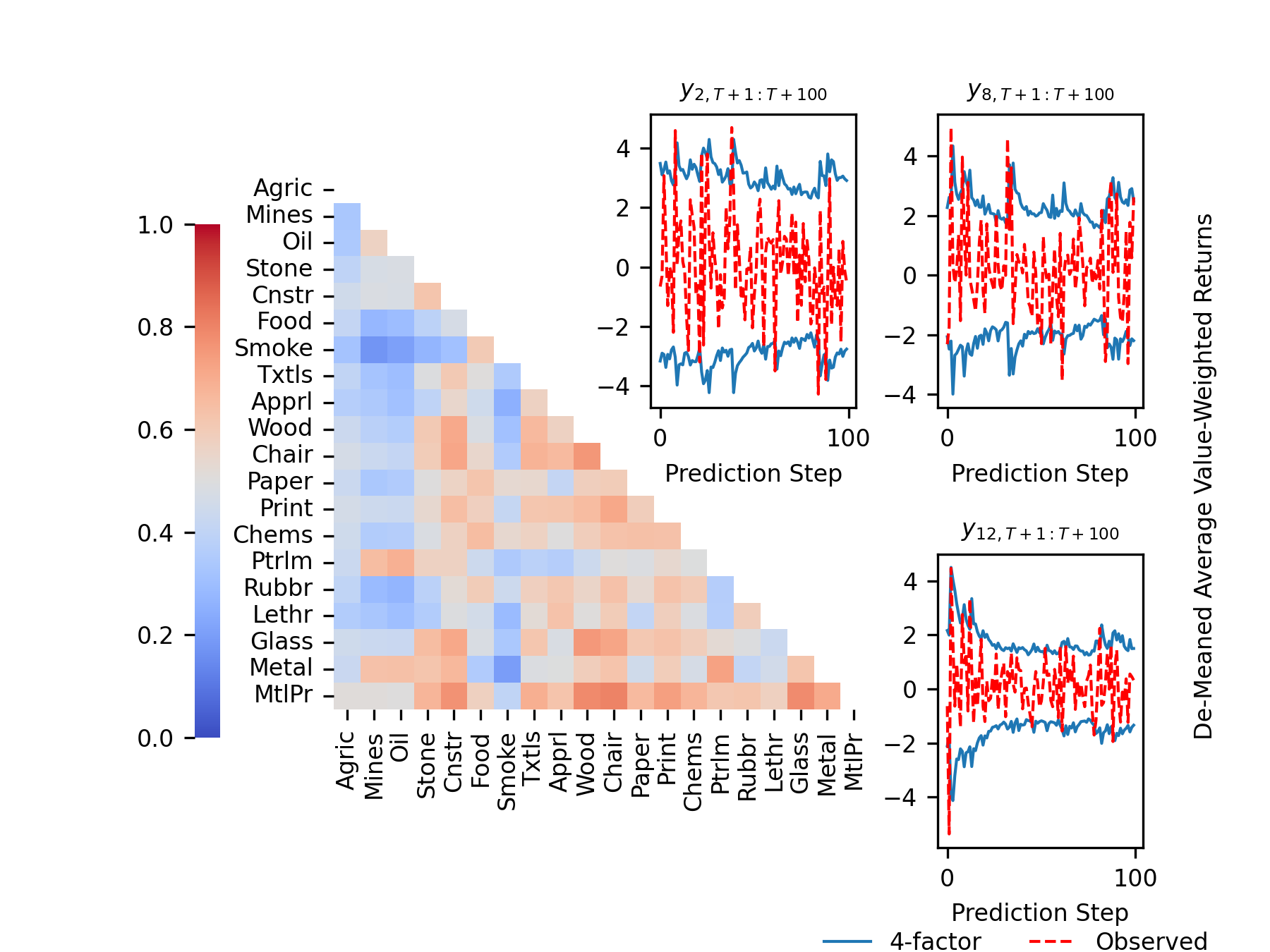}
    \caption{Left: estimated one-step-ahead correlation coefficients between each pair of the 20 industry portfolio stock returns obtained using the N-IFM method. 
    Right: the 90\% prediction intervals over {100 one-step-ahead forecasts from rolling origins} for the four-factor $t$ copula model with GARCH(1,1) marginals and Student's-$t$ errors, shown alongside the observed values for selected
industry data series.}
    \label{fig:stcop_heatmap_pred}
\end{figure}


\section{Conclusion \label{sec:Conclusion}}

In this paper, we develop the 
N-IFM approach for fitting time series copula models. We consider Gaussian and $t$ copula models with GARCH(1,1) marginals with Gaussian and Student's-$t$ errors. 
Two neural network models are developed. The first neural network maps the observations from each series to the parameters of the approximate posterior distribution of parameters of the GARCH(1,1) models. The second neural network maps the transformed copula ``data'' to the parameters of the approximate posterior distribution of parameters of the copula models. 
{Across all simulated and real applications, these approximated posteriors closely align with those obtained using HMC-IFM and to a lesser extent HMC.}
{We also found that introducing a rank-one factor covariance into the Gaussian copula posterior approximation did not improve accuracy relative to the mean-field approximation 
{for the one-factor Gaussian copula model.}
}
{A repeated-sampling calibration study further showed that the N-IFM posterior is well calibrated for the marginal parameters, whereas its approximation of the copula parameters becomes less accurate as the number of factors, and hence the number of copula parameters, increases.}

A key advantage of N-IFM is its substantial computational efficiency. The approximate posterior distributions are obtained in a fraction of the time required by {HMC and HMC-IFM}, rendering tasks such as real-time prediction, model selection and time series validation effectively costless {once the networks are trained} from a computational perspective, in contrast to the considerable burden imposed by {HMC and HMC-IFM}. In the simulation study, N-IFM and HMC-IFM agree in selecting the three-factor Gaussian copula with GARCH(1,1) marginals with Gaussian errors that generated the data.

When applied to the industry portfolio dataset, N-IFM identifies the four-factor $t$ copula with GARCH(1,1) marginals with Student's-$t$ errors as the preferred model. We then compare one-step-ahead predictions and the corresponding approximate posterior distributions of the model parameters obtained from N-IFM with those from {HMC and HMC-IFM}. Despite the dramatic reduction in computation time, N-IFM delivers highly comparable results in terms of predictive distributions. {The N-IFM posterior distributions align with those from HMC-IFM, although the HMC-based posterior distributions for the GARCH$(1,1)$ parameters are generally narrower. Overall, the results suggest that the N-IFM method provides an accurate and computationally efficient alternative to HMC and HMC-IFM.}

When conducting simulation experiments, we found that as the number of parameters increases, the quality of the posterior approximations tends to deteriorate. 
Future work will attempt to address 
{this issue} 
by exploring more flexible neural network architectures, which in turn will allow the use of more flexible approximations, such as normalising flows, to better represent true posterior distributions. We will also extend N-IFM to more flexible copula specifications, including vine copulas, that are able to accommodate complex and high-dimensional dependence structures. 


\section*{Acknowledgements}

This research is supported by the ARC ITRH for Transforming Energy Infrastructure through Digital Engineering (TIDE), which is led by The University of Western Australia (UWA), delivered with The University of Wollongong and other research partners, and funded by the Australian Research Council, INPEX Operations Australia, Shell Australia, Wood-side Energy, Fugro Australia Marine, Wood Group Kenny Australia, RPSGroup, Bureau Veritas, and Lloyd's Register Global Technology (grant No.IH200100009). The authors would like to thank Rapha\"el Huser for his helpful comments and suggestions.

\section*{Data availability statement}
The data that support the findings of this study are openly available in the Kenneth R.\ French Data Library at \url{https://mba.tuck.dartmouth.edu/pages/faculty/ken.french/index.html}, dataset: ``38 Industry Portfolios''.  
\begin{singlespace}
\bibliographystyle{apalike}
\bibliography{references_v1}
\end{singlespace}
\clearpage

\appendix

\renewcommand{\thesection}{S\arabic{section}}
\renewcommand{\thesubsection}{S\arabic{section}.\arabic{subsection}}

\setcounter{section}{0}
\setcounter{subsection}{0}

\section*{Supplementary Material}
We use the following notation in the online supplement. (1), Algorithm~1,
Section~1, etc, refer to the main paper, while (S1),
Algorithm~S1, Section~S1, etc, refer to the supplement. All the acronyms used without definition in the supplement, are defined in the main paper.

Section~\ref{sec:copula_deriv} outlines the derivation of the implicit copula PDF. 
Section~\ref{supp:datagen} outlines the {model} parameter transformations, prior calibration, {and the algorithm for simulating data.}
Section~\ref{supp:cnn} describes the convolutional neural network (CNN) architecture.
Section~\ref{supp:deepsets} describes the Deep Sets \citep{zaheer_deepsets} architecture.
Section~\ref{sec:nifm-training} provides additional implementation details on training the neural network within the N-IFM framework.
Section~\ref{supp:hmc_vs_hmcifm} provides additional results for the Gaussian copula simulation study from Section~\ref{sec:sim_gauscop} of the main paper.
Section~\ref{sec:computingtime} provides a summary of the computing time of {N-IFM, HMC and HMC-IFM.}
{Section~\ref{supp:gauscop_factor_q} of the online supplement investigates a Gaussian approximation with factor covariance structure for the parameters of the copula model, as an alternative to the mean-field approximation.}
{Section~\ref{sec:supp_gauscop} discusses the calibration metrics and provides the supporting figures and tables for the Gaussian copula calibration study of Section~\ref{sec:calib_gauscop} in the main paper.}
Section~\ref{sec:sim_stcop} presents a simulation study for the $t$ copula {with GARCH(1,1) marginals with Gaussian errors}.
{Section~\ref{sec:calib_stcop} presents the corresponding calibration study for the $t$ copula with GARCH(1,1) marginals with Student's-$t$ errors.}
Section~\ref{sec:stochterrors} presents a simulation study on model selection.
Section~\ref{sec:realdata_extrainfo} provides further information on the industry portfolio dataset {used in the real data application}.
Section~\ref{supp:realdata} contains additional results from the real data application in Section~\ref{sec:RealData} of the main paper.

\section{Implicit copula PDF derivation}\label{sec:copula_deriv}

Let $\boldsymbol{Z} \equiv (Z_1,\ldots,Z_D)^\top$ be a continuous random vector with joint
distribution function $F_{\boldsymbol{Z}}$ and marginal distribution functions
$F_{Z_1},\ldots,F_{Z_D}$. The implicit copula function is defined as
\begin{equation}
C(\boldsymbol{u})
=
F_{\boldsymbol{Z}}\!\left(
F_{Z_1}^{-1}(u_1), \ldots, F_{Z_D}^{-1}(u_D)
\right),
\qquad \boldsymbol{u} \in [0,1]^D .
\end{equation}
The copula density is obtained by differentiation:
\begin{align}
c(\boldsymbol{u})
&=
\frac{\partial^D}{\partial u_1 \cdots \partial u_D}
C(\boldsymbol{u}) \\
&=
\frac{\partial^D}{\partial u_1 \cdots \partial u_D}
F_{\boldsymbol{Z}}(\boldsymbol{z}),
\end{align}
where $\boldsymbol{z} \equiv (z_1, \dots, z_D)^\top$ denotes a realisation of $\boldsymbol{Z}$ and $z_d = F_{Z_d}^{-1}(u_d)$ for $d=1,\ldots,D$.
{Differentiating  $z_d = F_{Z_d}^{-1}(u_d)$ with respect to $u_d$, gives}
\[
\frac{\partial z_d}{\partial u_d}
=
\frac{1}{p_{Z_d}(z_d)},
\qquad d=1,\ldots,D .
\]
Since the transformation $\boldsymbol{u} \mapsto \boldsymbol{z}$ is componentwise,
the Jacobian determinant is
\[
\left|
\det\left(
\frac{\partial \boldsymbol{z}}{\partial \boldsymbol{u}}
\right)
\right|
=
\prod_{d=1}^D
\frac{1}{p_{Z_d}(z_d)} .
\]
Applying the multivariate change-of-variables formula yields
\begin{equation}
c(\boldsymbol{u})
=
p_{\boldsymbol{Z}}(\boldsymbol{z})
\left|
\det\left(
\frac{\partial \boldsymbol{z}}{\partial \boldsymbol{u}}
\right)
\right|
=
\frac{p_{\boldsymbol{Z}}(\boldsymbol{z})}
{\prod_{d=1}^D p_{Z_d}(z_d)} .
\end{equation}
Here,
\[
p_{\boldsymbol{Z}}(\boldsymbol{z})
=
\frac{\partial^D}{\partial z_1 \cdots \partial z_D}
F_{\boldsymbol{Z}}(\boldsymbol{z}),
\qquad
p_{Z_d}(z_d)
=
\frac{\partial}{\partial z_d}
F_{Z_d}(z_d),
\quad d=1,\ldots,D .
\]

\section{Parameter transformations, prior calibration, and data generation} \label{supp:datagen}
Section~\ref{sec:trangarch} outlines the transformations used to obtain the unrestricted GARCH(1,1) and degrees-of-freedom parameters. Section \ref{sec:prior_alg} outlines the algorithm {used} to obtain hyperparameters of the prior distributions of the model parameters. Section~\ref{supp:datagen_cop} outlines the data generation process for the copula model with GARCH(1,1) marginals. 

\subsection{Parameter transformations} \label{sec:trangarch}

We apply the parameterisation used by \cite{Pitt2015Jul} to obtain unrestricted GARCH(1,1) parameters. The transformation is carried out in two stages. In the first stage, we define
\begin{equation}\label{eqn:garchtran1}
\begin{aligned}
    & \psi_{1,d} = \alpha_{1,d} + \alpha_{2,d}, \\
    & \psi_{2,d} = \gamma_d / (1 - \psi_{1,d}), \\
    & \psi_{3,d} = \alpha_{1,d} / \psi_{1,d},
\end{aligned}
\end{equation}
for $d=1,...,D$.
In the second stage, we map these parameters to an unconstrained space as follows:
\begin{equation}\label{eqn:garchtran2}
\begin{aligned}
    & \varphi_{1,d} = \log \left( \psi_{1,d} / (1 - \psi_{1,d}) \right), \\
    & \varphi_{2,d} = \log(\psi_{2,d}), \\
    & \varphi_{3,d} = \log \left( \psi_{3,d} / (1 - \psi_{3,d}) \right),
\end{aligned}
\end{equation}
for $d=1,...,D$. The parameters \(\varphi_{1,d}\), \(\varphi_{2,d}\), and \(\varphi_{3,d}\) are unconstrained and ensure that the constrained GARCH(1,1) parameters satisfy the restrictions outlined in Section~\ref{sec:marginalmodel} of the main paper.

Then we consider the degrees-of-freedom parameters. This parameterisation applies to both the \(t\) copula degrees-of-freedom parameter \(\nu\) and the degrees-of-freedom parameters \(\widetilde{\nu}_d\) of the GARCH(1,1) marginals with Student's-\(t\) errors for \(d = 1, \dots, D\). We use the following transformations:
\begin{equation}\label{eqn:nutran}
\begin{aligned}
    & df = \log(\nu - 2), \\
    & \widetilde{df}_d = \log(\widetilde{\nu}_d - 2).
\end{aligned}
\end{equation}
The transformed parameters \(df\) and \(\widetilde{df}_d\) are unconstrained which ensures that \(\nu > 2\) and \(\widetilde{\nu}_d > 2\).

\subsection{Selecting hyperparameters for the prior distributions} \label{sec:prior_alg}

In this paper, we consider copula models with one to four factors under four specifications: (1) a Gaussian copula with GARCH(1,1) marginals with Gaussian errors; (2) a Gaussian copula with GARCH(1,1) marginals with Student's-$t$ errors; (3) a $t$ copula with GARCH(1,1) marginals with Gaussian errors; and (4) a $t$ copula with GARCH(1,1) marginals with Student's-$t$ errors. 


This section describes how we construct the prior hyperparameters for all model parameters. 
Algorithm~\ref{alg:priors} summarises the procedure used to calibrate the hyperparameters of the prior distributions using the historical industry stock return data of length $T^{*}=1000$, spanning 11 January 2011 to 31 December 2014; see Section \ref{sec:RealData} of the main paper.


\begin{algorithm}[t]
\caption{Algorithm to obtain hyperparameters of the prior distributions of the model parameters}
\label{alg:priors}
\begin{algorithmic}[1]
\Require Data $y_{d,t^*}$ for $d = 1, \dots, D$ and $t^* = 1, \dots, T^{*}$. 

\State \textbf{Estimate marginal (GARCH) parameters:}
\For{$d = 1$ to $D$}
    \State Estimate GARCH(1,1) parameters using HMC (alternatively, ML can be used) and compute posterior means
    \[ \widehat{\boldsymbol{\theta}}_d =
    (\widehat{\varphi}_{1,d},\,\widehat{\varphi}_{2,d},\,\widehat{\varphi}_{3,d})^{\top}.
    \]
    
\EndFor

\State \textbf{Compute pseudo-observations:}
\[
u_{d,t^*} = F_{Y_{d,t^*}}(y_{d,t^*}; 
\widehat{\boldsymbol{\theta}}_d),
\quad t^*=1,\dots,T^*  \text{~ and ~}  d=1,...,D.
\] 

\State \textbf{Estimate copula parameters}
\[
\textrm{vech}(\widetilde{\boldsymbol{G}})
\]
using HMC (alternatively, ML can be used) and compute their posterior means to obtain $\boldsymbol{\mu}^{*} $.

\State \textbf{Estimate prior hyperparameters:}
\Statex (1) Use the posterior mean estimates $\{\widehat{\alpha}_{1,d},\widehat{\alpha}_{2,d},\widehat{\gamma}_{d}\}_{d=1}^{D}$ as pseudo-observations. 
\Statex (2) Fit beta distributions to $\{\widehat{\alpha}_{1,d}\}_{d=1}^{D}$ and $\{\widehat{\alpha}_{2,d}\}_{d=1}^{D}$ using either HMC or ML to obtain the hyperparameters $(a_1,b_1)$ and $(a_2,b_2)$, respectively. 
\Statex (3) Fit a gamma distribution to $\{\widehat{\gamma}_{d}\}_{d=1}^{D}$ using either HMC or ML to obtain $(a_3,b_3)$.

\State \Return Hyperparameters of the prior distributions of the model parameters $(a_1,b_1,a_2,b_2,a_3,b_3,\boldsymbol{\mu}^{*\top})^\top $.
\end{algorithmic}
\end{algorithm}

%

Algorithm \ref{alg:priors} is now discussed. For each series $d=1,\ldots,D$, we assign beta priors, $\mathrm{beta}(a_1,b_1)$ and $\mathrm{beta}(a_2,b_2)$ for $\alpha_{1,d}$ and $\alpha_{2,d}$ respectively, and a gamma prior, $\mathrm{gamma}(a_3,b_3)$, to $\gamma_d$. 
To calibrate the hyperparameters $(a_1,b_1,a_2,b_2,a_3,b_3)$, we first use Hamiltonian Monte Carlo (HMC) to fit a GARCH$(1,1)$ model to each separate series of the historical data, applying the weakly informative priors in Table~\ref{tab:unif_priors} (alternatively, Maximum Likelihood (ML) can be used). 
This yields posterior mean estimates $\widehat{\alpha}_{1,d}$, $\widehat{\alpha}_{2,d}$, and $\widehat{\gamma}_d$ for $d=1,\ldots,D$. 
We then treat $\{\widehat{\alpha}_{1,d}\}_{d=1}^D$ as pseudo-observations from a $\mathrm{beta}(a_1,b_1)$ distribution and estimate $(a_1,b_1)$ using either HMC or ML. 
We apply the same approach to estimate $(a_2,b_2)$ from $\{\widehat{\alpha}_{2,d}\}_{d=1}^D$ and $(a_3,b_3)$ from $\{\widehat{\gamma}_d\}_{d=1}^D$. 
We use a truncated gamma prior, $\mathrm{gamma}(a_\nu,b_\nu)$ for the degrees-of-freedom for the $t$ copula, $\nu>2$. We choose the hyperparameters $(a_\nu,b_\nu)$ so that the prior mass is concentrated on empirically plausible values, with approximately 95\% of the distribution lying between $3$ and $20$. 

{For the GARCH$(1,1)$ model with Student's-$t$ errors, we adopt the same priors as for the GARCH$(1,1)$ model with Gaussian errors, and assign each marginal degrees-of-freedom parameter, $\widetilde{\nu}_d$, the same prior as the copula degrees-of-freedom parameter, $\nu$, as given in Table~\ref{tab:parameters}.}

We now specify the prior distribution for the copula parameters, $\textrm{vech}(\widetilde{\boldsymbol{G}})$. We first fit the Gaussian copula to the copula ``data'', constructed from the posterior means of the marginal model parameters, using HMC (although 
{ML} could also be used). Let $\boldsymbol{\mu}^*$ denote the resulting posterior mean estimate of the copula parameters. The dimension of $\textrm{vech}(\widetilde{\boldsymbol{G}})$ is given in Table~\ref{tab:num_gtilde}. We then assign the prior
\[
\textrm{vech}(\widetilde{\boldsymbol{G}}) \sim \mathcal{N}(\boldsymbol{\mu}^*, \boldsymbol{I}),
\]
where $\boldsymbol{I}$ denotes the identity matrix of appropriate dimension. The full prior specification is reported in Table~\ref{tab:parameters}.

\begin{table}[t]
\caption{Initial weakly informative priors used in HMC for historical industry stock return data from 11 January 2011 to 31 December 2014. 
    The final  column reports the $90\%$ probability intervals. The Gaussian ($\mathcal{N}$) prior is parameterised by its mean and standard deviation.}
    \centering
\begin{tabular}{lccc}
\hline \hline
Model & Parameter & Prior & 90\% probability intervals \\ \hline
GARCH(1,1) & $\varphi_{1,d}, \varphi_{2,d}, \varphi_{3,d}$ & $\mathcal{N}(0,1)$ & $(-1.64,\,1.64)$ \\

Gaussian copula & $\widetilde{{G}}_{ij}$ & $\mathcal{N}(0,1)$ & $(-1.64,\,1.64)$ \\
$t$ copula & $\widetilde{{G}}_{ij}$ & $\mathcal{N}(0,1)$ & $(-1.64,\,1.64)$ \\
& $df$ & $\mathcal{N}(0,1)$ & $(-1.64,\,1.64)$ \\
\hline
\end{tabular}
    \label{tab:unif_priors}
\end{table}


\begin{table}[t]
\caption{Summary of the prior distributions for all model parameters. The beta prior is parameterised by shape parameters $a$ and $b$; the gamma prior is parameterised by shape and rate; and the Gaussian ($\mathcal{N}$) prior is parameterised by its mean and standard deviation.}
    \centering
\begin{tabular}{p{4cm}ccc}
\hline \hline
Model component & Parameters & Prior & 90\% probability intervals \\ \hline
GARCH(1,1) & $\alpha_{1,d}$ & $\textrm{beta}(11.34, 85.12)$ & $(0.069,\;0.175)$ \\
& $\alpha_{2,d}$ & $\textrm{beta}(19.58, 4.62)$ & $(0.667,\;0.922)$ \\
& $\gamma_d$ & $\textrm{gamma}(4.69, 1/0.03)$ & $(0.053,\;0.262)$ \\[3pt]
Gaussian and $t$ copulas & $\textrm{vech}(\widetilde{\boldsymbol{G}})$ & $\mathcal{N}(\boldsymbol{\mu}^*,\boldsymbol{I})$ & $(\boldsymbol{\mu}^* - 1.64\cdot\boldsymbol{1},\;\boldsymbol{\mu}^* + 1.64\cdot\boldsymbol{1})$ \\[3pt]
Degrees of freedom$^*$ & $\nu$, $\widetilde{\nu}_d$ & $\textrm{gamma}(4.74, 1/2.03)$ & $(3.672,\;17.852)$ \\ \hline
\end{tabular}  
\vspace{4pt}
\noindent\parbox{\linewidth}{\footnotesize {$^{*}$$\widetilde{\nu}_d$ applies only to the GARCH(1,1) model with Student's-$t$ errors, and $\nu$ only to the $t$ copula.}}
\label{tab:parameters}
\end{table}

\begin{table}[t]
\caption{Number of non-zero elements in $\widetilde{\boldsymbol G}$ for one to four factor 20-dimensional Gaussian and $t$ copulas.}
    \centering
    \begin{tabular}{cc}
    \hline \hline
        Number of Factors & Number of elements in $\textrm{vech}(\widetilde{\boldsymbol G})$ \\
        \hline
        1 & 20\\
        2 & 39\\
        3 & 57\\
        4 & 74\\
        \hline
    \end{tabular}
    
    \label{tab:num_gtilde}
\end{table}

\subsection{Generating data from copula models with GARCH(1,1) margins} \label{supp:datagen_cop}

Algorithm~\ref{alg:copula_sim} summarises our procedure for generating simulated data from one to four factor Gaussian and $t$ copula models coupled with GARCH$(1,1)$ marginals under both Gaussian and Student's-$t$ innovations. 
We begin by sampling elements of the factor-loading matrix  $\textrm{vech}(\widetilde{\boldsymbol{G}})$ from their prior distributions. 
For the $t$ copula, we additionally draw the degrees-of-freedom parameter $\nu$ from its prior. 
Given $\textrm{vech}(\widetilde{\boldsymbol{G}})$, we construct the implied correlation matrix $\bar{\boldsymbol{\Omega}}$ following the factor-copula parameterisation in Section~\ref{sec:factorcopula} of the main paper.
Conditional on $\bar{\boldsymbol{\Omega}}$, we generate copula ``data'' and transform them to uniform distributions. 
For the Gaussian copula, we simulate $\boldsymbol{{z}}_{t} \sim \mathcal{N}_D(\boldsymbol{0},\bar{\boldsymbol{\Omega}})$ for $t=1,\ldots,T$ and set
\[
{u}_{d,t}=\Phi\!\left({z}_{d,t}\right), \qquad d=1,\ldots,D \textrm{ ~and~ } t=1,\ldots,T,
\]
where $\Phi(\cdot)$ denotes the standard normal CDF. 
For the $t$ copula, we simulate
\[
\boldsymbol z_t \sim t_D(\boldsymbol 0,\bar{\boldsymbol\Omega},\nu),
\qquad t=1,\ldots,T,
\]
via the scale-mixture representation
\[
\boldsymbol z_t
=
\frac{1}{\sqrt{s_t/\nu}}\boldsymbol x_t,
\qquad
\boldsymbol x_t \sim \mathcal{N}_D(\boldsymbol 0,\bar{\boldsymbol\Omega}),
\qquad
s_t \sim \chi^2_\nu,
\]
with $\boldsymbol x_t$ and $s_t$ independent. The copula data are then obtained by
\[
u_{d,t}=T_\nu(z_{d,t}),
\qquad d=1,\ldots,D \textrm{ ~and~ } t=1,\ldots,T,
\]
where $T_\nu$ is the CDF of the univariate Student's $t$ distribution with $\nu$ degrees of freedom.

For the marginal innovation process, we obtain the 
innovations
$\epsilon_{d,t}$ by transforming the copula ``data'' ${u}_{d,t}$.
Specifically, under Gaussian errors, we set
\[
\epsilon_{d,t}=\Phi^{-1}({u}_{d,t}),\qquad d=1,\ldots,D \textrm{ ~and~ } t=1,\ldots,T,
\]
where $\Phi^{-1}(\cdot)$ denotes the standard normal quantile function.
Under Student's-$t$ errors, we instead use
\[
\epsilon_{d,t}={\sqrt{\frac{\widetilde{\nu}_d-2}{\widetilde{\nu}_d}}\;}T^{-1}_{\widetilde{\nu}_d}({u}_{d,t}),\qquad d=1,\ldots,D \textrm{ ~and~ } t=1,\ldots,T,
\]
where $T^{-1}_{\widetilde{\nu}_d}(\cdot)$ is the quantile function of the
Student's $t$ distribution with $\widetilde{\nu}_d$ degrees of freedom.
{The scaling ensures $\mathbb{E}(\epsilon_{d,t})=0$ and $\textrm{Var}(\epsilon_{d,t})=1$.}
The degrees-of-freedom parameter is generated independently for each margin,
$\widetilde{\nu}_d \sim p(\widetilde{\nu}_d)$ {truncated to $\widetilde{\nu}_d>2$}, for $d=1,\ldots,D$.

Finally, for each marginal series, we draw the GARCH(1,1) parameters
$\alpha_{1,d}$, $\alpha_{2,d}$, and $\gamma_d$ from their prior distributions.
Given these draws and the innovations $\epsilon_{d,t}$, we recursively compute the conditional variances $\sigma^2_{d,t}$ and
 observations $y_{d,t}$  using \eqref{eqn:garch_var} and \eqref{eqn:garch_y}  from the main paper, respectively, for all
$d=1,\ldots,D$ and $t=1,\ldots,T$.

Algorithm~\ref{alg:copula_sim} can also be used in reduced form, depending on the simulation objective. 
For copula-only simulation, Steps~1--12 are sufficient, as they generate the copula parameters from their prior distributions, form $\bar{\boldsymbol\Omega}$, simulate $z_{d,t}$, and produce the uniform variables $u_{d,t}$, for $d=1,...,D$ and $t=1,...,T$. 
If interest instead lies solely in simulating the GARCH(1,1) marginals, then Steps~19--21 are sufficient, provided that the innovations $\epsilon_{d,t}$ are generated  independently from either a standard Gaussian distribution or a {standardised} Student's-$t$ distribution.

\begin{algorithm}[H]
\caption{Simulation of a copula with GARCH(1,1) marginals}
\label{alg:copula_sim}
\begin{algorithmic}[1]
\Require Dimension $D$, number of factors $k \in \{1,2,3,4\}$, series length $T = 1000$, copula type (Gaussian or $t$), prior distributions for the model parameters from Table~\ref{tab:parameters}.

\State Sample $\textrm{vech}(\widetilde{\boldsymbol G}) \sim p(\textrm{vech}(\widetilde{\boldsymbol G}))$.
\If{$t$ copula}
    \State Sample copula degrees of freedom $\nu \sim p(\nu)$.
\EndIf

\State Compute $\bar{\boldsymbol \Omega}$ following Section~\ref{sec:factorcopula} of the main paper.

\If{Gaussian copula}
    \State Generate $\boldsymbol z_t \sim \mathcal{N}_D(\boldsymbol 0, \bar{\boldsymbol \Omega})$ for $t=1,\dots,T$.
    \State Compute $u_{d,t} = \Phi(z_{d,t})$ for $d=1,\dots,D$ and $t=1,\dots,T$.
\Else \Comment{$t$ copula}
    \State Generate $\boldsymbol z_t \sim t_D(\boldsymbol 0, \bar{\boldsymbol \Omega}, \nu)$ for $t=1,\dots,T$.
    \State Compute $u_{d,t} = T_{\nu}(z_{d,t})$ for $d=1,\dots,D$ and $t=1,\dots,T$.
\EndIf

\Statex \textit{Explanation:} If the goal is only to simulate copula data, then Steps~1--12 are sufficient. These steps generate the copula parameters, construct $\bar{\boldsymbol \Omega}$, simulate $z_{d,t}$, and transform them to the uniform variables $u_{d,t}$.

\If{Gaussian errors}
    \State Set $\epsilon_{d,t} = \Phi^{-1}(u_{d,t})$ for $d=1,\dots,D$ and $t=1,\dots,T$.
\Else \Comment{Student's-$t$ errors}
    \State Sample $\widetilde{\nu}_{d} \sim p(\widetilde{\nu}_{d})$ for $d = 1,\dots,D$.
    \State Set $\epsilon_{d,t} = {\sqrt{\frac{\widetilde{\nu}_d-2}{\widetilde{\nu}_d}}\;}T^{-1}_{\widetilde{\nu}_{d}}(u_{d,t})$ for $d=1,\dots,D$ and $t=1,\dots,T$.
\EndIf

\State Sample GARCH parameters: $\alpha_{1,d} \sim p(\alpha_{1,d})$, $\alpha_{2,d} \sim p(\alpha_{2,d})$, and $\gamma_d \sim p(\gamma_d)$.
\State Compute $\sigma_{d,t}^2$ using \eqref{eqn:garch_var} for $d=1,\dots,D$ and $t=1,\dots,T$.
\State Compute $y_{d,t}$ using \eqref{eqn:garch_y} for $d=1,\dots,D$ and $t=1,\dots,T$.
\Statex \textit{Explanation:} If the goal is only to simulate the GARCH(1,1) marginals, then only Steps~19--21 are required, after generating innovations $\epsilon_{d,t}$ independently from either a standard Gaussian distribution or a Student's-$t$ distribution.

\State \Return the simulated time series, $y_{d,t}$, for $d=1,\dots,D$ and $t=1,\dots,T$.
\end{algorithmic}
\end{algorithm}

\section{Convolutional Neural Networks for GARCH(1,1) models } \label{supp:cnn}
Convolutional neural networks (CNNs) are most commonly used for image and other
spatially structured data \citep{Albawi2017CNN}. Here, we use a one-dimensional CNN
to learn a mapping from a univariate time series $\boldsymbol{y}_\textrm{mar}$ 
to the parameters of
the approximate posterior distribution, 
{$\boldsymbol{\lambda}_{\textrm{mar}}(\boldsymbol{y}_\textrm{mar}    ;\boldsymbol{\gamma}_{\textrm{mar}})$, following Algorithm \ref{alg:firststage} in Section \ref{sec:firststage} of the main paper}.



We now describe the 1D-CNN mechanism for the generic univariate time series $y_t$ for $t=1,...,1000$. The 1D-CNN applies a 1D-kernel to adjacent observations
in the time series, enabling the network to capture local temporal dependence. We use kernel size \(3\), stride \(2\), and padding \(1\) so that the sequence length is
halved at each convolutional layer.
The first convolutional operation can be written as
\[
x_{c,1} = k_{c,1}\cdot0  + k_{c,2}\,y_{1}  +  k_{c,3}\,y_{2} + b_c,
\]
where \(c = 1, \dots, 8\) indexes the output channels, \(k_{c,1}, k_{c,2}, k_{c,3}\) are
the kernel weights for channel \(c\), and \(b_c\) is the corresponding bias term.
The second convolutional operation shifts the kernel forward by two time steps,
reflecting the stride of \(2\).
Figure \ref{fig:cnn_kernel} shows the first (black) and second (orange) convolution operations for the
first 1D-CNN (\texttt{Conv1d}) layer in Table~\ref{supptab:garch_arch}, which summarises the network architecture for the GARCH(1,1) model with Gaussian errors implemented in \texttt{PyTorch} \citep{paszke2019pytorch}.

\begin{figure}[t]
\centering
\begin{tikzpicture}[
  >={Latex[length=2.2mm]},
  line/.style={line width=0.9pt},
  nodeR/.style={rectangle, rounded corners=2pt, draw, minimum height=9mm, minimum width=14mm, inner sep=2pt, font=\bfseries},
  dashedarrow/.style={->, dashed, thick, draw=black},
  goldarrow/.style={->, dashed, thick, color=orange!80!black}
]

\node[nodeR] (y1) {$0$};
\node[nodeR, right=0cm of y1] (y2) {$y_1$};
\node[nodeR, right=0cm of y2] (y3) {$y_2$};
\node[nodeR, right=0cm of y3] (y4) {$y_3$};
\node[nodeR, right=0cm of y4] (y5) {$y_4$};

\node[right=0.7cm of y5] (dots1) {$\dots$};

\node[nodeR, right=0.7cm of dots1] (y996) {$y_{997}$};
\node[nodeR, right=0cm of y996] (y997) {$y_{998}$};
\node[nodeR, right=0cm of y997] (y998) {$y_{999}$};
\node[nodeR, right=0cm of y998] (y999) {$y_{1000}$};
\node[nodeR, right=0cm of y999] (y1000) {$0$};

\node[nodeR, below=2cm of y2] (x1) {$x_{c,1}$};
\node[nodeR, right=0cm of x1] (x2) {$x_{c,2}$};
\node[nodeR, right=0cm of x2] (x3) {$x_{c,3}$};
\node[nodeR, right=0cm of x3] (x4) {$x_{c,4}$};

\node[right=0.7cm of x4] (dots2) {$\dots$};

\node[nodeR, right=0.7cm of dots2] (x497) {$x_{c,497}$};
\node[nodeR, right=0cm of x497] (x498) {$x_{c,498}$};
\node[nodeR, right=0cm of x498] (x499) {$x_{c,499}$};
\node[nodeR, right=0cm of x499] (x500) {$x_{c,500}$};

\draw[dashedarrow] (y1.south) -- node[midway, midway, fill=white] {$k_{c,1}$} (x1.north);
\draw[dashedarrow] (y2.south) -- node[midway, midway, fill=white] {$k_{c,2}$} (x1.north);
\draw[dashedarrow] (y3.south) -- node[midway, midway, fill=white] {$k_{c,3}$} (x1.north);

\draw[goldarrow] (y3.south) -- node[midway, midway, fill=white] {$k_{c,1}$} (x2.north);
\draw[goldarrow] (y4.south) -- node[midway, midway, fill=white] {$k_{c,2}$} (x2.north);
\draw[goldarrow] (y5.south) -- node[midway, midway, fill=white] {$k_{c,3}$} (x2.north);

\end{tikzpicture}
\caption{
    First (black) and second (orange) one-dimensional convolutional operations with 8 channels. Each channel applies a kernel of size 3 to the input time series ${y}_{t}$, for $t=1,...,1000$, producing the outputs ${x}_{c}$, for $c=1,...,8$. The second operation (orange) illustrates the shift induced by a stride of 2. The figure is adapted from \cite{Li2021Jan}.} 
    \label{fig:cnn_kernel}
\end{figure}
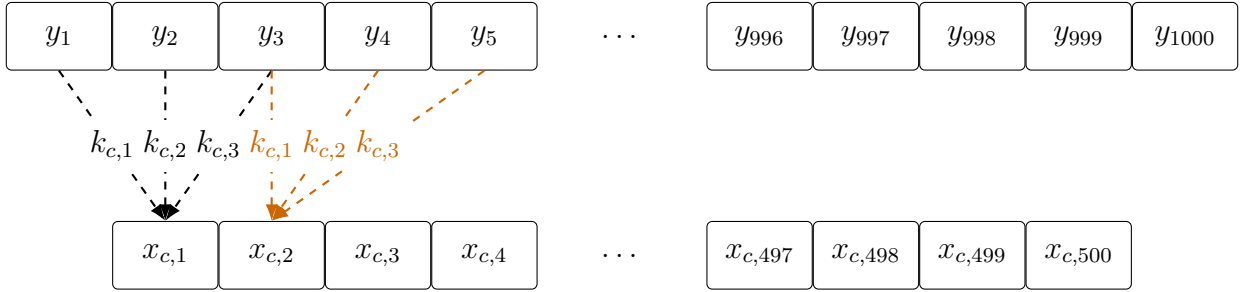

\begin{table}[t] 
\caption{Neural network architecture for the  GARCH(1,1) model with Gaussian errors, featuring a shared backbone and three output heads. The architecture is shown for a batch size of 32. \texttt{ReLU} activations are used throughout, except for the \texttt{Flatten} operation, which has no activation, and the final \texttt{Linear} layer of the middle output head, which uses a \texttt{SoftPlus} activation. The kernel size for the 1D-CNN (\texttt{Conv1d}) is 3.}
    \centering
    \begin{tabular}{llll}
    \hline \hline
        Layer & Input Shape & Output Shape & \# Parameters  \\ \hline
        Conv1d  & [32, 1, 1000] & [32, 8, 500] & 32 \\ 
        Conv1d & [32, 8, 500] & [32, 16, 250] & 400 \\ 
        Conv1d & [32, 16, 250] & [32, 32, 125] & 1,568 \\ 
        Flatten & [32, 32, 125] & [32, 4000] & - \\ \hline
        Linear (mean branch) & [32, 4000] & [32, 512] & 2,048,512 \\ 
        Linear & [32, 512] & [32, 256] & 131,328 \\ 
        Linear & [32, 256] & [32, 128] & 32,896 \\ 
        Linear & [32, 128] & [32, 3] & 387 \\ \hline
        Linear (diagonal branch) & [32, 4000] & [32, 512] & 2,048,512 \\ 
        Linear  & [32, 512] & [32, 256] & 131,328 \\ 
        Linear & [32, 256] & [32, 128] & 32,896 \\ 
        Linear & [32, 128] & [32, 3] & 387 \\ \hline
        Linear (lower-triangular) & [32, 4000] & [32, 512] & 2,048,512 \\ 
        Linear & [32, 512] & [32, 256] & 131,328 \\ 
        Linear & [32, 256] & [32, 128] & 32,896 \\ 
        Linear & [32, 128] & [32, 3] & 387 \\ \hline
        Total trainable parameters & ~ & ~ & 6,641,369 \\ \hline
    \end{tabular}
    \label{supptab:garch_arch}
\end{table}

By stacking multiple one-dimensional convolutional layers, the network learns temporal dependencies and produces informative summary statistics. These
features are flattened into a single vector and passed to three multilayer perceptron (MLP)
heads that output the parameters of a Gaussian approximate posterior distribution.

The Gaussian posterior approximation is parameterised by a mean vector and a covariance
matrix; see Section~\ref{sec:firststage} of the main paper. We reparameterise the covariance
matrix via its Cholesky factor, so that the network outputs (i) the mean vector,
(ii) the positive diagonal elements of the Cholesky factor, and (iii) the strictly
lower-triangular elements, respectively. Table~\ref{supptab:garch_arch} reports the full
architecture for the GARCH(1,1) model with Gaussian errors. For the GARCH(1,1) model with
Student's-\(t\) errors, we retain the same architecture but change the output dimensions of
the final layers of the three MLP heads to \([32,4]\), \([32,4]\), and \([32,6]\), respectively.



\section{Deep Sets for copula models} \label{supp:deepsets}

This section describes Deep Sets network architecture of \cite{zaheer_deepsets}, which is used to learn a mapping from copula ``data'' \(\boldsymbol{u}\)
to the parameters of the approximate posterior distribution,
\(\boldsymbol{\lambda}_{\textrm{cop}}(\boldsymbol{u};\boldsymbol{\gamma}_{\textrm{cop}})\),
following Algorithm~\ref{alg:secondstage} in Section~\ref{sec:secondstage} of the main paper.

During training, for each sample \(\textrm{vech}(\widetilde{\boldsymbol{G}}^{(n)})\),
\(n=1,\ldots,N\) with \(N=30,000\), we simulate \(T=1000\) i.i.d.\ realisations of $\boldsymbol z^{(n)}_t$ according to
the procedure in Section~\ref{supp:datagen_cop}. These replicates form the
exchangeable set \(\mathcal Z^{(n)} = \{\boldsymbol z^{(n)}_t\}_{t=1}^T\), and we compute the
corresponding copula ``data'' \(\mathcal U^{(n)} = \{\boldsymbol u^{(n)}_t\}_{t=1}^T\).
Because the replicates are exchangeable, any permutation of \(\mathcal U^{(n)}\) represents
the same information, which aligns naturally with the permutation-invariant Deep Sets
architecture.

\noindent The Deep Sets representation is
\begin{gather}
\boldsymbol{\lambda}_{\textrm{cop}}\!\left(\mathcal U^{(n)};\boldsymbol{\gamma}_{\textrm{cop}}\right)
=
\boldsymbol{\phi}\!\left(
\boldsymbol{\widetilde{T}}\!\left(\mathcal U^{(n)};\boldsymbol{\gamma}_\psi\right);\boldsymbol{\gamma}_\phi
\right), \label{eq:deepsets_inference}\\
\boldsymbol{\widetilde{T}}\!\left(\mathcal U^{(n)};\boldsymbol{\gamma}_\psi\right)
=
\boldsymbol{a}\!\left(
\left\{\boldsymbol{\psi}\!\left(\boldsymbol u^{(n)}_t;\boldsymbol{\gamma}_\psi\right)\right\}_{t=1}^T
\right),  \label{eq:deepsets_summary}
\end{gather}
{where \(\boldsymbol{\psi}(\cdot; \boldsymbol{\gamma}_\psi)\) and \(\boldsymbol{\phi}(\cdot; \boldsymbol{\gamma}_\phi)\) are neural networks and}
\(\boldsymbol{\gamma}_{\textrm{cop}} = (\boldsymbol{\gamma}_\phi^{\top}, \boldsymbol{\gamma}_\psi^{\top})^{\top}\)
collects all network parameters. Moreover, \(\boldsymbol{a}(\cdot)\) is a
permutation-invariant aggregation function, taken here to be the elementwise mean.
The learned summary statistic \(\boldsymbol{\widetilde{T}}\) is passed to the
inference network
{\(\boldsymbol{\phi}(\cdot; \boldsymbol{\gamma}_\phi)\),}
which outputs
\(\boldsymbol{\lambda}_{\textrm{cop}}\). 
Figure \ref{fig:deepsets} shows the schematic of the Deep Sets architecture.




\begin{figure}[t]
\centering
\tikzset{
  >={Latex[length=2.2mm]},
  line/.style={line width=0.9pt},
  nodeC/.style={circle, draw, minimum size=11mm, inner sep=0pt, font=\large},
  nodeR/.style={rectangle, rounded corners=2pt, draw, minimum height=11mm, minimum width=18mm, inner sep=2pt, font=\bfseries},
  bracebelow/.style={decorate, decoration={brace, amplitude=7pt, mirror}},
  labelsmall/.style={font=\small},
  annot/.style={font=\scriptsize, inner sep=1pt, fill=white, text opacity=1},
  dataClr/.style ={draw=black},
  netClr/.style  ={draw=black, fill=black!3},
  varClr/.style  ={draw=black},
  parClr/.style  ={draw=black},
  arrowClr/.style={draw=black},
  sampleClr/.style={draw=black}
}

\begin{tikzpicture}[node distance=1.8cm, baseline={(current bounding box.center)}]
  \node[nodeC, dataClr] (u1) {$\boldsymbol u^{(n)}_1$};
  \node[nodeC, below=1.3cm of u1] (uT) {$\boldsymbol u^{(n)}_T$};
  \node at ($(u1)!0.5!(uT)$) {\Large$\vdots$};
  
  \node[nodeR, right=1.2cm of u1, netClr] (psi1) {$\boldsymbol \psi(\cdot; \boldsymbol\gamma_\psi)$};
  \node[nodeR, right=1.2cm of uT, netClr] (psiT) {$\boldsymbol \psi(\cdot;\boldsymbol \gamma_\psi)$};

  \node[nodeR, right=1.5cm of $(psi1)!0.5!(psiT)$, netClr] (agg) {$\boldsymbol a(\cdot)$};
  \node[nodeR, right=1.2cm of agg, netClr] (phi) {$\boldsymbol \phi(\cdot;\boldsymbol \gamma_\phi)$};

  \node at ($(psi1)!0.5!(psiT)$) {\Large$\vdots$};
  
  \node[nodeC, right=1.5cm of phi, varClr] (lambda) {$\boldsymbol \lambda_{\textrm{cop}}$};

  \draw[line, arrowClr, ->] (u1) -- (psi1);
  \draw[line, arrowClr, ->] (uT) -- (psiT);

  \draw[line, arrowClr, ->] (psi1) -- (agg);
  \draw[line, arrowClr, ->] (psiT) -- (agg);

  \draw[line, arrowClr, ->] (agg) -- node[above, labelsmall] {$\widetilde{ \boldsymbol {T}}$} (phi);
  \draw[line, arrowClr, ->] (phi) -- (lambda);

\end{tikzpicture}

\caption{Schematic of the Deep Sets neural network architecture. Each independent realisation \( \boldsymbol u_t^{(n)} \) is transformed independently by the function \( \boldsymbol{\psi}(\cdot; \boldsymbol \gamma_\psi) \). The set of transformed inputs is then aggregated elementwise using a permutation-invariant function \( \boldsymbol{a}(\cdot) \), yielding the summary statistic \( \boldsymbol{\widetilde{T}} \). Finally, the summary statistic is mapped to parameter estimates \( \boldsymbol{\lambda}_{\textrm{cop}} \) by the function \( \boldsymbol{\phi}(\cdot; \boldsymbol \gamma_\phi) \).}
\label{fig:deepsets}
\end{figure}
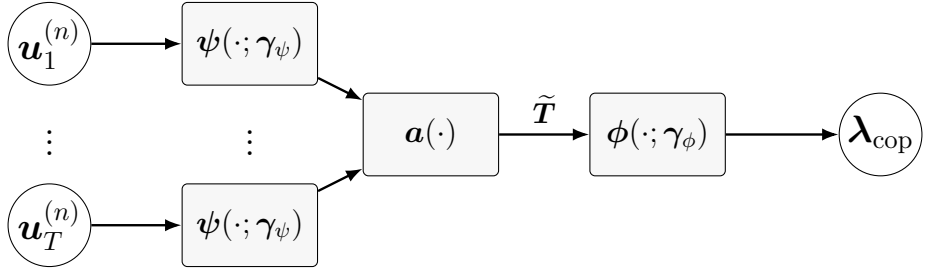

Table~\ref{supptab:copula_arch} summarises the Deep Sets neural network architecture
used for the one-factor Gaussian copula model, implemented in \texttt{PyTorch}. Each
copula observation
{\(\boldsymbol{u}_t^{(n)}\)}
is encoded independently using a shared sequence
of fully connected layers, denoted by {\(\boldsymbol{\psi}(\cdot; \boldsymbol{\gamma}_\psi)\)}. The resulting
variables are then aggregated by taking the mean over the set dimension
(the \texttt{Mean} layer, corresponding to the permutation-invariant function
\(\boldsymbol{a}(\cdot)\)), producing the summary statistic \(\boldsymbol{\widetilde{T}}\). Finally,
\(\boldsymbol{\widetilde{T}}\) is passed to two MLP output heads {\(\boldsymbol{\phi}(\cdot; \boldsymbol{\gamma}_\phi)\)},
implemented as fully connected layers with layer normalisation to stabilise training,
which output the parameters of the Gaussian approximate posterior distribution
\(\boldsymbol{\lambda}_{\textrm{cop}}\).

\begin{table}[t] 
 \caption{Neural network architecture of the one-factor $20$-dimensional Gaussian copula model, featuring a shared backbone and two output heads in a Deep Sets framework with a batch size of 32. \texttt{ReLU} activations are used throughout, except for the \texttt{Mean} operation (no activation) and the final \texttt{Linear} layer of the diagonal head (\texttt{SoftPlus} activation).}
    \centering
    \begin{tabular}{llll}
    \hline \hline
        Layer & Input Shape & Output Shape & \# Parameters \\ \hline
        Linear & [32, 1000, 20] & [32, 1000, 64] & 1,344 \\ 
        Linear & [32, 1000, 64] & [32, 1000, 128] & 8,320 \\ 
        Linear & [32, 1000, 128] & [32, 1000, 256] & 33,024 \\ 
        Linear & [32, 1000, 256] & [32, 1000, 512] & 131,584 \\ 
        Mean & [32, 1000, 512] & [32, 512] & -\\ \hline
        Linear (mean branch) & [32, 512] & [32, 1024] & 525,312 \\ 
        LayerNorm & [32, 1024] & [32, 1024] & 2,048 \\ 
        Linear & [32, 1024] & [32, 512] & 524,800 \\ 
        LayerNorm & [32, 512] & [32, 512] & 1,024 \\ 
        Linear & [32, 512] & [32, 256] & 131,328 \\ 
        LayerNorm & [32, 256] & [32, 256] & 512 \\ 
        Linear & [32, 256] & [32, 128] & 32,896 \\ 
        LayerNorm & [32, 128] & [32, 128] & 256 \\ 
        Linear & [32, 128] & [32, 20] & 2,580 \\ \hline
        Linear (diagonal branch) & [32, 512] & [32, 1024] & 525,312 \\ 
        LayerNorm & [32, 1024] & [32, 1024] & 2,048 \\ 
        Linear & [32, 1024] & [32, 512] & 524,800 \\ 
        LayerNorm & [32, 512] & [32, 512] & 1,024 \\ 
        Linear & [32, 512] & [32, 256] & 131,328 \\ 
        LayerNorm & [32, 256] & [32, 256] & 512 \\ 
        Linear & [32, 256] & [32, 128] & 32,896 \\ 
        LayerNorm & [32, 128] & [32, 128] & 256 \\ 
        Linear & [32, 128] & [32, 20] & 2,580 \\ \hline
        Total trainable parameters & & & 2,615,784 \\ \hline
    \end{tabular}
\label{supptab:copula_arch}
\end{table}

The Gaussian posterior approximation is parameterised by a mean vector and a diagonal
covariance matrix. Accordingly, the two output heads return (i) the mean vector and
(ii) the diagonal entries (variances) of the covariance matrix, constrained to be
positive via \texttt{SoftPlus}.

For higher-factor copula models, we replicate the network for each additional factor.
For example, in the two-factor model, the first network outputs the approximate posterior
parameters associated with the  {non-zero elements of the} first column of \(\widetilde{\boldsymbol{G}}\), while the
second network outputs the corresponding parameters for the second column. For the
four-factor copula model, we include an additional \texttt{Linear} layer in the summary
network to increase representational capacity when constructing the summary statistic. {Tables~\ref{tab:copula_f4_arch} and \ref{tab:factor_heads} show the full neural network architecture for the four-factor Gaussian copula.}

\begin{table}[ht]
\caption{{Neural network architecture for one of the four networks of the four-factor 20-dimensional Gaussian copula model.} 
The shared backbone maps the input set of copula
observations {$\boldsymbol{u}$} to a pooled summary representation. 
The four-factor model consists of four networks, one for each column of $\widetilde{\bs G}$, each with this architecture and its own weights; only the final output layers differ in size between factors
(Table~\ref{tab:factor_heads}). The total number of trainable parameters is $4\times 7{,}271{,}616$ plus the output layers in Table~\ref{tab:factor_heads}.
The architecture is shown for a batch size of 32. \texttt{ReLU} activations are used throughout, except for the \texttt{Mean} operation (no activation) and the output layers 
(no activation for the mean branch, \texttt{SoftPlus} for the diagonal branch).
}
\centering
\small
\begin{tabular}{llll}
\hline\hline
Layer & Input Shape & Output Shape & \# Parameters \\
\hline
Linear & [32, 1000, 20] & [32, 1000, 64] & 1,344 \\
Linear & [32, 1000, 64] & [32, 1000, 128] & 8,320 \\
Linear & [32, 1000, 128] & [32, 1000, 256] & 33,024 \\
Linear & [32, 1000, 256] & [32, 1000, 512] & 131,584 \\
Linear & [32, 1000, 512] & [32, 1000, 1024] & 525,312 \\ 
Mean & [32, 1000, 1024] & [32, 1024] & -- \\
\hline

Linear (mean branch) & [32, 1024] & [32, 1536] & 1,574,400 \\
LayerNorm & [32, 1536] & [32, 1536] & 3,072 \\
Linear & [32, 1536] & [32, 768] & 1,180,416 \\
LayerNorm & [32, 768] & [32, 768] & 1,536 \\
Linear & [32, 768] & [32, 512] & 393,728 \\
LayerNorm & [32, 512] & [32, 512] & 1,024 \\
Linear & [32, 512] & [32, 256] & 131,328 \\
LayerNorm & [32, 256] & [32, 256] & 512 \\
Linear & [32, 256] & Output (factor) & see Table \ref{tab:factor_heads} \\ \hline

Linear (diagonal branch) & [32, 1024] & [32, 1536] & 1,574,400 \\
LayerNorm & [32, 1536] & [32, 1536] & 3,072 \\
Linear & [32, 1536] & [32, 768] & 1,180,416 \\
LayerNorm & [32, 768] & [32, 768] & 1,536 \\
Linear & [32, 768] & [32, 512] & 393,728 \\
LayerNorm & [32, 512] & [32, 512] & 1,024 \\
Linear & [32, 512] & [32, 256] & 131,328 \\
LayerNorm & [32, 256] & [32, 256] & 512 \\
Linear & [32, 256] & Output (factor) & see Table \ref{tab:factor_heads} \\ \hline
Total trainable parameters & & & 7,271,616 + see Table~\ref{tab:factor_heads} \\ \hline

\end{tabular}

\label{tab:copula_f4_arch}
\end{table}

\begin{table}[t]
\caption{Factor-specific output heads for factors~1 to~4 in the four-factor, $20$-dimensional Gaussian copula model. These layers correspond to the ``Output (factor)'' components shown in Table~\ref{tab:copula_f4_arch}. The network is illustrated using a batch size of $32$.}
\centering
\small
\begin{tabular}{lllll}
\hline\hline
Factor &Head & Layer & Output Shape & \# Parameters \\
\hline
1 & Mean & Linear & [32, 20] & 5,140 \\
1 & Diagonal& Linear & [32, 20] & 5,140 \\
2 &Mean & Linear & [32, 19] & 4,883 \\
2 & Diagonal& Linear & [32, 19] & 4,883 \\
3 & Mean& Linear & [32, 18] & 4,626 \\
3 & Diagonal& Linear & [32, 18] & 4,626 \\
4 &Mean & Linear & [32, 17] & 4,369 \\
4 & Diagonal& Linear & [32, 17] & 4,369 \\
\hline
\end{tabular}

\label{tab:factor_heads}
\end{table}

\section{Training neural-network models} \label{sec:nifm-training}

This section provides additional implementation details on training the neural networks within the N-IFM framework, continuing on from Section~\ref{sec:trainingneuralnetwork} of the main paper. 
Figure~\ref{suppfig:gauscopd20f1_loss} displays the training and validation loss curves for the 20-dimensional one-factor Gaussian copula model in Section \ref{sec:sim_gauscop}. 
The loss decreases rapidly in the early steps, indicating that the network quickly captures the main structure in the training data, and then declines more gradually. 
The close agreement between the training and validation curves suggests good generalisation, with no evidence of overfitting.

\begin{figure}[t]
    \centering
    \includegraphics[width=\linewidth]{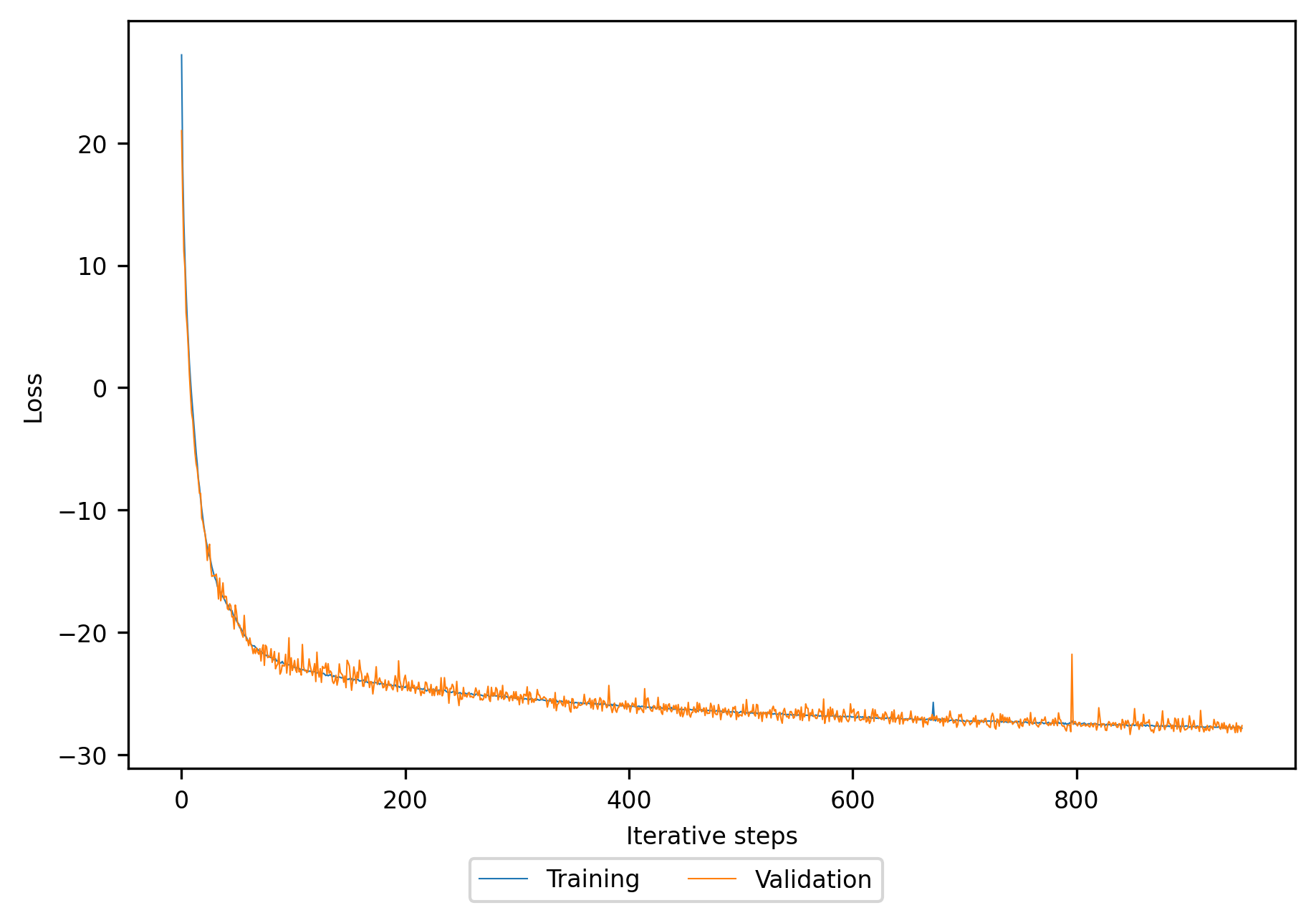}
    \caption{Training and validation loss curves for the 20-dimensional one-factor Gaussian copula over approximately 1000 iterations.}
    \label{suppfig:gauscopd20f1_loss}
\end{figure}

\clearpage
\section{Additional results for the Gaussian copula simulation study (Section~\ref{sec:sim_gauscop})} \label{supp:hmc_vs_hmcifm}

This section contains additional results for the Gaussian copula simulation study in Section \ref{sec:sim_gauscop} of the main paper. It presents additional figures and tables comparing N-IFM with HMC and HMC-IFM.


Similar to N-IFM, the HMC-IFM approach proceeds in two stages. 
First, it samples from the posterior distribution of the GARCH(1,1) parameters for each marginal series and computes their posterior means. 
Conditional on these posterior means, it then samples from the posterior distribution of the copula parameters. 
In contrast, HMC targets the joint posterior over all model parameters and updates the marginal and copula parameters simultaneously.

We consider HMC-IFM because full HMC can be computationally prohibitive for model selection, especially in sequential settings where each new data arrival  requires rerunning the sampler to obtain draws from the posterior distribution of the model parameters and to compute one-step-ahead LPDS.
To evaluate the accuracy of this two-stage approximation, we compare posterior estimates from N-IFM, HMC and HMC-IFM   using simulated data generated from the one-factor Gaussian copula model.

Figures~\ref{suppfig:alpha1_gauscop_f1} to \ref{suppfig:gamma_gauscop_f1} display the posterior distributions of the GARCH$(1,1)$ parameters and show near identical results across the three methods. 
Likewise, Figure~\ref{suppfig:gtilde_gauscop_f1} shows that the posterior distributions of the Gaussian copula parameters are also in close agreement. 
Overall, these results indicate that the two-stage N-IFM and HMC-IFM procedures closely approximate the full joint posterior obtained by standard HMC, supporting the use of HMC-IFM as a computationally efficient alternative to full HMC for model selection.

For the HMC implementation in \texttt{Stan}, standard diagnostics such as the effective sample size (\texttt{n\_eff}) and the potential scale reduction factor ($\widehat{\texttt{R}}$) are used to assess convergence. 
As reported in Table~\ref{tab:hmc_gaucsop_simf1}, \texttt{n\_eff} is large for all parameters and all $\widehat{\texttt{R}}$ values are close to 1, indicating good mixing and convergence across chains.
For HMC-IFM (not reported), all \(\widehat{\texttt{R}}\) values are close to 1, indicating good convergence, while \texttt{n\_eff} is slightly lower for the GARCH(1,1) parameters and slightly higher for the copula parameters relative to HMC.

\begin{figure}[t]
    \centering
    \includegraphics[width=\linewidth]{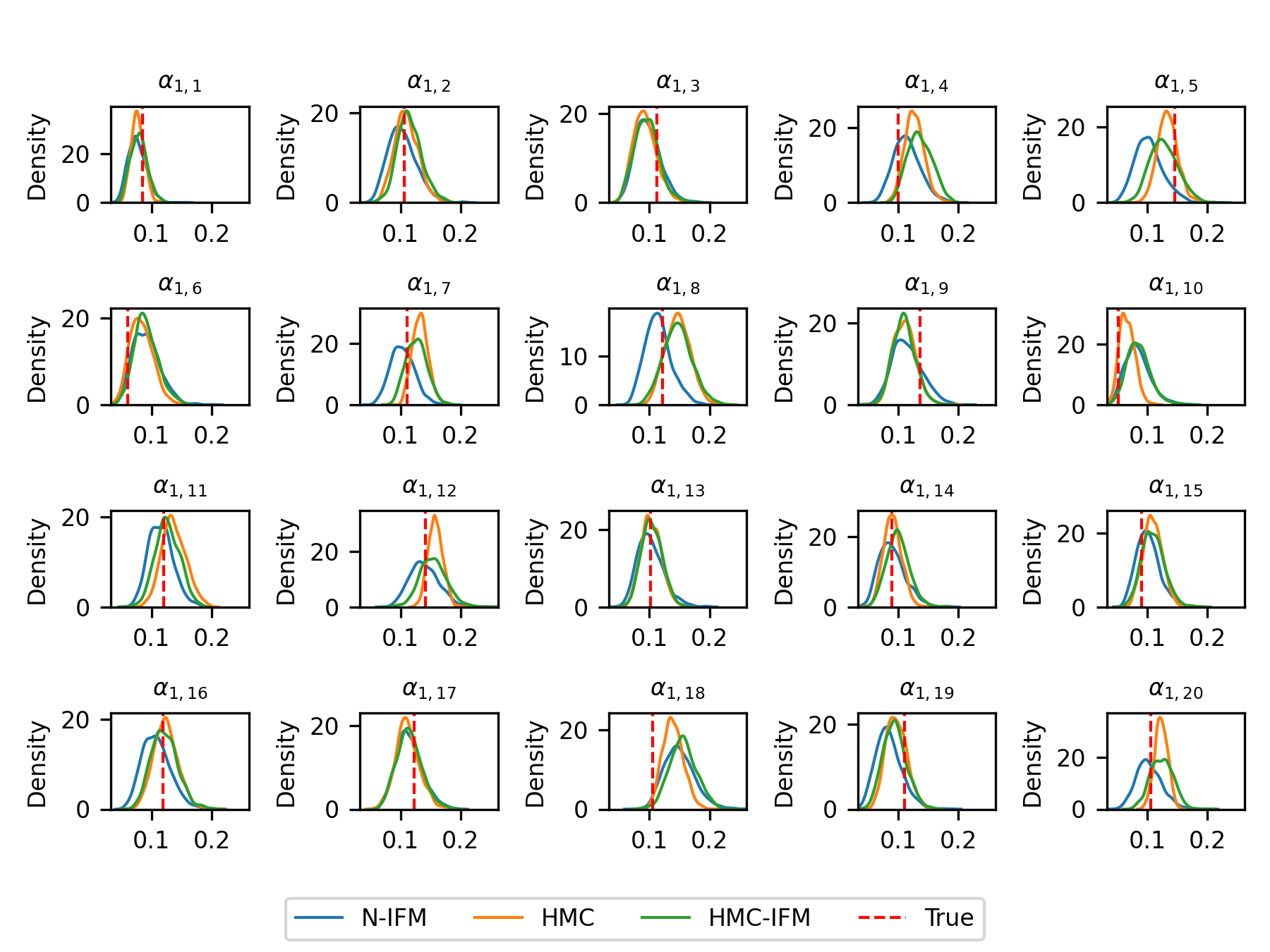}
        \caption{Posterior distributions of the GARCH$(1,1)$ parameters $\alpha_{1,d}$ for $d=1,\ldots,D$, comparing N-IFM, HMC and HMC-IFM with the true values. The results are based on simulated data from a 20 dimensional one-factor Gaussian copula with GARCH(1,1) marginals with Gaussian errors.}
    \label{suppfig:alpha1_gauscop_f1}
\end{figure}

\begin{figure}[t]
    \centering
    \includegraphics[width=\linewidth]{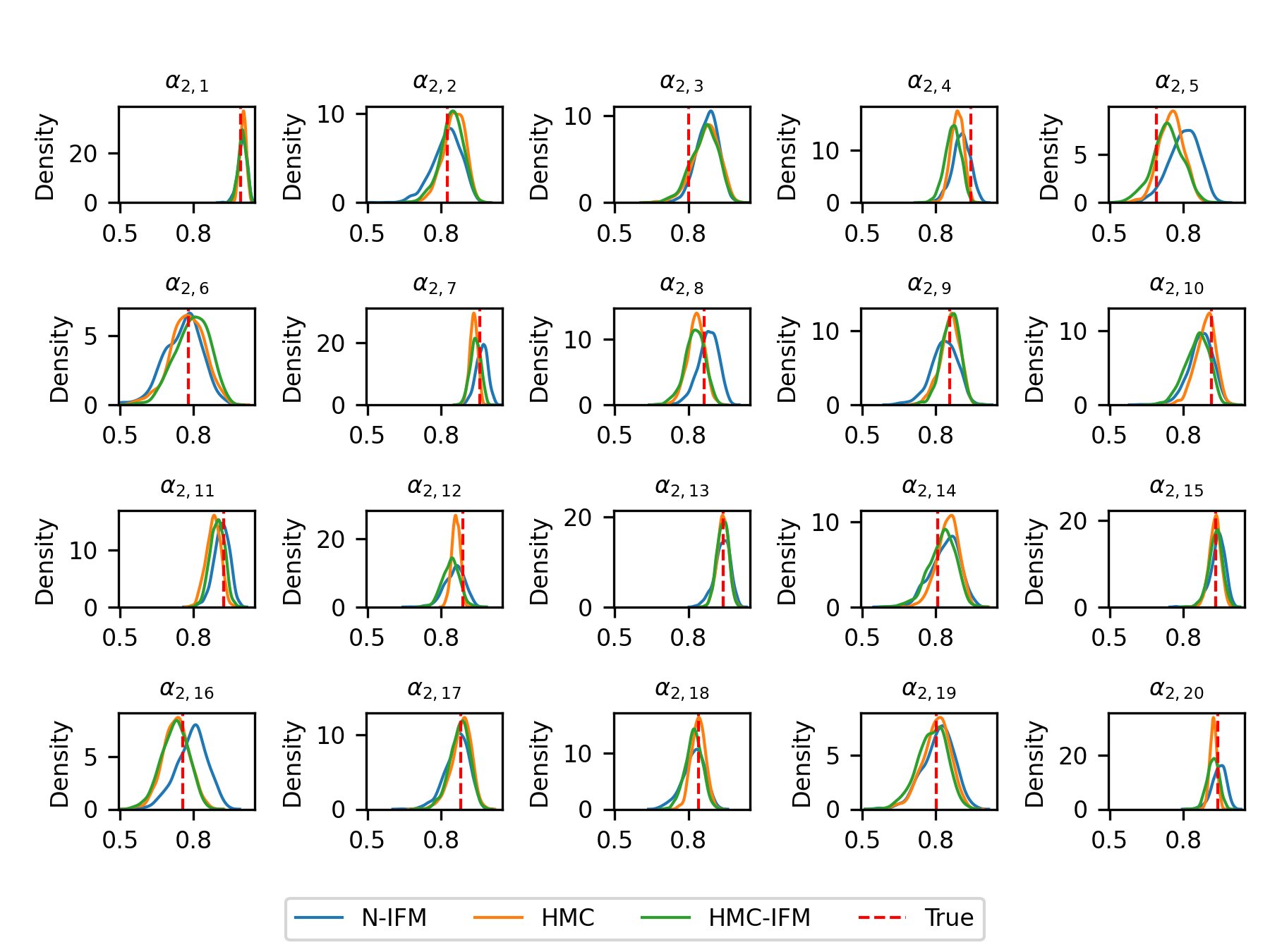}
    \caption{Posterior distributions of the GARCH$(1,1)$ parameters $\alpha_{2,d}$ for $d=1,\ldots,D$, comparing N-IFM, HMC and HMC-IFM with the true values. The results are based on simulated data from a 20 dimensional one-factor Gaussian copula with GARCH(1,1) marginals with Gaussian errors.}
    \label{suppfig:alpha2_gauscop_f1}
\end{figure}

\begin{figure}[t]
    \centering
    \includegraphics[width=\linewidth]{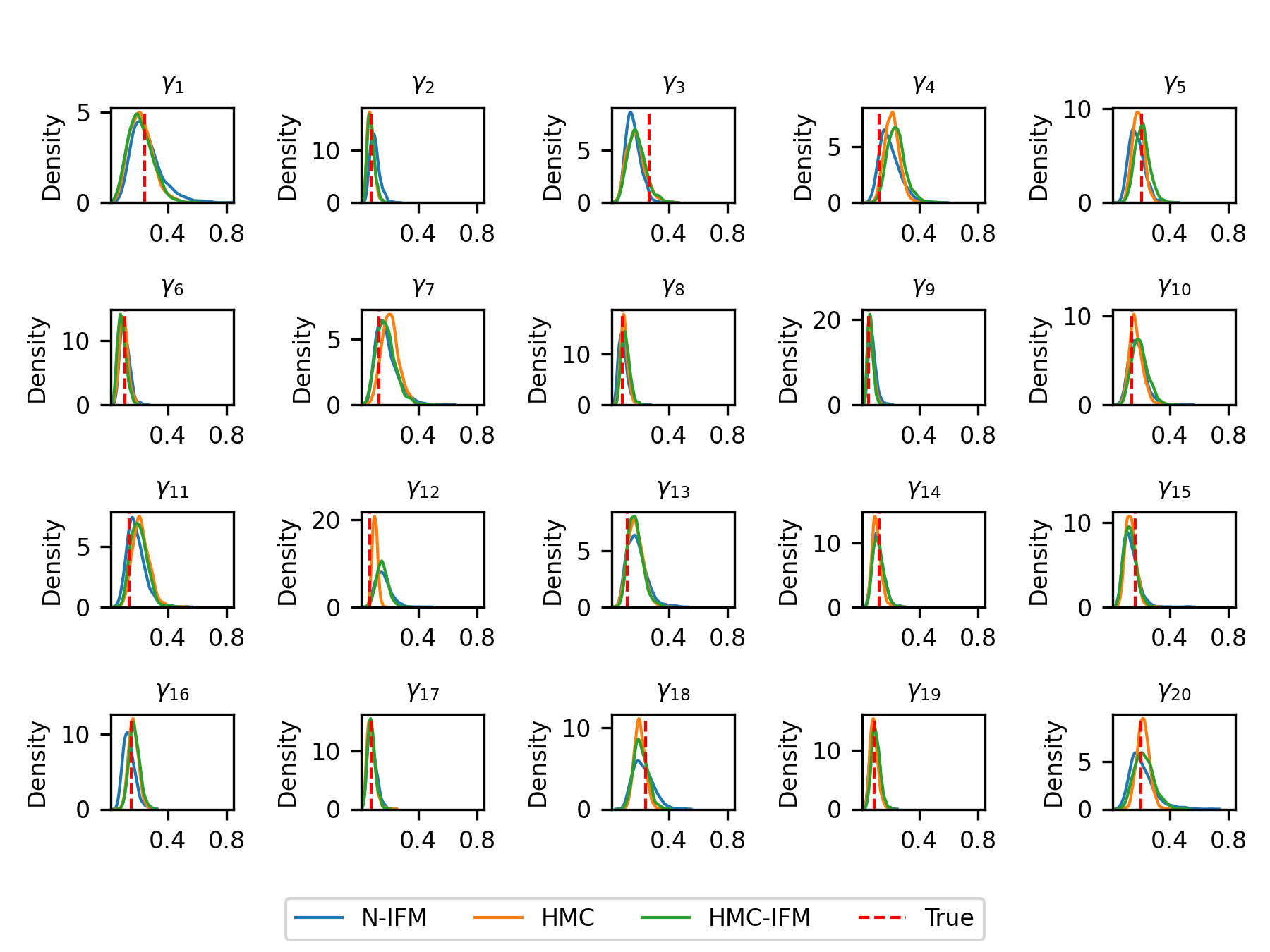}
    \caption{Posterior distributions of the GARCH$(1,1)$ parameters $\gamma_{d}$ for $d=1,\ldots,D$, comparing N-IFM, HMC and HMC-IFM with the true values. The results are based on simulated data from a 20 dimensional one-factor Gaussian copula with GARCH(1,1) marginals with Gaussian errors.}
    \label{suppfig:gamma_gauscop_f1}
\end{figure}

\begin{figure}[t]
    \centering
    \includegraphics[width=\linewidth]{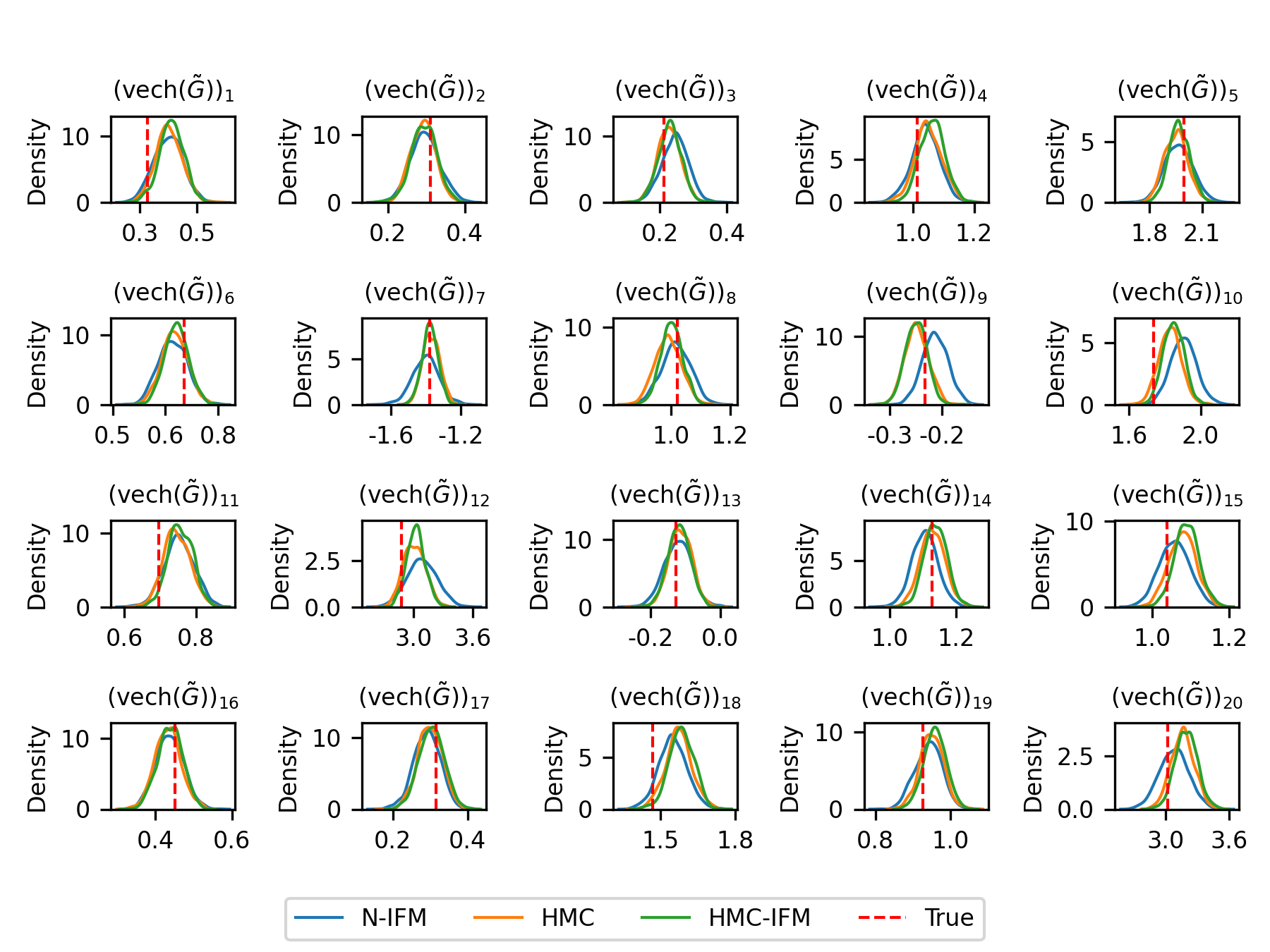}
    \caption{Posterior distributions of the copula parameters $\textrm{vech}(\widetilde{\boldsymbol{G}})$, comparing N-IFM, HMC and HMC-IFM with the true values. The results are based on  simulated data from a 20 dimensional one-factor Gaussian copula with GARCH(1,1) marginals with Gaussian errors.}
    \label{suppfig:gtilde_gauscop_f1}
\end{figure}

\newpage
\begin{table}[t]
\caption{Posterior diagnostics for model parameters estimated using HMC, based on simulated data from a 20-dimensional one-factor Gaussian copula model with GARCH$(1,1)$ marginals and Gaussian errors, including the effective sample size (\texttt{n\_eff}) and the potential scale reduction factor ($\widehat{\texttt{R}}$).}
\centering
{\small
\begin{tabular}{lll|lll}
\hline \hline
Parameter & \texttt{n\_eff} & {$\widehat{\texttt{R}}$} & Parameter & \texttt{n\_eff} & {$\widehat{\texttt{R}}$} \\
\hline
$\alpha_{1,1}$ & 7578 & 0.9997 & $\gamma_{1}$ & 5552 & 1.0004 \\
$\alpha_{1,2}$ & 5472 & 0.9999 & $\gamma_{2}$ & 4631 & 1.0000 \\
$\alpha_{1,3}$ & 5879 & 0.9998 & $\gamma_{3}$ & 4467 & 0.9996 \\
$\alpha_{1,4}$ & 4780 & 0.9995 & $\gamma_{4}$ & 5396 & 1.0000 \\
$\alpha_{1,5}$ & 5084 & 0.9996 & $\gamma_{5}$ & 4790 & 1.0014 \\
$\alpha_{1,6}$ & 6522 & 1.0000    & $\gamma_{6}$ & 5698 & 0.9996 \\
$\alpha_{1,7}$ & 6847 & 0.9997 & $\gamma_{7}$ & 4260 & 0.9996 \\
$\alpha_{1,8}$ & 4622 & 0.9996 & $\gamma_{8}$ & 4929 & 0.9997 \\
$\alpha_{1,9}$ & 5452 & 0.9999 & $\gamma_{9}$ & 5764 & 0.9998 \\
$\alpha_{1,10}$ & 5325 & 1.0000   & $\gamma_{10}$ & 5463 & 0.9997 \\
$\alpha_{1,11}$ & 5562 & 1.0002 & $\gamma_{11}$ & 4950 & 0.9997 \\
$\alpha_{1,12}$ & 3818 & 1.0003 & $\gamma_{12}$ & 5575 & 1.0000 \\
$\alpha_{1,13}$ & 5265 & 1.0000   & $\gamma_{13}$ & 4667 & 0.9996 \\
$\alpha_{1,14}$ & 5704 & 0.9999 & $\gamma_{14}$ & 5470 & 0.9997 \\
$\alpha_{1,15}$ & 5443 & 0.9999 & $\gamma_{15}$ & 4753 & 1.0000 \\
$\alpha_{1,16}$ & 6046 & 1.0003 & $\gamma_{16}$ & 6173 & 0.9999 \\
$\alpha_{1,17}$ & 3876 & 1.0001 & $\gamma_{17}$ & 3242 & 1.0014 \\
$\alpha_{1,18}$ & 4968 & 1.0003 & $\gamma_{18}$ & 5865 & 0.9996 \\
$\alpha_{1,19}$ & 6080 & 0.9998 & $\gamma_{19}$ & 6235 & 0.9996 \\
$\alpha_{1,20}$ & 3816 & 0.9997 & $\gamma_{20}$ & 4701 & 1.0001 \\
$\alpha_{2,1}$ & 8312 & 0.9996 & $(\textrm{vech}(\widetilde{G}))_{1}$ & 3564 & 0.9996 \\
$\alpha_{2,2}$ & 4743 & 0.9996 & $(\textrm{vech}(\widetilde{G}))_{2}$ & 6387 & 0.9999 \\
$\alpha_{2,3}$ & 4501 & 0.9997 & $(\textrm{vech}(\widetilde{G}))_{3}$ & 7076 & 0.9996 \\
$\alpha_{2,4}$ & 6233 & 0.9996 & $(\textrm{vech}(\widetilde{G}))_{4}$ & 2874 & 1.0001 \\
$\alpha_{2,5}$ & 4415 & 1.0010  & $(\textrm{vech}(\widetilde{G}))_{5}$ & 2822 & 1.0008 \\
$\alpha_{2,6}$ & 5462 & 0.9996 & $(\textrm{vech}(\widetilde{G}))_{6}$ & 4746 & 0.9997 \\
$\alpha_{2,7}$ & 7521 & 0.9997 & $(\textrm{vech}(\widetilde{G}))_{7}$ & 4313 & 1.0000 \\
$\alpha_{2,8}$ & 5995 & 0.9997 & $(\textrm{vech}(\widetilde{G}))_{8}$ & 3447 & 0.9996 \\
$\alpha_{2,9}$ & 5613 & 0.9998 & $(\textrm{vech}(\widetilde{G}))_{9}$ & 7620 & 0.9997 \\
$\alpha_{2,10}$ & 4480 & 0.9998 & $(\textrm{vech}(\widetilde{G}))_{10}$ & 2384 & 1.0000 \\
$\alpha_{2,11}$ & 7179 & 0.9996 & $(\textrm{vech}(\widetilde{G}))_{11}$ & 4010 & 0.9998 \\
$\alpha_{2,12}$ & 6056 & 1.0000    & $(\textrm{vech}(\widetilde{G}))_{12}$ & 3465 & 1.0000 \\
$\alpha_{2,13}$ & 7363 & 0.9996 & $(\textrm{vech}(\widetilde{G}))_{13}$ & 7219 & 0.9996 \\
$\alpha_{2,14}$ & 4964 & 0.9998 & $(\textrm{vech}(\widetilde{G}))_{14}$ & 3185 & 0.9998 \\
$\alpha_{2,15}$ & 7393 & 0.9999 & $(\textrm{vech}(\widetilde{G}))_{15}$ & 3425 & 0.9995 \\
$\alpha_{2,16}$ & 5786 & 0.9999 & $(\textrm{vech}(\widetilde{G}))_{16}$ & 5171 & 0.9996 \\
$\alpha_{2,17}$ & 4805 & 1.0003 & $(\textrm{vech}(\widetilde{G}))_{17}$ & 5264 & 1.0001 \\
$\alpha_{2,18}$ & 6604 & 1.0000    & $(\textrm{vech}(\widetilde{G}))_{18}$ & 2899 & 1.0000 \\
$\alpha_{2,19}$ & 5919 & 0.9996 & $(\textrm{vech}(\widetilde{G}))_{19}$ & 3228 & 0.9999 \\
$\alpha_{2,20}$ & 6502 & 1.0001 & $(\textrm{vech}(\widetilde{G}))_{20}$ & 3140 & 0.9998 \\
\hline
\end{tabular}
}

\label{tab:hmc_gaucsop_simf1}
\end{table}

\clearpage
\section{Computing time\label{sec:computingtime}}     

This section reports the computational time required for parameter estimation in the simulation studies and the real data application described in {Sections~\ref{sec:SimulationStudy}, \ref{sec:RealData}  and  \ref{sec:sim_stcop}}.
We compare three inference algorithms: N-IFM, HMC and HMC-IFM. The corresponding computing times are summarised in Table~\ref{tab:timings}.

\begin{table}[t]
 \caption{{Computation time (in seconds) for parameter estimation in the simulation studies and real-data application using N-IFM, HMC and HMC-IFM. {N-IFM computation times were obtained from 30 repetitions of the procedure and are reported as the mean and standard deviation.} }
 }
    \centering
    \begin{tabular}{p{10cm}ccc}
        \hline \hline
        {Study} & {N-IFM} & {HMC} & {HMC-IFM} \\
        \hline
        Simulation study (Gaussian copula) & 0.56 (0.05) & 862 & 515\\
        Simulation study ($t$ copula) & 0.61 (0.04) & 14,997 & 7,122 \\
        Real Data & 0.64 (0.03) & 272,780* & 194,086* \\
        \hline
    \end{tabular}
    \vspace{4pt}
\noindent\parbox{\linewidth}{\footnotesize $^*$The four-factor HMC and HMC-IFM models require longer Markov chains (20,000 iterations) due to poor mixing and a low effective sample size.}
    \label{tab:timings}
\end{table}


\section{Rank-one factor posterior approximation for the Gaussian copula} \label{supp:gauscop_factor_q}

{The copula posterior approximation adopted in Section~\ref{sec:secondstage} of the main paper is the mean-field Gaussian form~\eqref{eqn:cop_variationalform_mf}, which enforces posterior independence among the copula parameters. To assess whether this assumption is restrictive for the Gaussian copula, we retrain the copula network for the one-factor Gaussian copula simulation setting of Section~\ref{sec:sim_gauscop} using a Gaussian approximation with a rank-one factor covariance,
\begin{equation}\label{eqn:cop_variationalform_factor}
\begin{split}
q_{\textrm{cop}}\!\left(\boldsymbol{\theta}_{\textrm{cop}};\boldsymbol{\lambda}_{\textrm{cop}}(\boldsymbol{u};\boldsymbol{\gamma}_{\textrm{cop}})\right)
&= \mathcal{N}_{m_{\textrm{cop}}}\!\Big(
\boldsymbol{\theta}_{\textrm{cop}} \,\Big|\,
\boldsymbol{\mu}_{\textrm{cop}}\!\left(\boldsymbol{u};\boldsymbol{\gamma}_{\textrm{cop}}\right),\;
\boldsymbol{\Sigma}_{\textrm{cop}}\!\left(\boldsymbol{u};\boldsymbol{\gamma}_{\textrm{cop}}\right)
\Big), \\[4pt]
\boldsymbol{\Sigma}_{\textrm{cop}}
&= \operatorname{diag}\!\big(\boldsymbol{\sigma}^{2}_{\textrm{cop}}\big)
+ \boldsymbol{b}_{\textrm{cop}}\,\boldsymbol{b}_{\textrm{cop}}^{\top},
\end{split}
\end{equation}
where $\bs{b}_{\textrm{cop}} = \bs{b}_{\textrm{cop}}(\bs{u};\bs{\gamma}_{\textrm{cop}}) \in \mathbb{R}^{m_{\textrm{cop}}}$ is a loading vector. Setting $\bs{b}_{\textrm{cop}} = \bs{0}$ recovers the mean-field form~\eqref{eqn:cop_variationalform_mf}. The density $-\log q_{\textrm{cop}}$ required for the training objective~\eqref{eq:LossFunction} is evaluated using the Woodbury identity and the matrix-determinant lemma, and the network is trained with the same prior and Adam protocol as the mean-field model. 
The architecture is the same as that described in Table~\ref{supptab:copula_arch} with an additional output head, which is the same in structure as that of the others, and that returns the loading vector $\boldsymbol{b}_{\textrm{cop}}$.
}

\begin{table}[t]
\caption{Neural network architecture of the one-factor $20$-dimensional Gaussian copula model with the rank-one factor Gaussian posterior approximation~\eqref{eqn:cop_variationalform_factor}, featuring a shared backbone and three output heads in a Deep Sets framework with a batch size of 32. The structure matches Table~\ref{supptab:copula_arch} with an additional head that returns the loading vector $\boldsymbol{b}_{\textrm{cop}}$. \texttt{ReLU} activations are used throughout, except for the \texttt{Mean} operation (no activation) and the final \texttt{Linear} layer of the diagonal head (\texttt{SoftPlus} activation); the mean and factor-loading heads have no final activation.}
\centering
{\small
\begin{tabular}{llll}
\hline \hline
Layer & Input Shape & Output Shape & \# Parameters \\ \hline
Linear & [32, 1000, 20] & [32, 1000, 64] & 1,344 \\
Linear & [32, 1000, 64] & [32, 1000, 128] & 8,320 \\
Linear & [32, 1000, 128] & [32, 1000, 256] & 33,024 \\
Linear & [32, 1000, 256] & [32, 1000, 512] & 131,584 \\
Mean & [32, 1000, 512] & [32, 512] & -\\ \hline
Linear (mean branch) & [32, 512] & [32, 1024] & 525,312 \\
LayerNorm & [32, 1024] & [32, 1024] & 2,048 \\
Linear & [32, 1024] & [32, 512] & 524,800 \\
LayerNorm & [32, 512] & [32, 512] & 1,024 \\
Linear & [32, 512] & [32, 256] & 131,328 \\
LayerNorm & [32, 256] & [32, 256] & 512 \\
Linear & [32, 256] & [32, 128] & 32,896 \\
LayerNorm & [32, 128] & [32, 128] & 256 \\
Linear & [32, 128] & [32, 20] & 2,580 \\ \hline
Linear (diagonal branch) & [32, 512] & [32, 1024] & 525,312 \\
LayerNorm & [32, 1024] & [32, 1024] & 2,048 \\
Linear & [32, 1024] & [32, 512] & 524,800 \\
LayerNorm & [32, 512] & [32, 512] & 1,024 \\
Linear & [32, 512] & [32, 256] & 131,328 \\
LayerNorm & [32, 256] & [32, 256] & 512 \\
Linear & [32, 256] & [32, 128] & 32,896 \\
LayerNorm & [32, 128] & [32, 128] & 256 \\
Linear & [32, 128] & [32, 20] & 2,580 \\ \hline
Linear (factor loading branch) & [32, 512] & [32, 1024] & 525,312 \\
LayerNorm & [32, 1024] & [32, 1024] & 2,048 \\
Linear & [32, 1024] & [32, 512] & 524,800 \\
LayerNorm & [32, 512] & [32, 512] & 1,024 \\
Linear & [32, 512] & [32, 256] & 131,328 \\
LayerNorm & [32, 256] & [32, 256] & 512 \\
Linear & [32, 256] & [32, 128] & 32,896 \\
LayerNorm & [32, 128] & [32, 128] & 256 \\
Linear & [32, 128] & [32, 20] & 2,580 \\ \hline
Total trainable parameters & & & 3,836,540 \\ \hline
\end{tabular}
}
\label{supptab:copula_f1_qfactor_arch}
\end{table}


{Figure~\ref{suppfig:gtilde_gauscop_f1_qfactor} shows the marginal posterior distributions of the copula parameters $\textrm{vech}(\widetilde{\boldsymbol{G}})$ for the rank-one factor and mean-field N-IFM approximations, together with HMC and HMC-IFM, for the dataset used in Section~\ref{sec:sim_gauscop} ($D = 20$, $T = 1000$). The two N-IFM approximations are close to each other and to HMC and HMC-IFM. Table~\ref{tab:gauscop_f1_qfactor_wass} reports the mean univariate Wasserstein-1 distance (Section~\ref{sec:supp_gauscop}) between the marginal posteriors, averaged over the $D = 20$ components. The rank-one factor and mean-field approximations sit at similar distances from HMC ($0.0375$ and $0.0370$) and from HMC-IFM ($0.0448$ and $0.0455$); the difference between them is negligible.}

\begin{figure}[t]
    \centering
    \includegraphics[width=\linewidth]{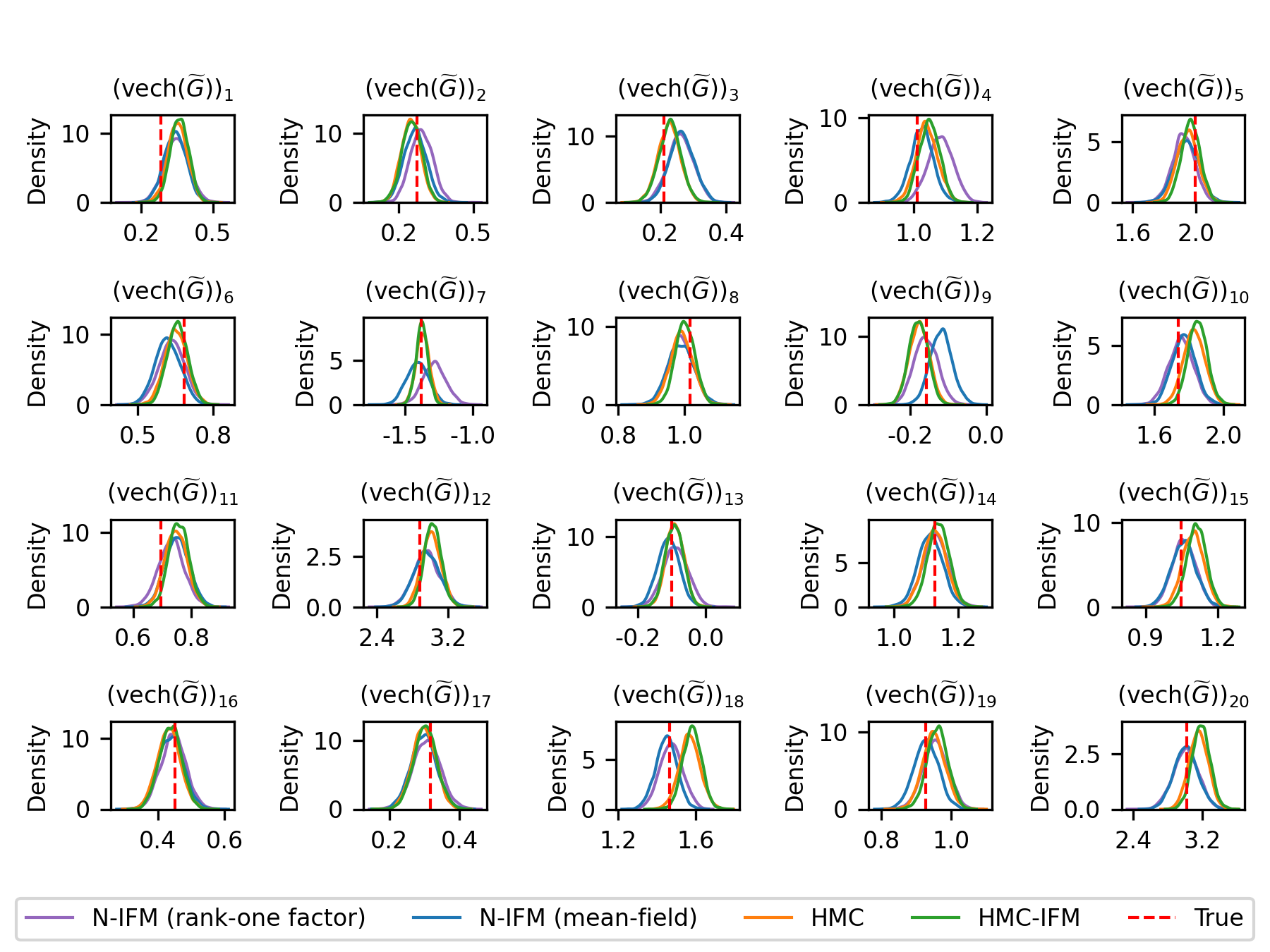}
    \caption{Posterior distributions of the copula parameters $\textrm{vech}(\widetilde{\boldsymbol{G}})$ for the rank-one factor and mean-field Gaussian approximations, compared with HMC and HMC-IFM and with the true values (vertical dashed lines). The results are based on simulated data from a 20-dimensional one-factor Gaussian copula with GARCH$(1,1)$ marginals with Gaussian errors.}
    \label{suppfig:gtilde_gauscop_f1_qfactor}
\end{figure}

\begin{table}[t]
\caption{Averaged Wasserstein-1 distance between the posterior distributions of the Gaussian copula parameters $\textrm{vech}(\widetilde{\boldsymbol{G}})$
for the rank-one factor and mean-field Gaussian approximations against HMC and HMC-IFM on simulated data from a 20-dimensional one-factor Gaussian copula model with GARCH$(1,1)$ marginals and Gaussian errors.}
\centering
\begin{tabular}{lcc}
\hline \hline
 & HMC & HMC-IFM \\
\hline
N-IFM (mean-field)          & 0.0370 & 0.0455\\
N-IFM (rank-one factor)     & 0.0375 & 0.0448\\
\hline
\end{tabular}
\label{tab:gauscop_f1_qfactor_wass}
\end{table}

{The negligible role of the factor term is not specific to this dataset. 
For each of 1000 datasets simulated from copula parameters drawn from the prior $(D = 20$, $T = 1000)$, the loading vector satisfies $\lVert \boldsymbol{b}_{\textrm{cop}}(\boldsymbol{u}) \rVert < 0.06$.
These summaries are reported in Table~\ref{tab:gauscop_f1_qfactor_bnorm}. Under the chosen priors and the copula model, the rank-one factor covariance is therefore effectively diagonal, and~\eqref{eqn:cop_variationalform_factor} reduces to the mean-field form~\eqref{eqn:cop_variationalform_mf}.}


\begin{table}[t]
\caption{The Euclidean norm of the rank-one factor loading $\boldsymbol{b}_{\textrm{cop}}(\boldsymbol{u})$ of the Gaussian posterior approximation across $1{,}000$ datasets, each simulated from the prior of the copula parameters used in training ($D = 20$, $T = 1000$). 
}
\centering
\begin{tabular}{lcccc}
\hline \hline
 & Mean & Median & 95th pct. & Max \\
\hline
$\lVert \boldsymbol{b}_{\textrm{cop}}(\boldsymbol{u}) \rVert$ & 0.033 & 0.033 & 0.038 & 0.056 \\
\hline
\end{tabular}
\label{tab:gauscop_f1_qfactor_bnorm}
\end{table}

{Taken together, these results indicate that the posterior distribution of $\textrm{vech}(\widetilde{\boldsymbol{G}})$ under the one-factor Gaussian copula exhibits negligible cross-parameter dependence at this sample size. Relaxing the mean-field assumption to a rank-one factor covariance produces no material improvement in the agreement with HMC, so the mean-field approximation~\eqref{eqn:cop_variationalform_mf} is adequate for this model.}


\clearpage
\section{Additional results for the Gaussian copula calibration study}\label{sec:supp_gauscop}

This section defines the metrics used to compare the posterior approximations obtained by N-IFM, HMC, and HMC-IFM, and presents additional figures and tables supporting the Gaussian copula calibration study in Section~\ref{sec:calib_gauscop}. The same metrics and evaluation criteria are used for the $t$ copula calibration study in Section~\ref{sec:calib_stcop}.

\subsection{Metric definitions}\label{sec:metrics}

{
Five summary measures are used to assess calibration performance. Bias and RMSE evaluate the accuracy of the point estimates, while empirical coverage and mean interval width (IW) assess the performance of the corresponding interval estimates. 
}
{
For a scalar parameter with true value $\theta$, let
$\{\theta^{(r,j)}\}_{j=1}^{J}$ denote the $J=1000$ draws from the N-IFM
posterior obtained for simulation replication $r$, where $r=1,\ldots,R$ and
$R=1000$. For replication $r$, define the posterior mean as
\[
\widehat{\theta}^{(r)}
=
\frac{1}{J}\sum_{j=1}^{J}\theta^{(r,j)},
\]
and let $q_{0.025}^{(r)}$ and $q_{0.975}^{(r)}$ denote the $2.5\%$ and
$97.5\%$ empirical quantiles, respectively, of
$\{\theta^{(r,j)}\}_{j=1}^{J}$. The resulting $95\%$ credible interval is
therefore
\[
\left[q_{0.025}^{(r)},q_{0.975}^{(r)}\right].
\]
}
{
Across the $R$ simulation replications, we assess the accuracy of the
posterior mean using the bias and root mean squared error (RMSE) where
\[
\mathrm{Bias}
=
\frac{1}{R}
\sum_{r=1}^{R}
\left(
\widehat{\theta}^{(r)}-\theta
\right),
\]
and
\[
\mathrm{RMSE}
=
\left[
\frac{1}{R}
\sum_{r=1}^{R}
\left(
\widehat{\theta}^{(r)}-\theta
\right)^2
\right]^{1/2}.
\]
The calibration and precision of the posterior credible intervals are
assessed using the empirical coverage probability and mean interval width
(IW),
\[
\mathrm{Coverage}
=
\frac{1}{R}
\sum_{r=1}^{R}
\mathbbm{1}
\left\{
q_{0.025}^{(r)}
\leq
\theta
\leq
q_{0.975}^{(r)}
\right\},
\]
and
\[
\mathrm{IW}
=
\frac{1}{R}
\sum_{r=1}^{R}
\left(
q_{0.975}^{(r)}-q_{0.025}^{(r)}
\right).
\]
}
{
These quantities are computed separately for each parameter considered:
$\alpha_{1,d}$, $\alpha_{2,d}$, $\gamma_d$, and, where applicable,
$\widetilde{\nu}_d$ for each series $d$; each free element of
$\mathrm{vech}(\widetilde{\boldsymbol{G}})$; and the scalar copula parameter
$\nu$. Bias and coverage are reported across the individual marginal or
copula parameters, whereas RMSE and IW are averaged over the elements of each parameter group, as specified in the corresponding tables and figures.
}

{
The agreement between posterior distributions obtained by different methods
is assessed using the univariate Wasserstein-1 distance
\citep{villani2009optimal}. For two one-dimensional distributions with
cumulative distribution functions (CDFs) $F$ and $G$ and finite first
moments, the Wasserstein-1 distance can be written as
\[
W_1(F,G)
=
\int_{-\infty}^{\infty}
\left|F(x)-G(x)\right|\,\mathrm{d}x,
\]
which corresponds to the area between the two CDFs. Smaller values therefore
indicate closer agreement between the two distributions, with $W_1(F,G)=0$
if and only if they are identical.}

{
To evaluate posterior agreement, we fit N-IFM, HMC-IFM, and HMC to a
dataset. For each parameter, $F$ and $G$ are taken to be the empirical
posterior distributions obtained from a given pair of methods, and the
corresponding Wasserstein-1 distance is computed. We consider all three
pairwise comparisons: N-IFM versus HMC-IFM, N-IFM versus HMC, and HMC versus
HMC-IFM. For parameter groups, the reported Wasserstein-1 distance is the
average of the element-wise distances within that group.}

\subsection{{Additional results}}

{
Figure~\ref{fig:calib_bias_gauscop} presents the distribution of 
biases in the N-IFM posterior means across the different number of factors. The biases of the GARCH(1,1) parameters remain centred close to
zero, with no systematic deterioration as the number of factors increases.
The biases of the copula parameters,
$\operatorname{vech}(\widetilde{\boldsymbol{G}})$, are also centred near zero,
but their dispersion becomes substantially greater in the four-factor model.
Table~\ref{tab:calib_rmse_gauscop} supports this interpretation. The average RMSEs of
the marginal parameters 
are similar across the four settings, whereas the average RMSE of
$\operatorname{vech}(\widetilde{\boldsymbol{G}})$ increases from $0.130$ in
the three-factor model to $0.244$ in the four-factor model.}
{

Figure~\ref{fig:calib_cov_gauscop} reports the empirical coverage of the
nominal $95\%$ credible intervals for each parameter. For the GARCH(1,1)
parameters, the median coverage is generally close to or above the nominal
level, although the dispersion becomes greater for some parameters in the
four-factor model. For
$\operatorname{vech}(\widetilde{\boldsymbol{G}})$, the median coverage is below
the nominal level in the one-factor model and at or above it with two or more factors. Nevertheless, several individual parameters have
coverages below the nominal level. As shown in
Table~\ref{tab:lowcov}, five parameters in the Gaussian-copula experiment have coverage below $50\%$: three copula parameters in $\operatorname{vech}(\widetilde{\boldsymbol{G}})$, one in the one-factor and two in the four-factor model, and $\gamma_d$ and $\alpha_{2,d}$ of series $d=2$ in the four-factor model. Most of these involve true values drawn from the tails of their prior distributions, or, for series $d=2$, a small $\gamma_d$ combined with GARCH persistence close to the stationarity boundary ($\alpha_{1,d}+\alpha_{2,d}=0.954$). Such configurations are necessarily represented relatively infrequently in a finite training sample and are therefore particularly challenging for amortised neural inference.}

{
Finally, Table~\ref{tab:calib_wass_gauscop} compares the complete univariate
posterior distributions using the Wasserstein-1 distance. HMC and HMC-IFM
remain in close agreement for
$\operatorname{vech}(\widetilde{\boldsymbol{G}})$ across all factor
specifications, with distances between $0.011$ and $0.029$. By contrast, the
distance between N-IFM and each HMC-based method increases sharply in the
four-factor model: it reaches $0.262$ relative to HMC-IFM and $0.256$ relative
to HMC. No comparable deterioration is observed for the marginal GARCH(1,1)
parameters.}

\begin{figure}[t]
    \centering
    \includegraphics[width=\linewidth]{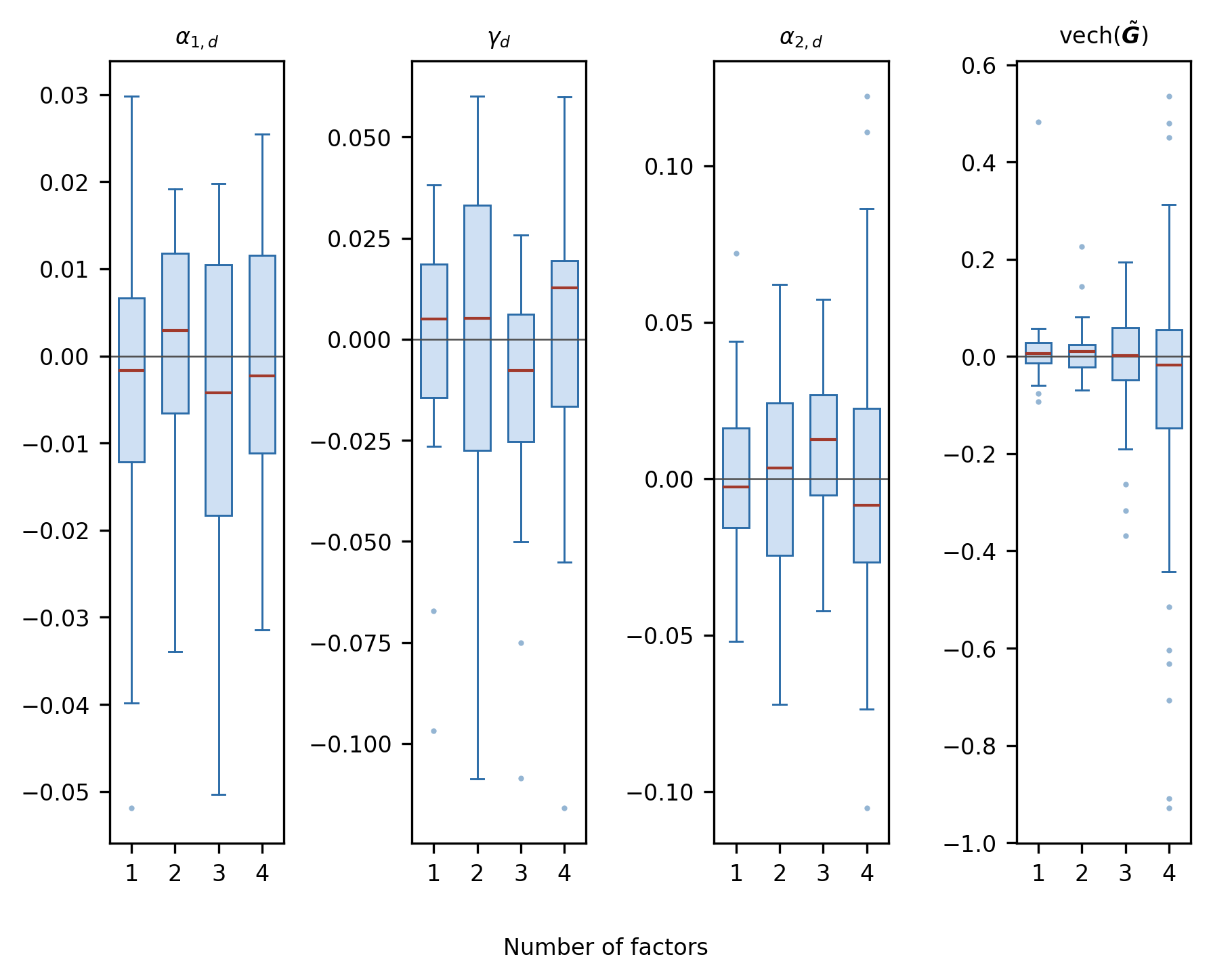}
    \caption{Bias of the N-IFM posterior mean for marginal-series and copula parameters in the Gaussian copula model with GARCH(1,1) marginals and Gaussian errors, shown separately for models with one, two, three, and four copula factors.
    }     \label{fig:calib_bias_gauscop} 
    \end{figure}

\begin{table}[t]
\caption{Average RMSE of the N-IFM posterior mean for the marginal and copula
parameters in the Gaussian copula model with GARCH(1,1) marginals and Gaussian
errors, reported for models with one to four copula factors. Here,
$\bs\alpha_{1,1:20}=(\alpha_{1,1},\ldots,\alpha_{1,20})$,
$\bs\gamma_{1:20}=(\gamma_1,\ldots,\gamma_{20})$, and
$\bs\alpha_{2,1:20}=(\alpha_{2,1},\ldots,\alpha_{2,20})$ denote the corresponding
GARCH(1,1) parameters across the $20$ marginal series. The notation
$\textrm{ave}(\cdot)$ indicates that the reported RMSE is averaged over the
number of parameters in the corresponding parameter group.}
\centering
\begin{tabular}{lcccc}
\hline\hline
Parameter & 1-factor & 2-factor & 3-factor & 4-factor \\
\hline
$\textrm{ave}(\boldsymbol\alpha_{1,1:20})$ & 0.022 & 0.018 & 0.022 & 0.020 \\
$\textrm{ave}(\boldsymbol\gamma_{1:20})$   & 0.035 & 0.043 & 0.038 & 0.039 \\
$\textrm{ave}(\boldsymbol\alpha_{2,1:20})$ & 0.033 & 0.036 & 0.029 & 0.047 \\
$\operatorname{ave}\!\left\{
\operatorname{vech}(\widetilde{\boldsymbol{G}})
\right\}$
                                     & 0.109 & 0.103 & 0.130 & 0.244 \\
\hline
\end{tabular}
\label{tab:calib_rmse_gauscop}
\end{table}

\begin{table}[t]
\caption{All parameter elements for which the nominal $95\%$ credible
intervals produced by N-IFM have empirical coverage below $50\%$ in the
Gaussian-copula calibration studies, across models with one to four
copula factors. For the GARCH(1,1) parameters, persistence is defined as
$\alpha_{1,d}+\alpha_{2,d}$ for the corresponding marginal series. The prior
quantile is the cumulative prior probability evaluated at the true parameter
value. A dash indicates that persistence is not applicable.}
\centering
\setlength{\tabcolsep}{4pt}
\begin{tabular}{llcccccc}
\hline\hline
Copula & Factor & Parameter &  $d$ & Coverage & True value & Persistence & Prior quantile \\
\hline
Gaussian & 1 & $\textrm{vech}(\widetilde{\boldsymbol{G}})$ & 12 & 0.334 & -2.108 & -- & 0.000 \\
Gaussian & 4 & $\gamma_d$ & 2 & 0.401 & 0.048 & 0.954 & 0.035 \\
Gaussian & 4 & $\alpha_{2,d}$ & 2 & 0.439 & 0.869 & 0.954 & 0.759 \\
Gaussian & 4 & $\textrm{vech}(\widetilde{\boldsymbol{G}})$ & 14 & 0.432 & 1.420 & -- & 0.794 \\
Gaussian & 4 & $\textrm{vech}(\widetilde{\boldsymbol{G}})$ & 58 & 0.387 & 1.292 & -- & 0.944 \\
\hline
\end{tabular}
\label{tab:lowcov}
\end{table}

\begin{table}[t]
\centering
\caption{Average univariate Wasserstein-1 distances between the posterior
distributions obtained using HMC, HMC-IFM, and N-IFM for the Gaussian copula
model with GARCH(1,1) marginals and Gaussian errors, reported for models with
one to four copula factors. Here,
$\boldsymbol{\alpha}_{1,1:20}
=(\alpha_{1,1},\ldots,\alpha_{1,20})$,
$\boldsymbol{\gamma}_{1:20}
=(\gamma_1,\ldots,\gamma_{20})$, and
$\boldsymbol{\alpha}_{2,1:20}
=(\alpha_{2,1},\ldots,\alpha_{2,20})$
denote the corresponding GARCH(1,1) parameters across the $20$ marginal
series. The notation $\operatorname{ave}(\cdot)$ indicates that the
Wasserstein-1 distance is computed separately for each element of the
parameter group and then averaged across its elements.}
\label{tab:calib_wass_gauscop}

\resizebox{\textwidth}{!}{%
\begin{tabular}{lcccccccccccc}
\toprule
&
\multicolumn{4}{c}{HMC vs.\ HMC-IFM}
&
\multicolumn{4}{c}{N-IFM vs.\ HMC-IFM}
&
\multicolumn{4}{c}{N-IFM vs.\ HMC}
\\
\cmidrule(lr){2-5}
\cmidrule(lr){6-9}
\cmidrule(lr){10-13}
Parameter
& 1 & 2 & 3 & 4
& 1 & 2 & 3 & 4
& 1 & 2 & 3 & 4
\\
\midrule
$\operatorname{ave}(\boldsymbol\alpha_{1,1:20})$
& 0.010 & 0.013 & 0.012 & 0.009
& 0.009 & 0.009 & 0.010 & 0.007
& 0.012 & 0.014 & 0.017& 0.013
\\

$\operatorname{ave}(\boldsymbol\gamma_{1:20})$
& 0.018 & 0.030 & 0.027 & 0.015
& 0.012 & 0.017 & 0.017 & 0.010
& 0.023 & 0.040& 0.032 & 0.021
\\

$\operatorname{ave}(\boldsymbol\alpha_{2,1:20})$
& 0.010 & 0.017 & 0.020 & 0.019& 0.014 & 0.017 & 0.016 & 0.015
& 0.019 & 0.028 & 0.022 & 0.028
\\

$\operatorname{ave}\!\left\{
\operatorname{vech}(\widetilde{\boldsymbol{G}})
\right\}$
& 0.011 & 0.029 & 0.017 & 0.013& 0.068 & 0.066 & 0.114 & 0.262 & 0.072 & 0.068 & 0.113 & 0.256\\
\bottomrule
\end{tabular}%
}
\end{table}

\clearpage

\clearpage
\section{Simulation study: $t$ copula} \label{sec:sim_stcop}

This section compares the accuracy of N-IFM, HMC, and HMC-IFM in fitting the \(t\)-copula models with GARCH(1,1) marginals and Gaussian errors.
The simulated data is generated from a one-factor $t$ copula model with GARCH(1,1)
marginals with Gaussian errors, comprising $D = 20$ series of length $T = 1000$, {following Algorithm \ref{alg:copula_sim} in Section \ref{supp:datagen_cop}. 


{
Figures~\ref{suppfig:alpha1_stcop_f1}--\ref{suppfig:nu_stcop_f1} show posterior
distributions of GARCH(1,1) parameters and $t$ copula parameters estimated using the N-IFM, HMC and HMC-IFM methods for a one-factor $t$ copula model with GARCH(1,1) marginals with Gaussian errors, comprising $D = 20$ series of length $T = 1000$. The figures show that N-IFM and HMC-IFM produce very similar posterior distributions, with only small differences.
For the copula parameters, the N-IFM and HMC-IFM results also closely match those from full HMC. By contrast, full HMC gives tighter posterior distributions for the GARCH(1,1) marginal parameters than N-IFM and HMC-IFM. This appears to be because HMC estimates the marginal and copula parameters jointly, whereas N-IFM and HMC-IFM estimate them in two steps.
{For the degrees-of-freedom parameter $\nu$ of the $t$ copula, the N-IFM posterior is wider than the HMC and HMC-IFM posteriors, but its centre is closer to the true value.}
}

\begin{figure}[t]
    \centering
    \includegraphics[width=\linewidth]{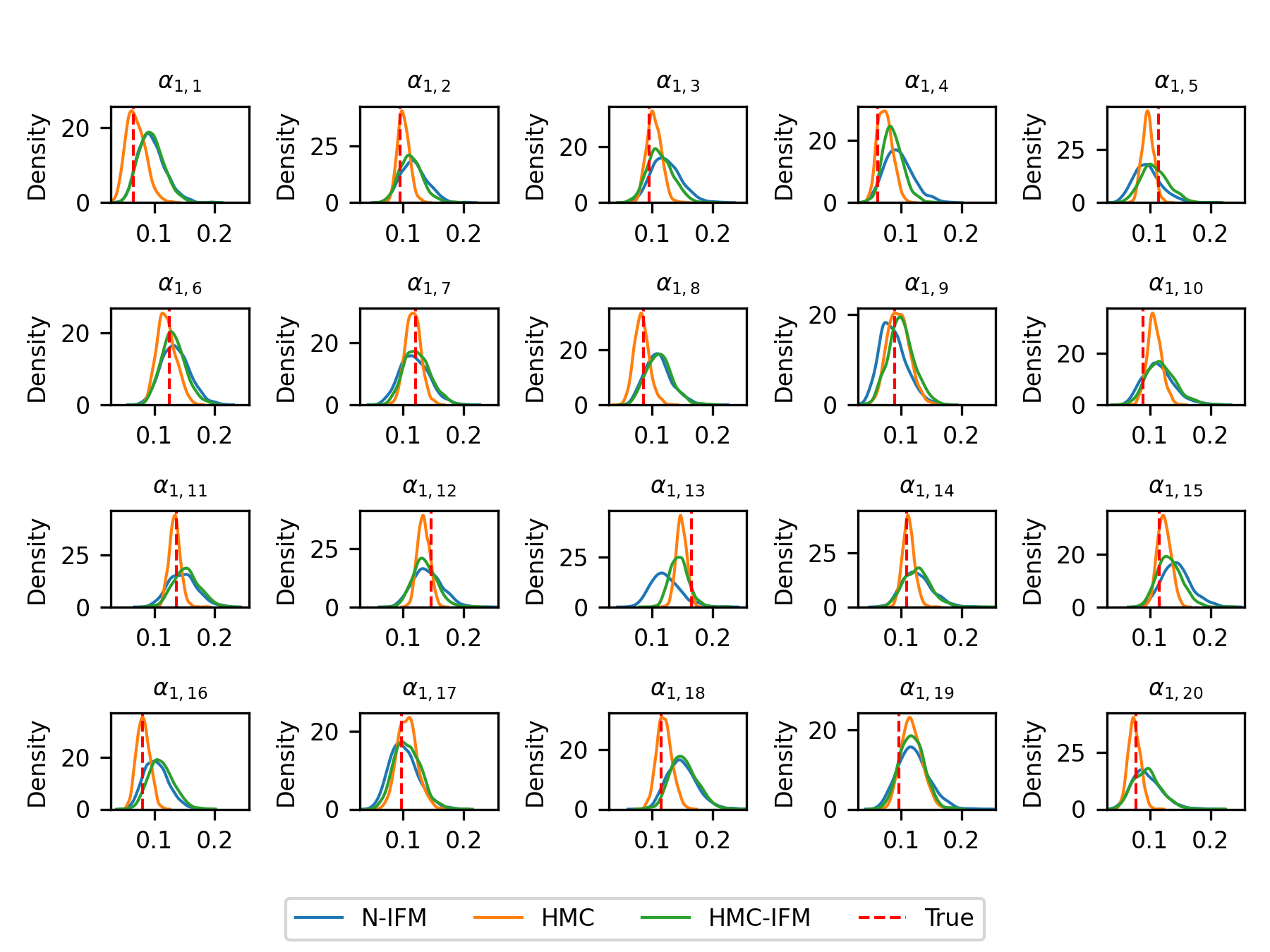}
    \caption{Posterior distributions of the GARCH$(1,1)$ parameters $\alpha_{1,d}$ for $d=1,\ldots,D$, comparing N-IFM, HMC and HMC-IFM with the true values. The results are based on simulated data from a 20 dimensional one-factor $t$ copula with GARCH(1,1) marginals with Gaussian errors.}
    \label{suppfig:alpha1_stcop_f1}
\end{figure}

Figure~\ref{fig:stcop_bipredplot} shows selected univariate and bivariate one-step-ahead posterior predictive densities obtained from {N-IFM, HMC and HMC-IFM} for a one-factor $t$ copula model
with GARCH(1,1) marginals with Gaussian errors. 
The predictive densities obtained from {N-IFM, HMC and HMC-IFM} are very similar. The lower triangular
panels display the corresponding bivariate predictive distributions, which exhibit strong
agreement across the three methods. 
{In particular, the distributions derived from {N-IFM, HMC and HMC-IFM} are very similar to one another and tilt in the direction of the true correlation.}
In terms of computational efficiency, as shown in  Table \ref{tab:timings} in Section \ref{sec:computingtime}, N-IFM is approximately 24,500 times faster than HMC and 11,500 times faster than HMC-IFM post-training, while delivering posterior estimates of comparable
accuracy.

For the HMC implementation in \texttt{Stan}, standard diagnostics such as the effective sample size (\texttt{n\_eff}) and the potential scale reduction factor ($\widehat{\texttt{R}}$) are used to assess convergence. 
As reported in Table~\ref{tab:hmc_stcop_simf1}, \texttt{n\_eff} is large for all parameters and all $\widehat{\texttt{R}}$ values are close to 1, indicating good mixing and convergence across chains. 
For HMC-IFM (not reported), all \(\widehat{\texttt{R}}\) values are close to 1, indicating good convergence, while \texttt{n\_eff} is slightly lower for the GARCH(1,1) parameters and slightly higher for the copula parameters relative to HMC.



\begin{figure}[t]
    \centering
    \includegraphics[width=\linewidth]{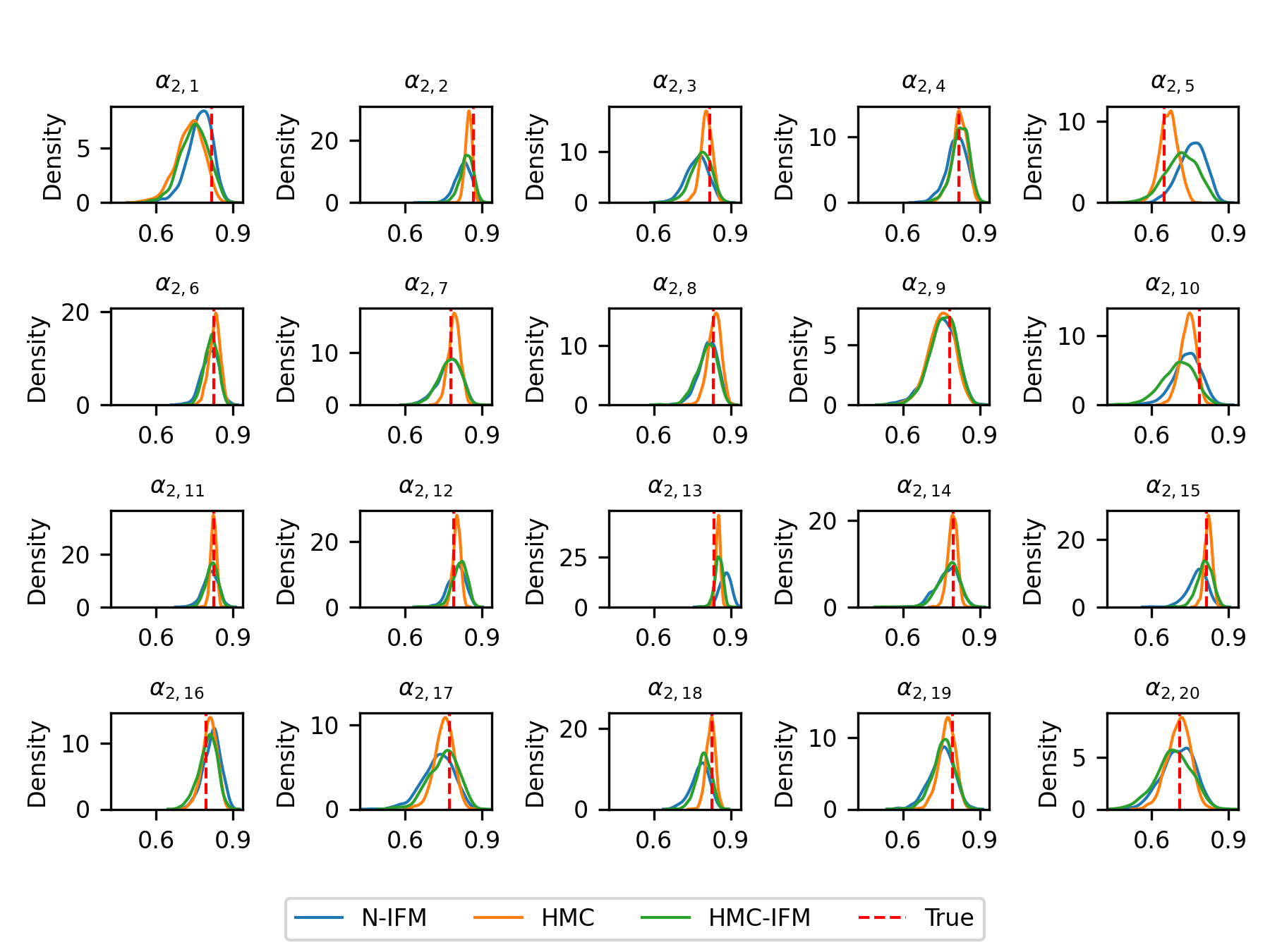}
    \caption{Posterior distributions of the GARCH$(1,1)$ parameters $\alpha_{2,d}$ for $d=1,\ldots,D$, comparing N-IFM, HMC and HMC-IFM with the true values. The results are based on simulated data from a 20 dimensional one-factor $t$ copula with GARCH(1,1) marginals with Gaussian errors.}
    \label{suppfig:alpha2_stcop_f1}
\end{figure}

\begin{figure}[t]
    \centering
    \includegraphics[width=\linewidth]{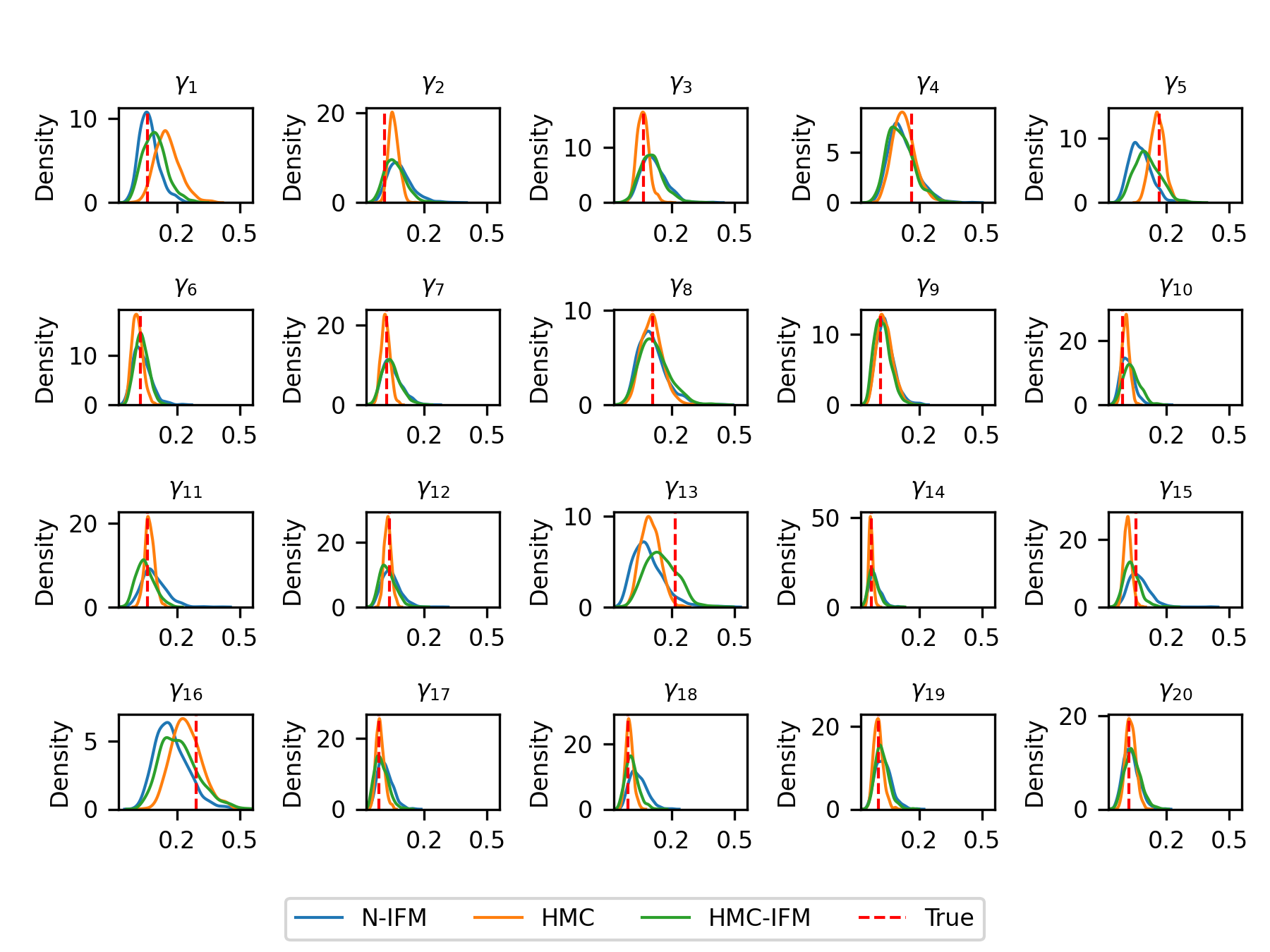}
    \caption{Posterior distributions of the GARCH$(1,1)$ parameters $\gamma_{d}$ for $d=1,\ldots,D$, comparing N-IFM, HMC and HMC-IFM with the true values. The results are based on simulated data from a 20 dimensional one-factor $t$ copula with GARCH(1,1) marginals with Gaussian errors.}
    \label{suppfig:gamma_stcop_f1}
\end{figure}

\begin{figure}[t]
    \centering
    \includegraphics[width=\linewidth]{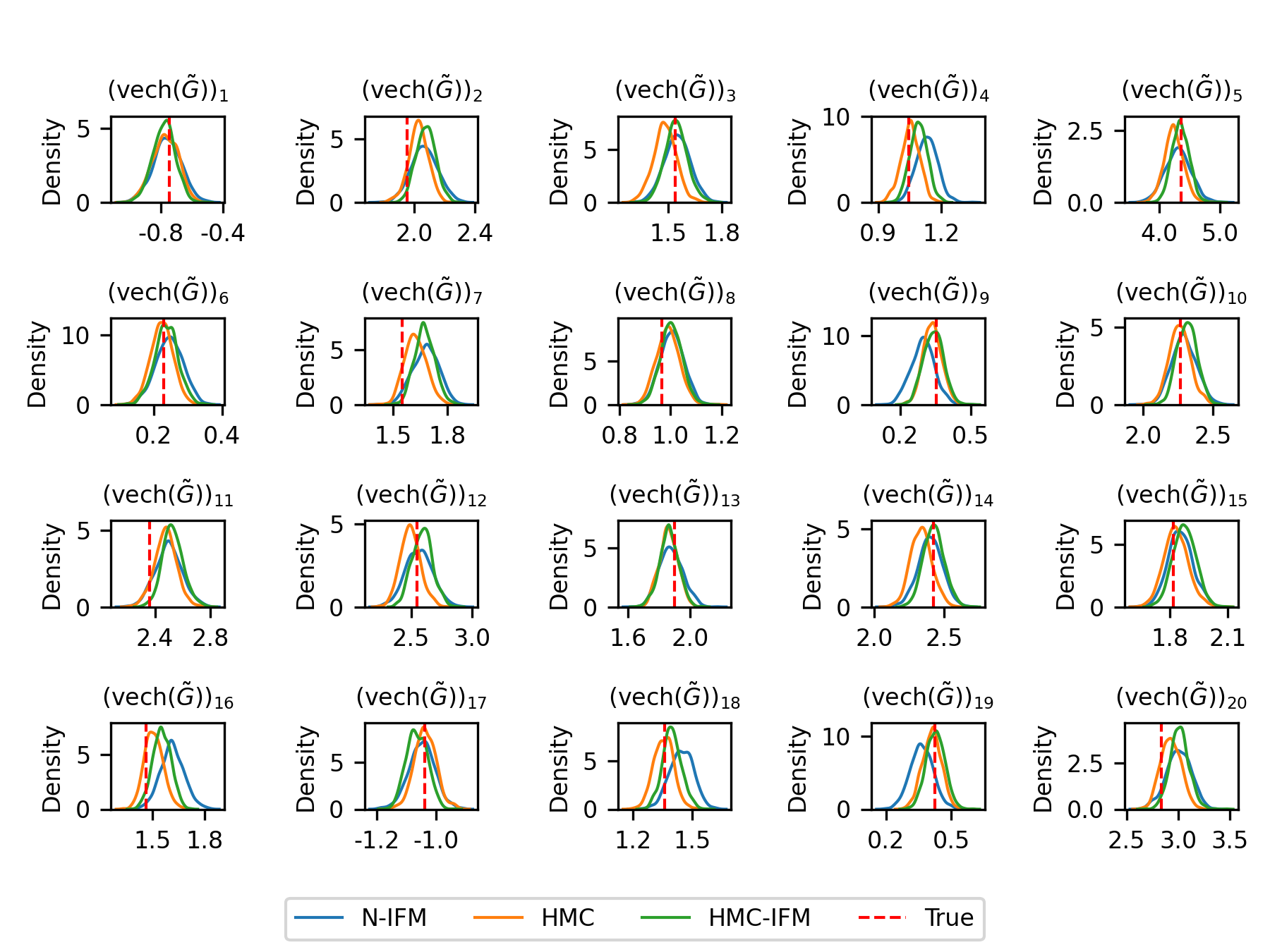}
    \caption{Posterior distributions of the $t$ copula parameters $\textrm{vech}(\widetilde{\boldsymbol{G}})$, comparing N-IFM, HMC and HMC-IFM with the true values. The results are based on simulated data from a 20 dimensional one-factor $t$ copula with GARCH(1,1) marginals with Gaussian errors.}
    \label{suppfig:gtilde_stcop_f1}
\end{figure}

\begin{figure}[t]
    \centering
    \includegraphics[width=\linewidth]{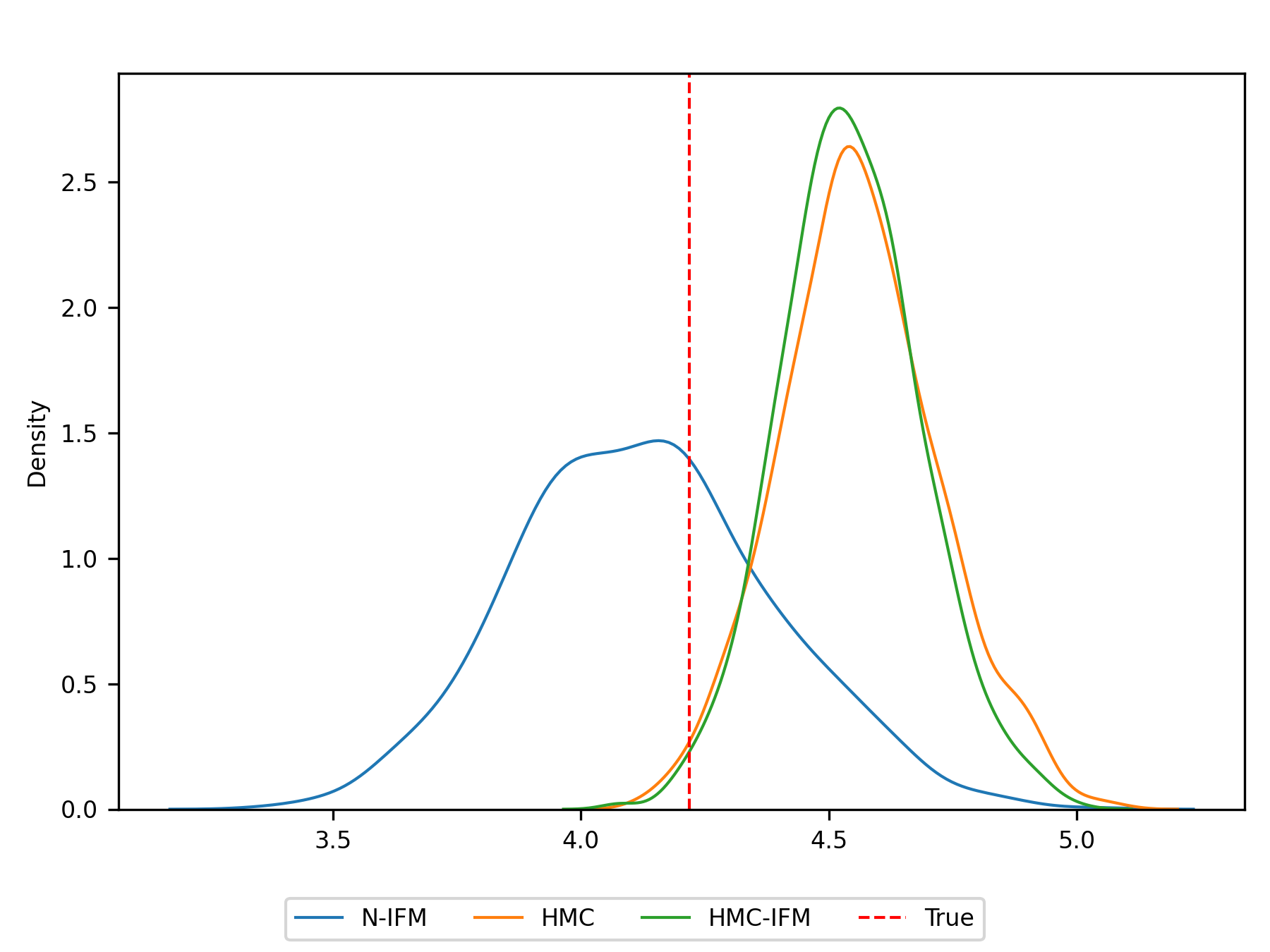}
    \caption{Posterior distributions of the $t$ copula degrees-of-freedom parameter $\nu$, comparing N-IFM, HMC and HMC-IFM with the true values. The results are based on simulated data from a 20 dimensional one-factor $t$ copula with GARCH(1,1) marginals with Gaussian errors.}
    \label{suppfig:nu_stcop_f1}
\end{figure}

\begin{figure}[t]
    \centering
    \includegraphics[width=\linewidth]{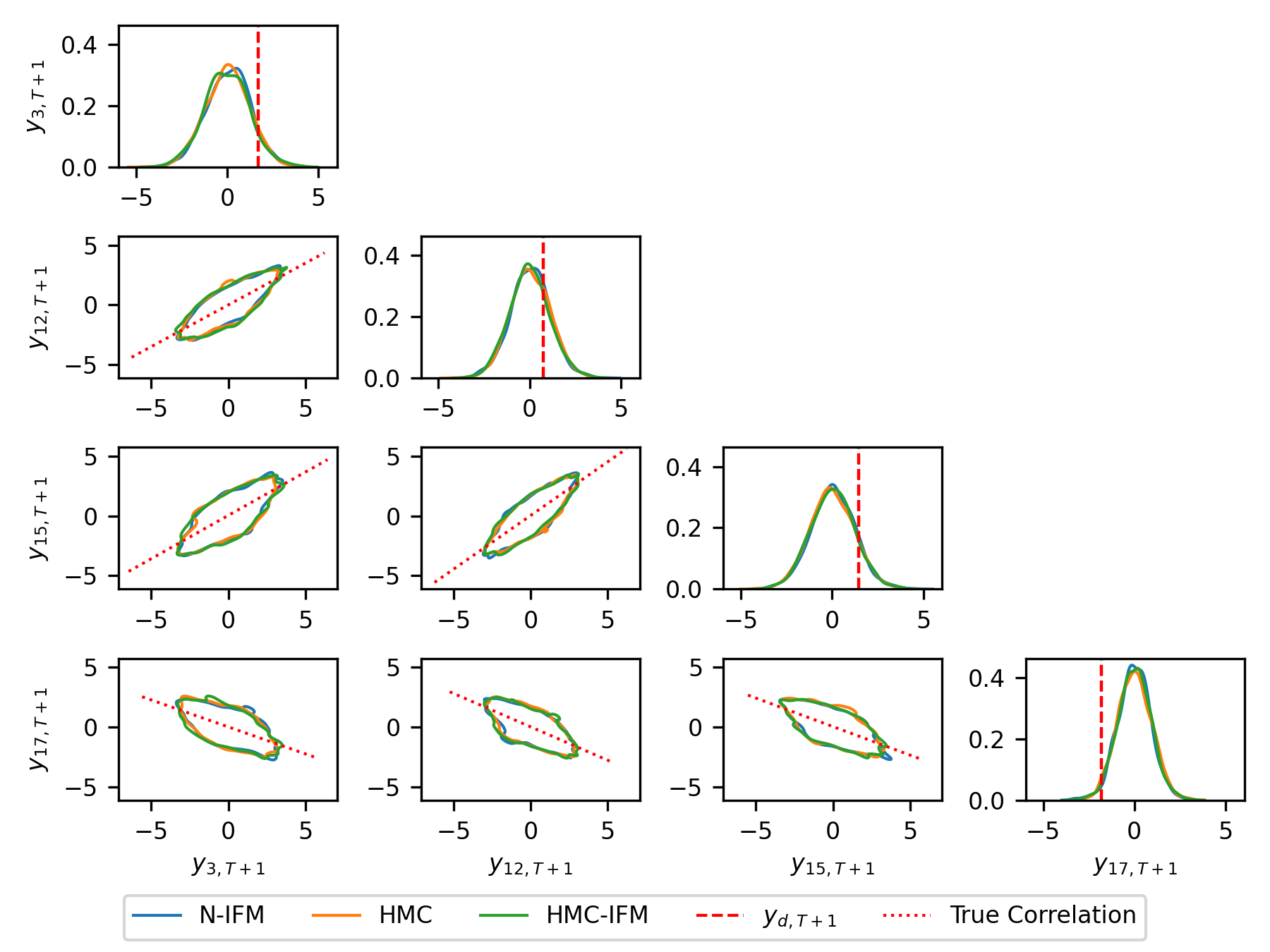}
    \caption{Selected one-step-ahead posterior predictive densities obtained from {N-IFM, HMC and HMC-IFM} for data simulated from a one-factor $t$ copula model
    with GARCH(1,1) marginals with Gaussian errors, comprising $D = 20$ series of length $T = 1000$. Diagonal panels show the marginal predictive distributions, while
    off-diagonal panels display the corresponding bivariate predictive distributions, 
    {shown as 95\% highest-density contours obtained from kernel density estimates of the predictive draws.}
    In the diagonal panels, the vertical lines indicate the realised $y_{d,T+1}$, while in the off-diagonal panels, 
    {the dotted lines have slope $\rho_{ij}\,\sigma_i/\sigma_j$, where $\rho_{ij}$ is the true correlation between series $i$ and $j$ and $\sigma_i$, $\sigma_j$ are the standard deviations of the predictive draws.}
    }
    \label{fig:stcop_bipredplot}
\end{figure}

\clearpage

\begin{table}[h]
\caption{Posterior diagnostics for model parameters estimated using HMC, based on simulated data from a 20 dimensional one factor $t$ copula model with GARCH$(1,1)$ marginals with Gaussian errors, including the effective sample size (\texttt{n\_eff}) and the potential scale reduction factor ($\widehat{\texttt{R}}$).}
\centering
{\small
\begin{tabular}{lll|lll}
\hline \hline
Parameter & \texttt{n\_eff} & {$\widehat{\texttt{R}}$} & Parameter & \texttt{n\_eff} & {$\widehat{\texttt{R}}$} \\
\hline
$\alpha_{1,1}$ & 7160 & 0.9996 & $\gamma_{2}$ & 6135 & 0.9996  \\
$\alpha_{1,2}$ & 5100 & 1.0002 & $\gamma_{3}$ & 5215 & 0.9996  \\
$\alpha_{1,3}$ & 5379 & 0.9997 & $\gamma_{4}$ & 5483 & 0.9995 \\
$\alpha_{1,4}$ & 6330 & 0.9997 & $\gamma_{5}$ & 4254 & 0.9997 \\
$\alpha_{1,5}$ & 4126 & 0.9998 & $\gamma_{6}$ & 6012 & 1.0000 \\
$\alpha_{1,6}$ & 4749 & 1.0006 & $\gamma_{7}$ & 5275 & 0.9996 \\
$\alpha_{1,7}$ & 5092 & 0.9996   & $\gamma_{8}$ & 5983 & 0.9999 \\
$\alpha_{1,8}$ & 5954 & 0.9999   & $\gamma_{9}$ & 5936 & 0.9996 \\
$\alpha_{1,9}$ & 6360 & 0.9997   & $\gamma_{10}$ & 4410 & 0.9996 \\
$\alpha_{1,10}$ & 4734 & 0.9997  & $\gamma_{11}$ & 5376 & 1.0008 \\
$\alpha_{1,11}$ & 2928 & 1.0009  & $\gamma_{12}$ & 5265 & 1.0000 \\
$\alpha_{1,12}$ & 3708 & 1.0008  & $\gamma_{13}$ & 2031 & 1.0024 \\
$\alpha_{1,13}$ & 5887 & 0.9997  & $\gamma_{14}$ & 5123 & 0.9999 \\
$\alpha_{1,14}$ & 3876 & 1.0000  & $\gamma_{15}$ & 5802 & 0.9996 \\
$\alpha_{1,15}$ & 4919 & 1.0002  & $\gamma_{16}$ & 4660 & 0.9996 \\
$\alpha_{1,16}$ & 4915 & 0.9996  & $\gamma_{17}$ & 4529 & 0.9997 \\
$\alpha_{1,17}$ & 4255 & 0.9997   & $\gamma_{18}$ & 5715 & 1.0002 \\
$\alpha_{1,18}$ & 4907 & 1.0000   & $\gamma_{19}$ & 5803 & 0.9999 \\
$\alpha_{1,19}$ & 5667 & 0.9998   & $\gamma_{20}$ & 3611 & 1.0006 \\
$\alpha_{1,20}$ & 4423 & 1.0000   & $(\textrm{vech}(\widetilde{G}))_{1}$ & 6554 & 0.9999 \\
$\alpha_{2,1}$ & 5929 & 0.9999    & $(\textrm{vech}(\widetilde{G}))_{2}$ & 2504 & 1.0006 \\
$\alpha_{2,2}$ & 5773 & 1.0002    & $(\textrm{vech}(\widetilde{G}))_{3}$ & 2310 & 1.0023 \\
$\alpha_{2,3}$ & 5075 & 0.9998    & $(\textrm{vech}(\widetilde{G}))_{4}$ & 3187 & 1.0009 \\
$\alpha_{2,4}$ & 5194 & 0.9996    & $(\textrm{vech}(\widetilde{G}))_{5}$ & 3163 & 1.0011 \\
$\alpha_{2,5}$ & 4696 & 0.9998 & $(\textrm{vech}(\widetilde{G}))_{6}$ & 8707 & 0.9997 \\
$\alpha_{2,6}$ & 7175 & 0.9996 & $(\textrm{vech}(\widetilde{G}))_{7}$ & 2540 & 1.0016 \\
$\alpha_{2,7}$ & 5006 & 0.9997 & $(\textrm{vech}(\widetilde{G}))_{8}$ & 3187 & 1.0004 \\
$\alpha_{2,8}$ & 5790 & 0.9999 & $(\textrm{vech}(\widetilde{G}))_{9}$ & 5728 & 0.9997 \\
$\alpha_{2,9}$ & 5708 & 0.9996 & $(\textrm{vech}(\widetilde{G}))_{10}$ & 2201 & 1.0029 \\
$\alpha_{2,10}$ & 4464 & 1.0003 & $(\textrm{vech}(\widetilde{G}))_{11}$ & 2336 & 1.0036 \\
$\alpha_{2,11}$ & 6942 & 0.9995 & $(\textrm{vech}(\widetilde{G}))_{12}$ & 2509 & 1.0017 \\
$\alpha_{2,12}$ & 6531 & 0.9997 & $(\textrm{vech}(\widetilde{G}))_{13}$ & 4066 & 1.0016 \\
$\alpha_{2,13}$ & 5903 & 0.9997 & $(\textrm{vech}(\widetilde{G}))_{14}$ & 2018 & 1.0030 \\
$\alpha_{2,14}$ & 5107 & 0.9995 & $(\textrm{vech}(\widetilde{G}))_{15}$ & 2484 & 1.0012 \\
$\alpha_{2,15}$ & 6412 & 0.9997 & $(\textrm{vech}(\widetilde{G}))_{16}$ & 2755 & 1.0026 \\
$\alpha_{2,16}$ & 4510 & 0.9999 & $(\textrm{vech}(\widetilde{G}))_{17}$ & 2882 & 1.0011 \\
$\alpha_{2,17}$ & 4387 & 0.9995 & $(\textrm{vech}(\widetilde{G}))_{18}$ & 2861 & 1.0018 \\
$\alpha_{2,18}$ & 6062 & 1.0005 & $(\textrm{vech}(\widetilde{G}))_{19}$ & 6286 & 1.0000 \\
$\alpha_{2,19}$ & 6072 & 0.9998 & $(\textrm{vech}(\widetilde{G}))_{20}$ & 2139 & 1.0022 \\
$\alpha_{2,20}$ & 3771 & 0.9999 & $\nu$ & 1798 & 1.0015 \\
$\gamma_{1}$ & 5777 & 1.0000 & & & \\

\hline
\end{tabular}
}

\label{tab:hmc_stcop_simf1}

\end{table}

\clearpage
\section{{Calibration study: $t$} copula}\label{sec:calib_stcop}

{This section presents the calibration study for the $t$ copula model with
GARCH(1,1) marginals and Student-$t$ errors. We use the same simulation design
as in the Gaussian-copula study in Section \ref{sec:supp_gauscop}. For each model with one to four copula
factors, we generate $1{,}000$ datasets using fixed true parameter values, with
each dataset comprising $20$ series of length $T=1{,}000$. Performance is
evaluated using the same five metrics defined in
Section~\ref{sec:metrics}. Compared with the Gaussian-copula model, the
$t$ copula model contains 21 additional parameters: the series-specific
marginal degrees-of-freedom parameter $\widetilde{\nu}_d$ and the copula
degrees-of-freedom parameter $\nu$.}

{As in the Gaussian-copula study, Figure~\ref{fig:calib_bias_stcop} shows that the biases of the GARCH(1,1) parameters $\alpha_{1,d}$ and $\alpha_{2,d}$ are small and centred close to zero across the four factor specifications, whereas $\gamma_d$ and $\widetilde{\nu}_d$ show a modest positive bias, with $\widetilde{\nu}_d$ also having a few large negative outliers. The biases of the copula-loading parameters, $\operatorname{vech}(\widetilde{\boldsymbol{G}})$, remain centred near zero but become increasingly dispersed in the three- and four-factor models. The $t$ copula degrees-of-freedom parameter $\nu$ is underestimated for two to four factors, most markedly for the two-factor model, where its true value ($\nu = 19.714$) is much larger than the other settings.
}

{Table~\ref{tab:calib_rmse_tcop} shows that, among the GARCH(1,1) parameters, the RMSEs of $\alpha_{1,d}$, $\gamma_d$ and $\alpha_{2,d}$ are small and similar across the four settings, ranging from $0.020$ to $0.052$. The GARCH(1,1) degrees-of-freedom parameters $\widetilde{\nu}_d$ are much harder to estimate, with an average RMSE between $2.273$ and $3.294$ across the four settings; the higher values for the two- to four-factor models are partly due to the outliers mentioned above. For the copula loadings, the average RMSE of $\operatorname{vech}(\widetilde{\boldsymbol{G}})$ increases steadily with the number of factors, from $0.085$ in the one-factor model to $0.114$, $0.202$ and $0.336$ in the two-, three- and four-factor models, respectively, so estimating the loadings becomes less precise as the factor specification becomes more complex. Finally, the RMSE of the $t$ copula degrees-of-freedom parameter $\nu$ is below one for the one-, three- and four-factor models, but rises to $3.831$ in the two-factor model, reflecting its large negative bias there.}


\begin{figure}[t]
    \centering
    \includegraphics[width=\linewidth]{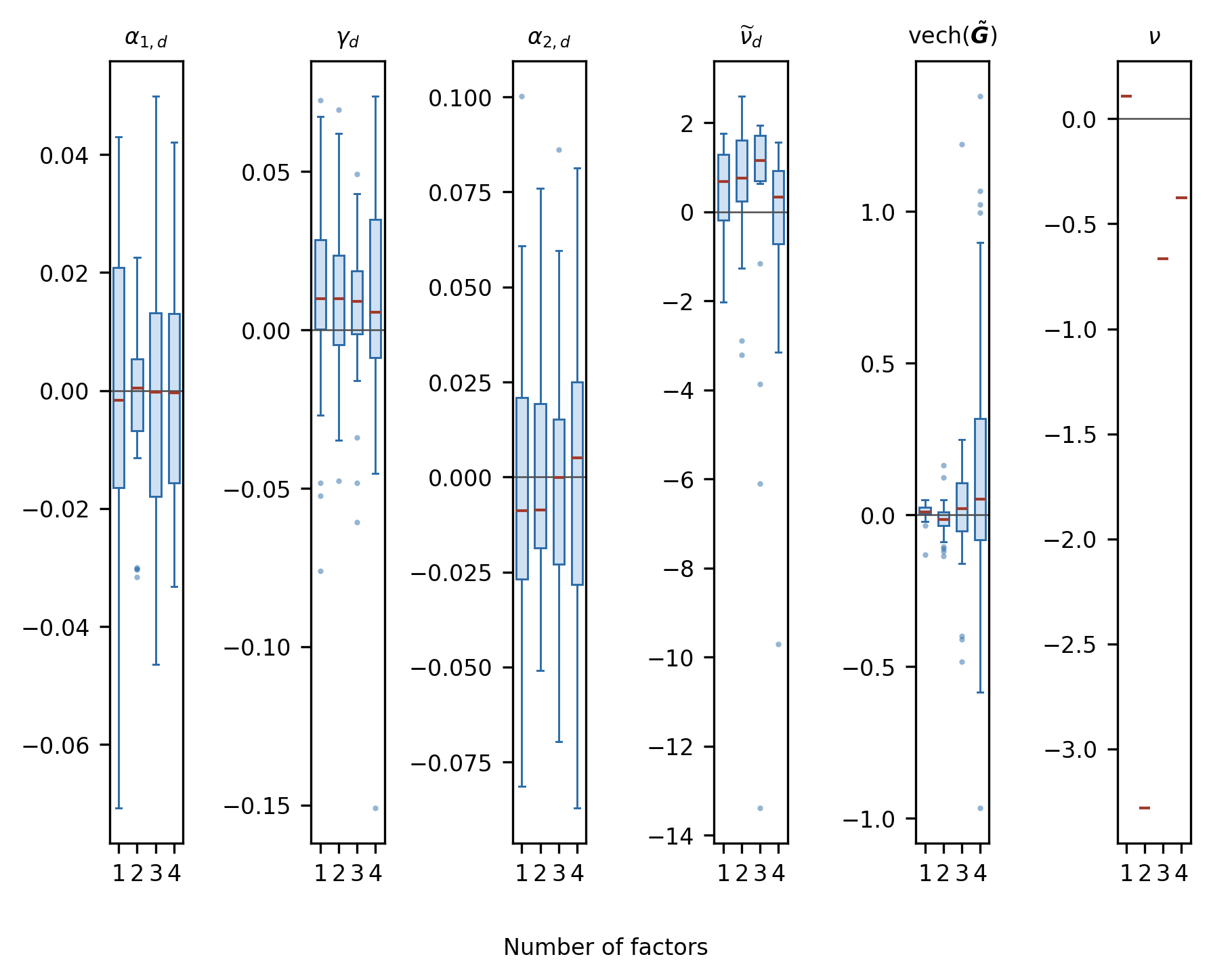}
    \caption{Bias of the N-IFM posterior mean for marginal-series and copula parameters for the $t$ copula model with GARCH(1,1) marginals and Student's-$t$ errors, shown separately for models with one, two, three, and four copula factors.}
    \label{fig:calib_bias_stcop}
\end{figure}

\begin{table}[t]
\centering
\caption{Average RMSE of the N-IFM posterior mean for the marginal and copula
parameters in the $t$ copula model with GARCH(1,1) marginals and Student-$t$
errors, reported for models with one to four copula factors. Here,
$\boldsymbol{\alpha}_{1,1:20}
=(\alpha_{1,1},\ldots,\alpha_{1,20})$,
$\boldsymbol{\gamma}_{1:20}
=(\gamma_1,\ldots,\gamma_{20})$,
$\boldsymbol{\alpha}_{2,1:20}
=(\alpha_{2,1},\ldots,\alpha_{2,20})$, and
$\widetilde{\boldsymbol{\nu}}_{1:20}
=(\widetilde{\nu}_1,\ldots,\widetilde{\nu}_{20})$
denote the corresponding marginal parameters across the $20$ series. The
notation $\operatorname{ave}(\cdot)$ indicates that the reported value is the
average of the element-specific RMSEs within the corresponding parameter
group. Since the copula degrees-of-freedom parameter $\nu$ is scalar, its RMSE
is reported directly.}
\label{tab:calib_rmse_tcop}

\begin{tabular}{lcccc}
\toprule
Parameter
& 1-factor
& 2-factor
& 3-factor
& 4-factor
\\
\midrule

$\textrm{ave}(\boldsymbol{\alpha}_{1,1:20})$
& 0.028 & 0.020 & 0.025 & 0.023
\\

$\operatorname{ave}(\boldsymbol\gamma_{1:20})$
& 0.039 & 0.052 & 0.034 & 0.045
\\

$\operatorname{ave}(\boldsymbol\alpha_{2,1:20})$
& 0.041 & 0.031 & 0.034 & 0.038
\\

$\operatorname{ave}(\widetilde{\boldsymbol\nu}_{1:20})$
& 2.273 & 3.294 & 3.110 & 2.626
\\

$\operatorname{ave}\!\left\{
\operatorname{vech}(\widetilde{\boldsymbol{G}})
\right\}$
& 0.085 & 0.114 & 0.202 & 0.336
\\

$\nu$
& 0.898 & 3.831 & 0.995 & 0.857
\\

\bottomrule
\end{tabular}
\end{table}

\begin{figure}[t]
    \centering
    \includegraphics[width=\linewidth]{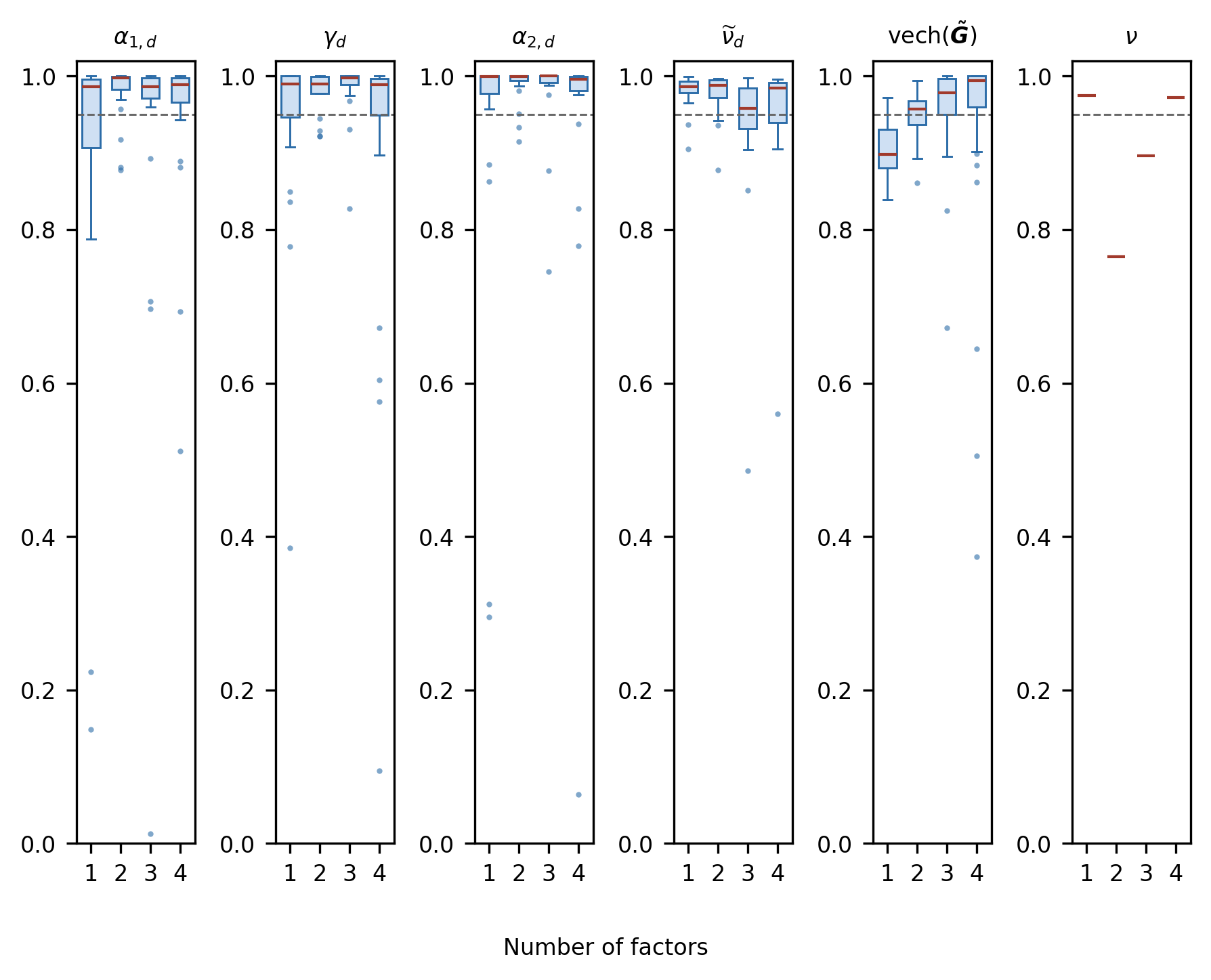}
    \caption{Empirical coverage probabilities of the nominal $95\%$ credible
    intervals produced by N-IFM for the $t$ copula model with GARCH(1,1) marginals
    and Student-$t$ errors. For each parameter group and number of copula factors,
    the boxplot summarises the parameter-specific coverage probabilities. The dashed
    horizontal line indicates the nominal coverage level of $0.95$.}
    \label{fig:calib_cov_stcop}
\end{figure}

{Figure~\ref{fig:calib_cov_stcop} shows that the median empirical coverage of the GARCH(1,1) parameters is close to or above the nominal $95\%$ level across all four factor specifications, with $\alpha_{2,d}$ remaining close to $1.00$ throughout, and the median coverage of $\widetilde{\nu}_d$ only dipping to about $95\%$ in the three-factor model. For the copula-loading parameters, $\operatorname{vech}(\widetilde{\boldsymbol{G}})$, the median coverage is below the nominal level in the one-factor model (about $90\%$) but rises to, or above, $95\%$ as the number of factors increases. The coverage of the $t$ copula degrees-of-freedom parameter $\nu$ is close to nominal for the one- and four-factor models, but falls to about $77\%$ and $90\%$ for the two- and three-factor models, consistent with its negative bias there. Several individual parameters have coverages below the nominal level.}
As shown in Table~\ref{tab:lowcov_t}, ten parameters in the $t$-copula experiment have coverage below $50\%$: five in the one-factor model, two in the three-factor model and three in the four-factor model, with none in the two-factor model. All but one are marginal parameters; the exception is a single copula parameter in $\operatorname{vech}(\widetilde{\boldsymbol{G}})$ in the four-factor model. Most of these involve true values drawn from the tails of their prior distributions (prior quantiles of $0.002$, $0.989$ and $0.001$ for $\alpha_{1,d}$, and $0.997$ for $\widetilde{\nu}_d$), or, for series $d=16$ in the one-factor model and $d=9$ in the four-factor model, a small $\gamma_d$ combined with GARCH persistence close to the stationarity boundary ($\alpha_{1,d}+\alpha_{2,d}=0.967$ and $0.987$). Such configurations are necessarily represented relatively infrequently in a finite training sample and are therefore particularly challenging for amortised neural inference.

{
Table~\ref{tab:calib_width_tcop} reports the average width of the $95\%$ credible intervals. For the GARCH(1,1) parameters, the intervals for $\alpha_{1,d}$, $\gamma_d$ and $\alpha_{2,d}$ are narrow and stable across the four settings (between $0.095$ and $0.190$), whereas those for $\widetilde{\nu}_d$ are wide, with average widths between $10.384$ and $12.322$. 
For comparison, the central $95\%$ range of the Gamma prior with shape $4.74$ and scale $2.03$ has a width of approximately $17$, so the N-IFM posterior for $\widetilde{\nu}_d$ narrows the prior range by roughly a third.
Its near-nominal coverage therefore partly reflects the width of these intervals rather than precise identification from the data. 
The intervals for $\operatorname{vech}(\widetilde{\boldsymbol{G}})$ widen steadily with the number of factors, from $0.283$ in the one-factor model to $0.462$, $0.896$ and $1.454$ in the two-, three- and four-factor models, in line with the increase in RMSE. Finally, the intervals for the $t$ copula degrees-of-freedom parameter $\nu$ have widths of about $4$ in the one-, three- and four-factor models, but widen to $9.790$ in the two-factor model, where the true value of $\nu$ is large.}

\begin{table}[t]
\caption{All parameter elements for which the nominal $95\%$ credible
intervals produced by N-IFM have empirical coverage below $50\%$ in the $t$-copula calibration studies, across models with one to four copula factors. For the GARCH(1,1) parameters, persistence is defined as
$\alpha_{1,d}+\alpha_{2,d}$ for the corresponding marginal series. The prior
quantile is the cumulative prior probability evaluated at the true parameter
value. A dash indicates that persistence is not applicable.}
\centering
\setlength{\tabcolsep}{4pt}
\begin{tabular}{llcccccc}
\hline\hline
Copula & Factor & Parameter &  $d$ & Coverage & True value & Persistence & Prior quantile \\
\hline
$t$ & 1 & $\alpha_{1,d}$ & 15 & 0.149 & 0.044 & 0.796 & 0.002 \\
$t$ & 1 & $\alpha_{1,d}$ & 18 & 0.224 & 0.203 & 0.891 & 0.989 \\
$t$ & 1 & $\gamma_d$ & 16 & 0.385 & 0.062 & 0.967 & 0.083 \\
$t$ & 1 & $\alpha_{2,d}$ & 16 & 0.312 & 0.885 & 0.967 & 0.832 \\
$t$ & 1 & $\alpha_{2,d}$ & 18 & 0.295 & 0.688 & 0.891 & 0.075 \\
$t$ & 3 & $\alpha_{1,d}$ & 6 & 0.013 & 0.037 & 0.843 & 0.001 \\
$t$ & 3 & $\widetilde\nu_d$ & 5 & 0.486 & 26.571 & -- & 0.997 \\
$t$ & 4 & $\gamma_d$ & 9 & 0.095 & 0.036 & 0.987 & 0.013 \\
$t$ & 4 & $\alpha_{2,d}$ & 9 & 0.064 & 0.914 & 0.987 & 0.930 \\
$t$ & 4 & $\textrm{vech}(\widetilde{\boldsymbol{G}})$ & 62 & 0.374 & -0.607 & -- & 0.418 \\
\hline
\end{tabular}
\label{tab:lowcov_t}
\end{table}

\begin{table}[t]
\centering
\caption{Average widths of the nominal $95\%$ credible intervals produced by
N-IFM for the marginal and copula parameters in the $t$ copula model with
GARCH(1,1) marginals and Student-$t$ errors, reported for models with one to
four copula factors. Here,
$\boldsymbol{\alpha}_{1,1:20}
=(\alpha_{1,1},\ldots,\alpha_{1,20})$,
$\boldsymbol{\gamma}_{1:20}
=(\gamma_1,\ldots,\gamma_{20})$,
$\boldsymbol{\alpha}_{2,1:20}
=(\alpha_{2,1},\ldots,\alpha_{2,20})$, and
$\widetilde{\boldsymbol{\nu}}_{1:20}
=(\widetilde{\nu}_1,\ldots,\widetilde{\nu}_{20})$
denote the corresponding marginal parameters across the $20$ series. The
notation $\operatorname{ave}(\cdot)$ indicates that the interval widths are
averaged over the $1{,}000$ simulated datasets and the elements within the
corresponding parameter group. Since the copula degrees-of-freedom parameter
$\nu$ is scalar, its interval width is averaged only over the simulated
datasets.}
\label{tab:calib_width_tcop}

\begin{tabular}{lcccc}
\toprule
Parameter
& 1-factor
& 2-factor
& 3-factor
& 4-factor
\\
\midrule

$\operatorname{ave}(\boldsymbol\alpha_{1,1:20})$
& 0.095 & 0.096 & 0.096 & 0.096
\\

$\operatorname{ave}(\boldsymbol\gamma_{1:20})$
& 0.161 & 0.190 & 0.173 & 0.178
\\

$\operatorname{ave}(\boldsymbol\alpha_{2,1:20})$
& 0.171 & 0.159 & 0.172 & 0.163
\\

$\operatorname{ave}(\widetilde{\boldsymbol\nu}_{1:20})$
& 12.322 & 11.502 & 10.384 & 11.442
\\

$\operatorname{ave}\!\left\{
\operatorname{vech}(\widetilde{\boldsymbol{G}})
\right\}$
& 0.283 & 0.462 & 0.896 & 1.454
\\

$\nu$
& 4.005 & 9.790 & 3.501 & 4.017
\\

\bottomrule
\end{tabular}
\end{table}

Table~\ref{tab:calib_wass_stcop} reports the Wasserstein-1 distance. For the GARCH(1,1) parameters all three methods agree closely, with distances of at most $0.031$. The two HMC-based methods also agree closely on $\textrm{vech}(\widetilde{\boldsymbol{G}})$ at each factor ($\le 0.024$), while the distance between N-IFM and either HMC method grows from $0.036$--$0.043$ with one factor to $0.176$--$0.234$ with three and four factors, mirroring the pattern seen for the Gaussian copula. For $\widetilde\nu_d$ the distances are large for all three pairs, including between the two HMC methods ($0.942$--$1.704$); the N-IFM versus HMC distance ($1.478$--$2.230$) exceeds this for each factor, whereas the N-IFM versus HMC-IFM distance ($0.943$--$1.613$) is smaller at one and four factors. For the copula $\nu$, N-IFM is further from both HMC methods than they are from each other with two to four factors ($0.607$--$1.622$ versus $0.307$--$0.665$), while with one factor N-IFM is close to HMC ($0.282$) and it is HMC-IFM that differs from both. 

\begin{table}[t]
\centering
\caption{Average univariate Wasserstein-1 distances between the posterior
distributions obtained using HMC, HMC-IFM, and N-IFM for the $t$ copula model
with GARCH(1,1) marginals and Student-$t$ errors, reported for models with one
to four copula factors. Here,
$\boldsymbol{\alpha}_{1,1:20}
=(\alpha_{1,1},\ldots,\alpha_{1,20})$,
$\boldsymbol{\gamma}_{1:20}
=(\gamma_1,\ldots,\gamma_{20})$,
$\boldsymbol{\alpha}_{2,1:20}
=(\alpha_{2,1},\ldots,\alpha_{2,20})$, and
$\widetilde{\boldsymbol{\nu}}_{1:20}
=(\widetilde{\nu}_1,\ldots,\widetilde{\nu}_{20})$
denote the corresponding marginal parameters across the $20$ series. The
notation $\operatorname{ave}(\cdot)$ indicates that the Wasserstein-1 distance
is computed separately for each element of the parameter group and then
averaged across its elements. Since the copula degrees-of-freedom parameter
$\nu$ is scalar, its Wasserstein-1 distance is reported directly.}
\label{tab:calib_wass_stcop}

\resizebox{\textwidth}{!}{%
\begin{tabular}{lcccccccccccc}
\toprule
&
\multicolumn{4}{c}{HMC vs.\ HMC-IFM}
&
\multicolumn{4}{c}{N-IFM vs.\ HMC-IFM}
&
\multicolumn{4}{c}{N-IFM vs.\ HMC}
\\
\cmidrule(lr){2-5}
\cmidrule(lr){6-9}
\cmidrule(lr){10-13}
Parameter
& 1 & 2 & 3 & 4
& 1 & 2 & 3 & 4
& 1 & 2 & 3 & 4
\\
\midrule
$\operatorname{ave}(\boldsymbol\alpha_{1,1:20})$
& 0.012 & 0.013 & 0.017 & 0.014
& 0.010 & 0.007 & 0.011 & 0.011
& 0.019 & 0.015 & 0.021 & 0.020
\\

$\operatorname{ave}(\boldsymbol\gamma_{1:20})$
& 0.018 & 0.021 & 0.017 & 0.020
& 0.014 & 0.017 & 0.015 & 0.016
& 0.028 & 0.031 & 0.027 & 0.030
\\

$\operatorname{ave}(\boldsymbol\alpha_{2,1:20})$
& 0.020 & 0.016 & 0.024 & 0.017
& 0.015 & 0.015 & 0.017 & 0.016
& 0.030 & 0.026 & 0.030 & 0.027
\\

$\operatorname{ave}(\widetilde{\boldsymbol\nu}_{1:20})$
& 1.704 & 1.027 & 0.942 & 1.002
& 1.422 & 1.613 & 1.133 & 0.943
& 2.115 & 2.230 & 1.800 & 1.478
\\

$\operatorname{ave}\!\left\{
\operatorname{vech}(\widetilde{\boldsymbol{G}})
\right\}$
& 0.015 & 0.021 & 0.015 & 0.024
& 0.036 & 0.079 & 0.177 & 0.234
& 0.043 & 0.086 & 0.176 & 0.231
\\

$\nu$
& 1.049 & 0.307 & 0.665 & 0.398
& 1.210 & 1.427 & 1.622 & 0.949
& 0.282 & 1.248 & 0.967 & 0.607
\\
\bottomrule
\end{tabular}%
}
\end{table}

{Overall, the calibration of the $t$ copula model is similar to that of the Gaussian model for the GARCH(1,1) and copula parameters, while the two degrees-of-freedom parameters are estimated with some biases and weakly identified intervals.}

{Overall, the calibration of the $t$ copula model is similar to that of the Gaussian model for the GARCH(1,1) and copula parameters, while the two degrees-of-freedom parameters are estimated with some biases and weakly identified intervals.}

\clearpage
\section{Simulation study 3: model selection} \label{sec:stochterrors}

We investigate the performance of the N-IFM method to
identify the true data-generating model, using the one-step-ahead log predictive density score (LPDS), computed under a rolling-window scheme with window length $T=1000$, as in Section~\ref{sec:sim_gauscop} of the main paper. The model achieving the highest LPDS is selected as the preferred model. 
We consider copula models with four factors under four specifications: (C1) a Gaussian copula with GARCH(1,1) marginals with Gaussian errors; (C2) a Gaussian copula with GARCH(1,1) marginals with Student's-$t$ errors; (C3) a $t$ copula with GARCH(1,1) marginals with Gaussian errors; and (C4) a $t$ copula with GARCH(1,1) marginals with Student's-$t$ errors. 

We generate $D=20$ dimensional multivariate time series of length $T + K=1100$ from each of the four models described above, and denote them as $M_{(j)}$ for data generated from model $j$.
For each simulated dataset $M_{(j)}$, $j=1,\ldots,4$, we compare the four candidate models C1--C4 using the one-step-ahead LPDS.
{Table~\ref{tab:factorselection_sim_t} reports the mean and standard deviation of the LPDS and computation time for 100 one-step-ahead predictions, based on 30 independent repetitions of the prediction procedure.} 
It compares the LPDS {and computation times} across the four candidate models C1--C4 for each simulated dataset $M_{(j)}$, for $j=1,\ldots,4$. 

{In three of the four cases, the highest LPDS is attained by the candidate model matching the true data-generating process. For $M_{(3)}$, however, the highest LPDS is instead attained by $C_1$ rather than the true DGP, $C_3$. 
{For $M_{(3)}$, the true degrees-of-freedom parameter of the $t$ copula is relatively large ($\nu \approx 12.7$). As $\nu$ increases the $t$ copula approaches the Gaussian copula, so the two are difficult to distinguish. In this case the Gaussian copula gives the better predictive fit: both Gaussian-copula candidates (C1 and C2) attain a higher LPDS than both $t$-copula candidates (C3 and C4). The choice of copula, rather than the marginal error distribution, determines the ranking for this dataset; within each copula family the LPDS differences are small.}
{Computational time is generally lower for simpler models, with C1 typically exhibiting the shortest runtimes and C4 the longest.}

\begin{table}[t]
 \caption{Comparison of LPDS values for 100 one-step-ahead forecasts across the four candidate models C1--C4 evaluated on the four simulated datasets $ M_{(1)},\ldots,M_{(4)}$. The values in bold indicate the best model.
  {Values are reported as the mean (SD) obtained by repeating the computation 30 times.}}
    \centering
    \begin{tabular}{cccccc}
\hline \hline
Data Simulation & Criteria & C1 & C2 & C3 & C4 \\
\hline \noalign{\vskip 4pt}
$M_{(1)}$ & LPDS & \textbf{-2051.81} & -2062.79 & -2057.43 & -2063.00 \\
         &  & \textbf{(0.96)} & (1.05) & (1.28) & (1.29) \\[4pt]
         & Time (s) & 98.6 & 105.8 & 116.6 & 123.3 \\
         &  & (0.7) & (0.8) & (0.8) & (0.8) \\[4pt]
        $M_{(2)}$ & LPDS & -2553.53 & \textbf{-2428.32} & -2585.83 & -2453.20 \\
         &  & (7.32) & \textbf{(5.78)} & (6.54) & (4.89) \\[4pt]
         & Time (s) & 98.8 & 106.3 & 116.6 & 123.1 \\
         &  & (0.5) & (1.5) & (0.9) & (0.9) \\[4pt]
        $M_{(3)}$ & LPDS & \textbf{-2502.97} & -2515.91 & -2549.36 & -2546.44 \\
         &  & \textbf{(5.00)} & (5.24) & (4.16) & (4.22) \\[4pt]
         & Time (s) & 99.9 & 105.2 & 117.9 & 123.8 \\
         &  & (0.3) & (0.3) & (0.6) & (0.8) \\[4pt]
        $M_{(4)}$ & LPDS & -1913.17 & -1895.27 & -1785.66 & \textbf{-1709.83} \\
         &  & (9.65) & (10.05) & (3.04) & \textbf{(2.23)} \\[4pt]
         & Time (s) & 115.5 & 105.1 & 117.6 & 123.7 \\
         &  & (35.0) & (0.7) & (0.8) & (0.6) \\[4pt]
\hline
\end{tabular}
    \label{tab:factorselection_sim_t}
\end{table}

\clearpage
\section{Additional details on the industry portfolio dataset} \label{sec:realdata_extrainfo}

This section provides additional details on the industry portfolio dataset. 
Table~\ref{tab:siccodes} lists the industry sectors. 
The first $d=1,\ldots,20$ sectors correspond to the 20 industry portfolios analysed in this study.



\begin{table}[htbp]
\caption{List of industry portfolios.}
\centering
\begin{tabular}{lll}
\hline \hline
$d$ &Code & Industry Name  \\ 
\hline
1& Agric & Agriculture, forestry, and fishing \\
2& Mines & Mining \\
3&Oil & Oil and Gas Extraction \\
4&Stone & Nonmetallic Minerals Except Fuels \\
5&Cnstr & Construction \\
6&Food & Food and Kindred Products \\
7&Smoke & Tobacco Products \\
8&Txtls & Textile Mill Products \\
9&Apprl & Apparel and Other Textile Products \\
10&Wood & Lumber and Wood Products \\
11&Chair & Furniture and Fixtures \\
12&Paper & Paper and Allied Products \\
13&Print & Printing and Publishing \\
14&Chems & Chemicals and Allied Products\\
15&Ptrlm & Petroleum and Coal Products \\
16&Rubbr & Rubber and Miscellaneous Plastics Products \\
17&Lethr & Leather and Leather Products \\
18&Glass & Stone, Clay and Glass Products \\
19&Metal & Primary Metal Industries \\
20&MtlPr & Fabricated Metal Products \\
21&Machn & Machinery, Except Electrical  \\
22&Elctr & Electrical and Electronic Equipment \\
23&Cars & Transportation Equipment\\
24&Instr & Instruments and Related Products  \\
25&Manuf &Miscellaneous Manufacturing Industries \\
26&Trans & Transportation  \\
27&Phone & Telephone and Telegraph Communication  \\
28&TV & Radio and Television Broadcasting \\
29&Utils & Electric, Gas, and Water Supply \\
30&Garbg & Sanitary Services \\
31&Steam &Steam Supply  \\
32&Water & Irrigation Systems \\
33&Whlsl & Wholesale \\
34&Rtail & Retail Stores\\
35&Money & Finance, Insurance, and Real Estate  \\
36&Srvc & Services \\
37&Govt & Public Administration\\
38&Other & Almost Nothing   \\

\hline
\end{tabular}

\label{tab:siccodes}
\end{table}

{
Figure~\ref{fig:realdata_all} displays the de-meaned, average value-weighted returns for the 20 industry portfolios from 2 January 2015 to 20 December 2018. The agricultural industry 
is labelled as $\boldsymbol y_1$.
The $\boldsymbol{y}_1$ series has a minimum (-6.42) and maximum (7.66) return and shows the presence of some extreme movements and occasional large shocks. 
Several other series  exhibit extreme movements and pronounced shocks, most notably $\boldsymbol{y}_{7}$ and $\boldsymbol{y}_{8}$. 
The series $\boldsymbol{y}_{1}$ and $\boldsymbol{y}_{8}$ show substantial excess kurtosis (8.23 and 33.16 respectively, compared with 0 under a Gaussian distribution), indicating heavy tails and a higher frequency of extreme observations.
In addition, series $\boldsymbol{y}_{1}$ illustrates clear volatility clustering, highlighting the heteroskedastic nature of the data.  
These features support the use of GARCH-type volatility models with Student's-$t$ errors rather than Gaussian errors.}

Figure~\ref{fig:realdata_all_hist} presents histograms of the de-meaned, average value-weighted returns for all the $20$ industry portfolios from 2 January 2015 to 20 December 2018. 
The series $\boldsymbol{y}_{1}$, $\boldsymbol{y}{6}$, and $\boldsymbol{y}_{14}$ exhibit the most pronounced spikes around zero, which is consistent with Figure~\ref{fig:realdata_all}. 
Overall, the histograms suggest leptokurtic return distributions, supporting the use of Student's-$t$ innovations with relatively small degrees of freedom.

\begin{figure}[t]
    \centering
    \includegraphics[width=\linewidth]{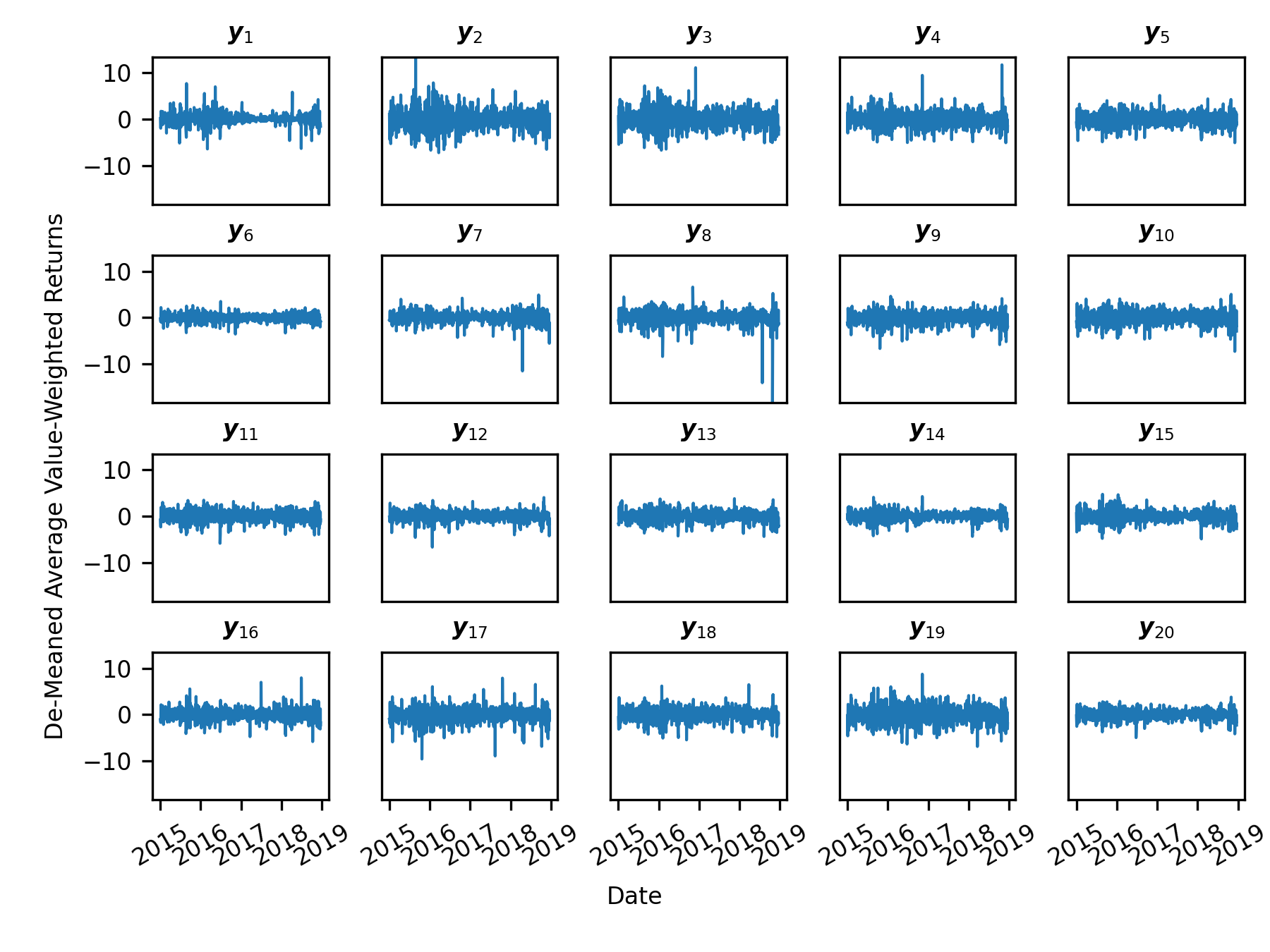}
    \caption{De-meaned, average value-weighted returns for the 20 industry portfolios from 2 January 2015 to 20 December 2018.}
    \label{fig:realdata_all}
\end{figure}

\begin{figure}[t]
    \centering
    \includegraphics[width=\linewidth]{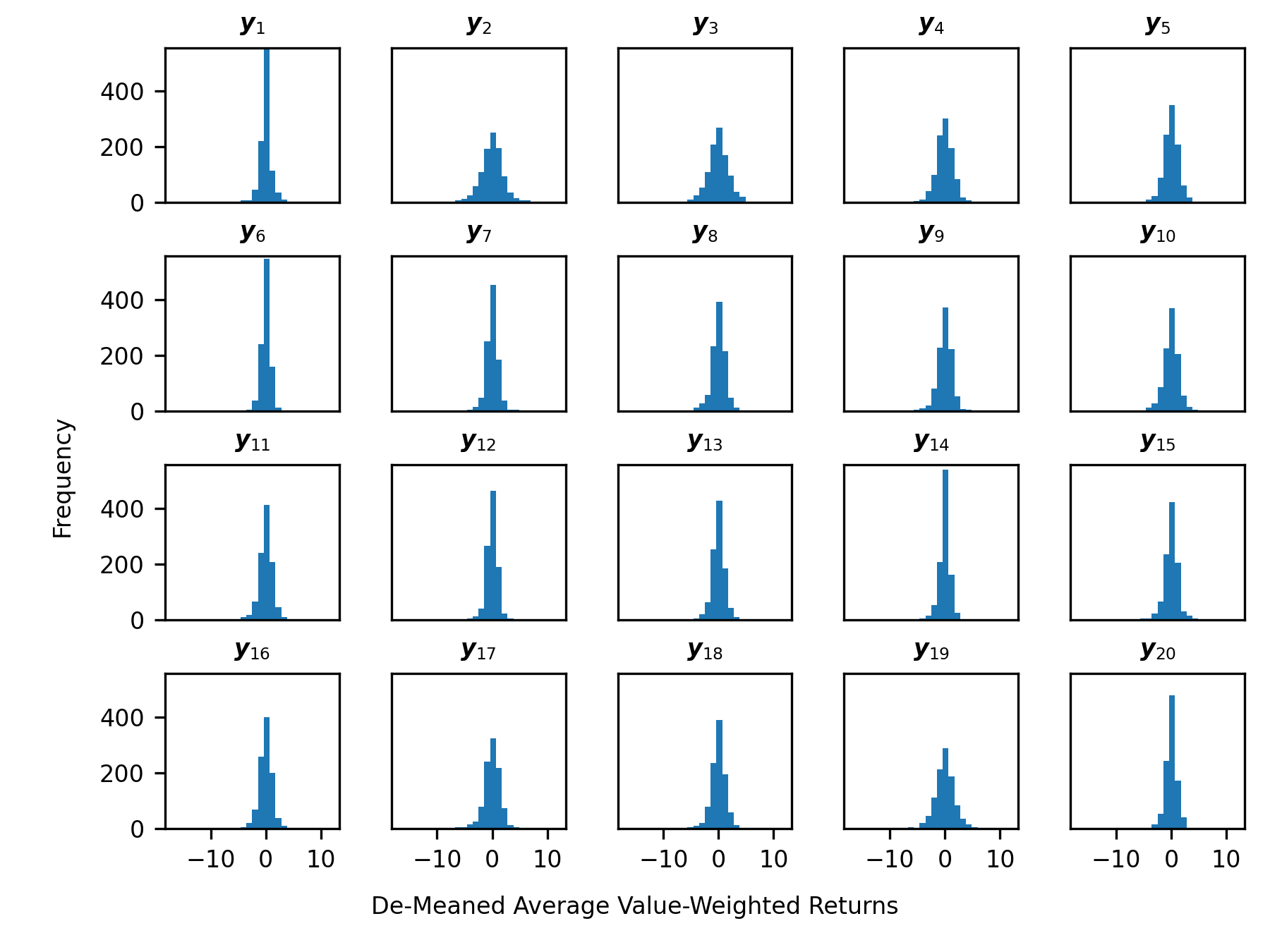}
    \caption{Histograms of de-meaned, average value-weighted returns for the 20 industry portfolios from 2 January 2015 to 20 December 2018.}
    \label{fig:realdata_all_hist}
\end{figure}


\clearpage 

\section{Additional results for the real data application (Section~\ref{sec:RealData})}\label{supp:realdata}

In Section \ref{sec:RealData} of the main paper, we find that the best model on the real dataset is the four-factor $t$ copula with GARCH(1,1) marginals with Student's-$t$ errors. 
The posterior distributions of the parameters of the GARCH(1,1) model with Student's-$t$ errors, obtained using N-IFM, HMC and HMC-IFM, are shown in Figures~\ref{suppfig:alpha1_stcop_f4} to \ref{suppfig:nu_stcop_f4}. Overall, the methods produce broadly similar posteriors. 
However, noticeable discrepancies appear for some marginal model parameters. {In particular, the N-IFM approximation for some $\alpha_{1,d}$ parameters, and to a lesser extent for $\alpha_{2,d}$ and $\gamma_d$, differs from the corresponding HMC, while still having strong agreement with HMC-IFM, agreeing with findings from the simulation studies.} {Some of the differences between the N-IFM and HMC posterior distributions can be attributed to cases in which the HMC posterior means lie in the tails of the corresponding prior distributions.} 

A noticeable difference occurs for the degrees-of-freedom parameter of the GARCH(1,1) errors.
Figure~\ref{suppfig:nu_stcop_f4} shows the posterior distribution of the degrees-of-freedom parameters.

{The N-IFM posterior for the first degree-of-freedom parameter, $\widetilde{\nu}_1$, lies to the right of both the HMC and HMC-IFM posteriors. The HMC posterior mean of $\widetilde{\nu}_1$ falls at the 0.02 quantile of the prior for $\widetilde{\nu}_d$. }
The posterior distributions for the $t$ copula parameters are displayed in Figures~\ref{suppfig:gtilde_stcop_f4_1} to \ref{suppfig:copnu_stcop_f4}. These results, except in a few cases, agree between N-IFM, HMC and HMC-IFM.
{For the degrees-of-freedom parameter $\nu$ of the $t$ copula, the N-IFM posterior is wider than the HMC posterior and, like HMC-IFM, is centred at larger values (around $21$ compared with about $17$ for HMC), overlapping the HMC posterior only in its
lower tail.}
{

For the HMC implementation in \texttt{Stan}, we assess posterior sampling quality using standard MCMC diagnostics, including the effective sample size (\texttt{n\_eff}) and the potential scale reduction factor ($\widehat{\texttt{R}}$), which compares within-chain and between-chain variability for each parameter. These diagnostics are reported in Table~\ref{tab:hmc_stcop_realf4}. Overall, \texttt{n\_eff} is large across parameters and the $\widehat{\texttt{R}}$ values are very close to 1, indicating that the chains have mixed well. 
For HMC-IFM (results not reported), the $\widehat{\texttt{R}}$ values are close to 1; 
\texttt{n\_eff} is lower for the GARCH(1,1) parameters, whose separate fits use $4{,}000$ iterations per chain, and similar to HMC for the copula parameters. HMC and HMC-IFM were both run with $20{,}000$ iterations per chain for the copula model to obtain an adequate \texttt{n\_eff}.
{Identifiability of the factor loadings is a well-known issue in this class of models \citep{Kastner:2017}; we address it by constraining the loading matrix to be lower-triangular with a positive diagonal, which rules out label switching by construction. Nonetheless, repeated HMC and HMC-IFM runs for one- to four-factor models continue to show convergence problems when some of the diagonal elements are close to zero.}


\begin{figure}[t]
    \centering
    \includegraphics[width=\linewidth]{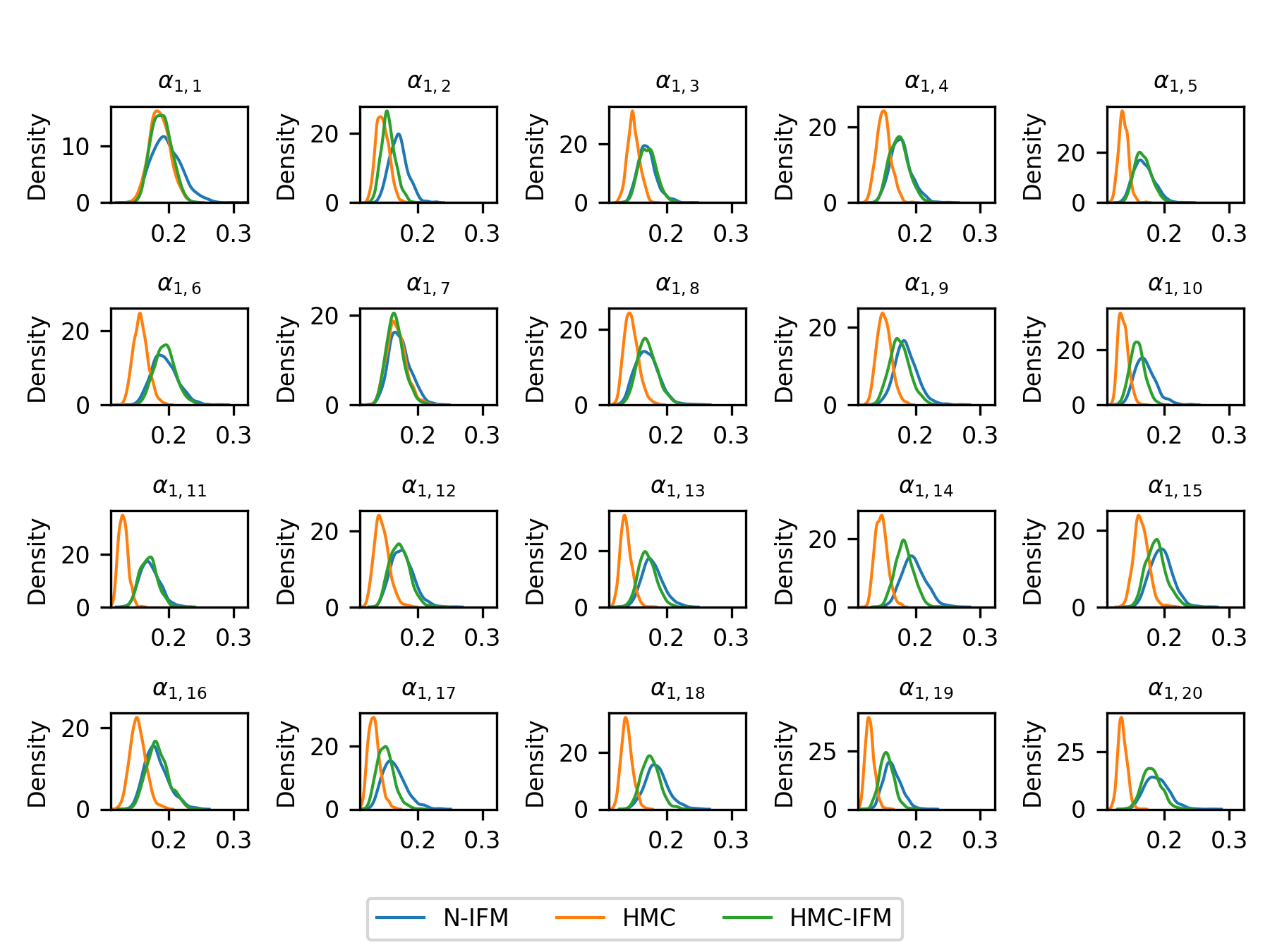}
    \caption{Posterior distributions of the GARCH$(1,1)$ parameters $\alpha_{1,d}$ for $d=1,\ldots,D$, comparing N-IFM with HMC and HMC-IFM for the industry portfolio data.}
    \label{suppfig:alpha1_stcop_f4}
\end{figure}

\begin{figure}[t]
    \centering
    \includegraphics[width=\linewidth]{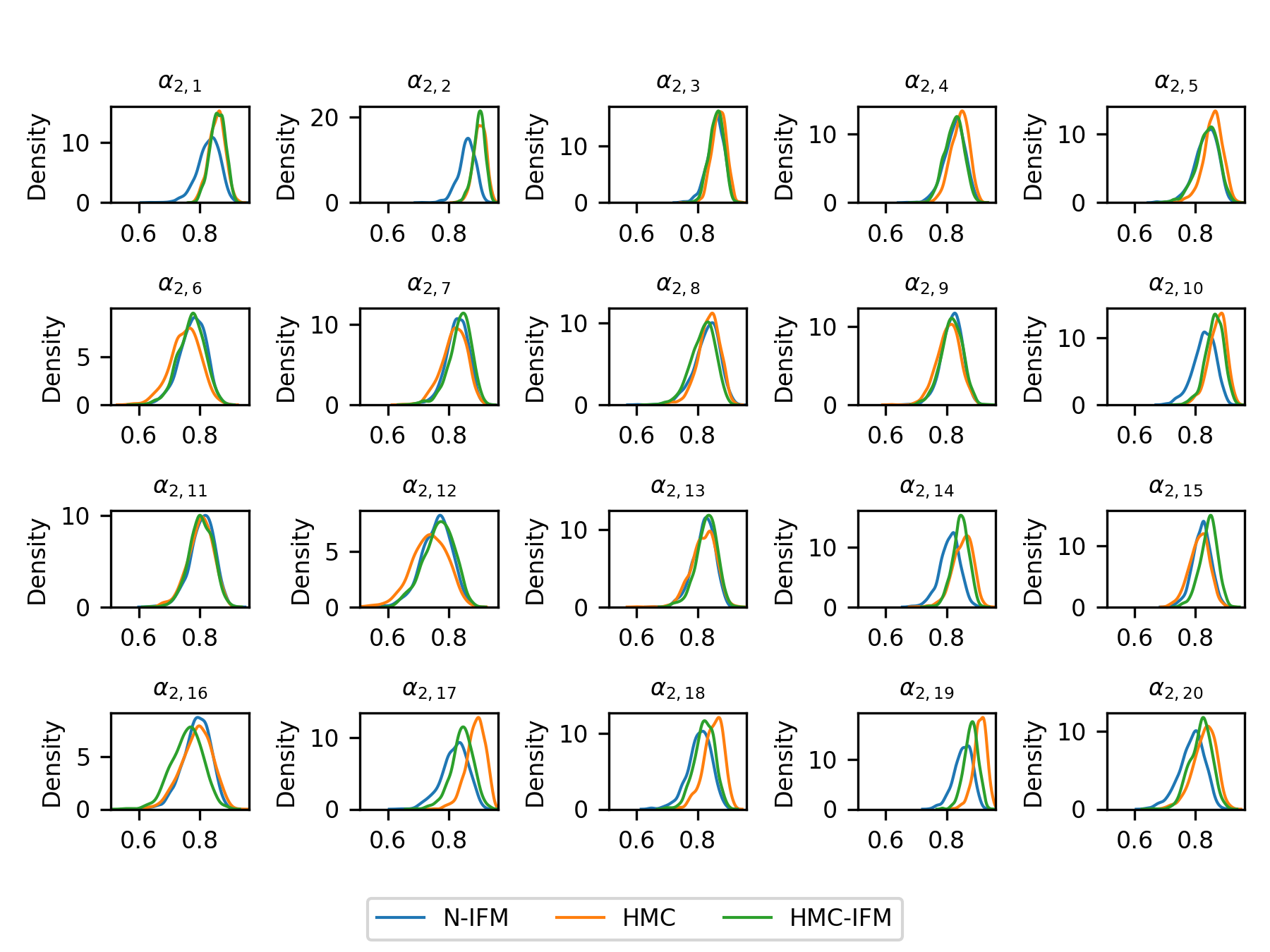}
    \caption{Posterior distributions of the GARCH$(1,1)$ parameters $\alpha_{2,d}$ for $d=1,\ldots,D$, comparing N-IFM with HMC and HMC-IFM for the industry portfolio data.}
    \label{suppfig:alpha2_stcop_f4}
\end{figure}

\begin{figure}[t]
    \centering
    \includegraphics[width=\linewidth]{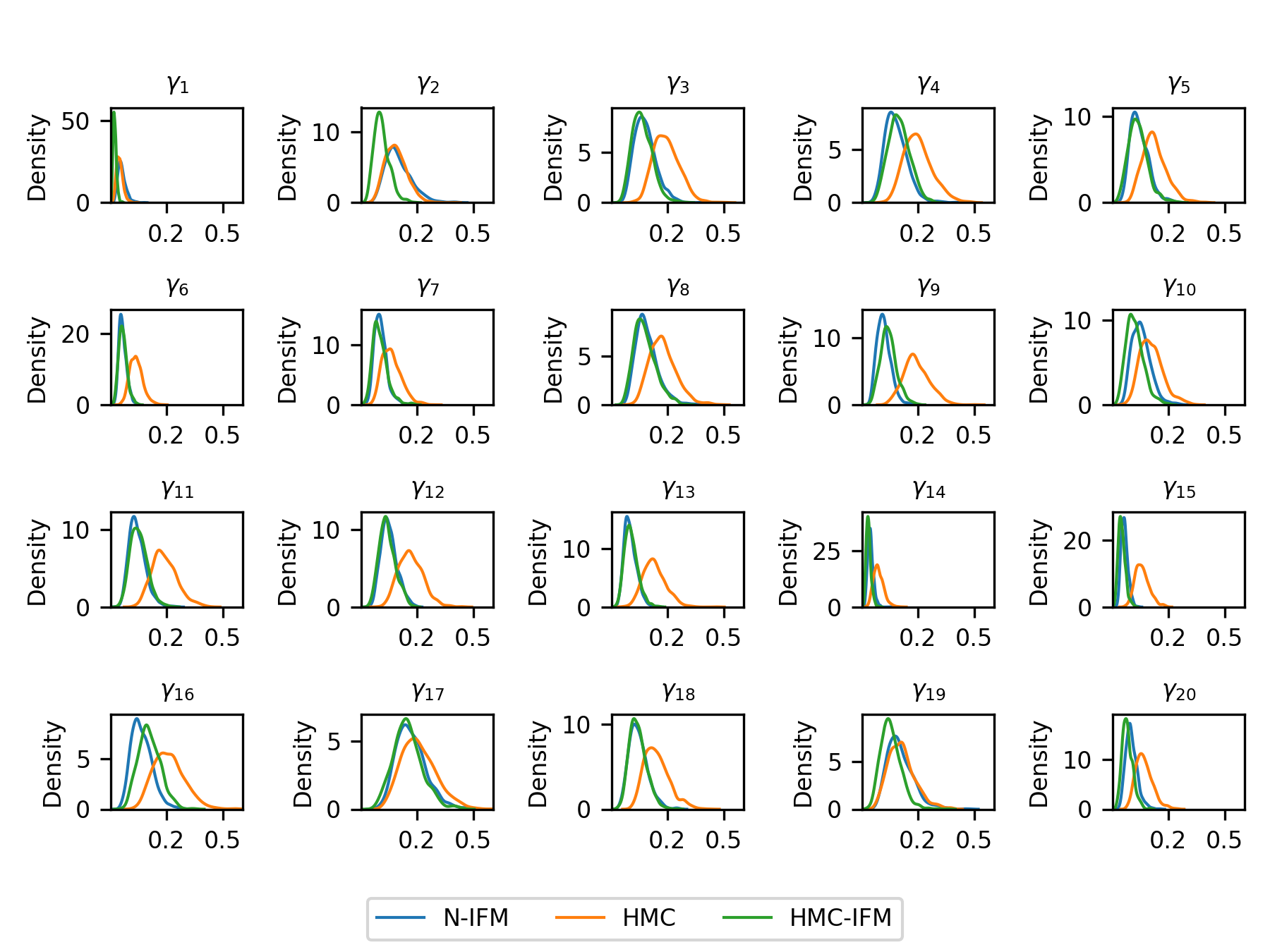}
    \caption{Posterior distributions of the GARCH$(1,1)$ parameters $\gamma_{d}$ for $d=1,\ldots,D$, comparing N-IFM with HMC and HMC-IFM for the industry portfolio data.}
    \label{suppfig:gamma_stcop_f4}
\end{figure}

\begin{figure}[t]
    \centering
    \includegraphics[width=\linewidth]{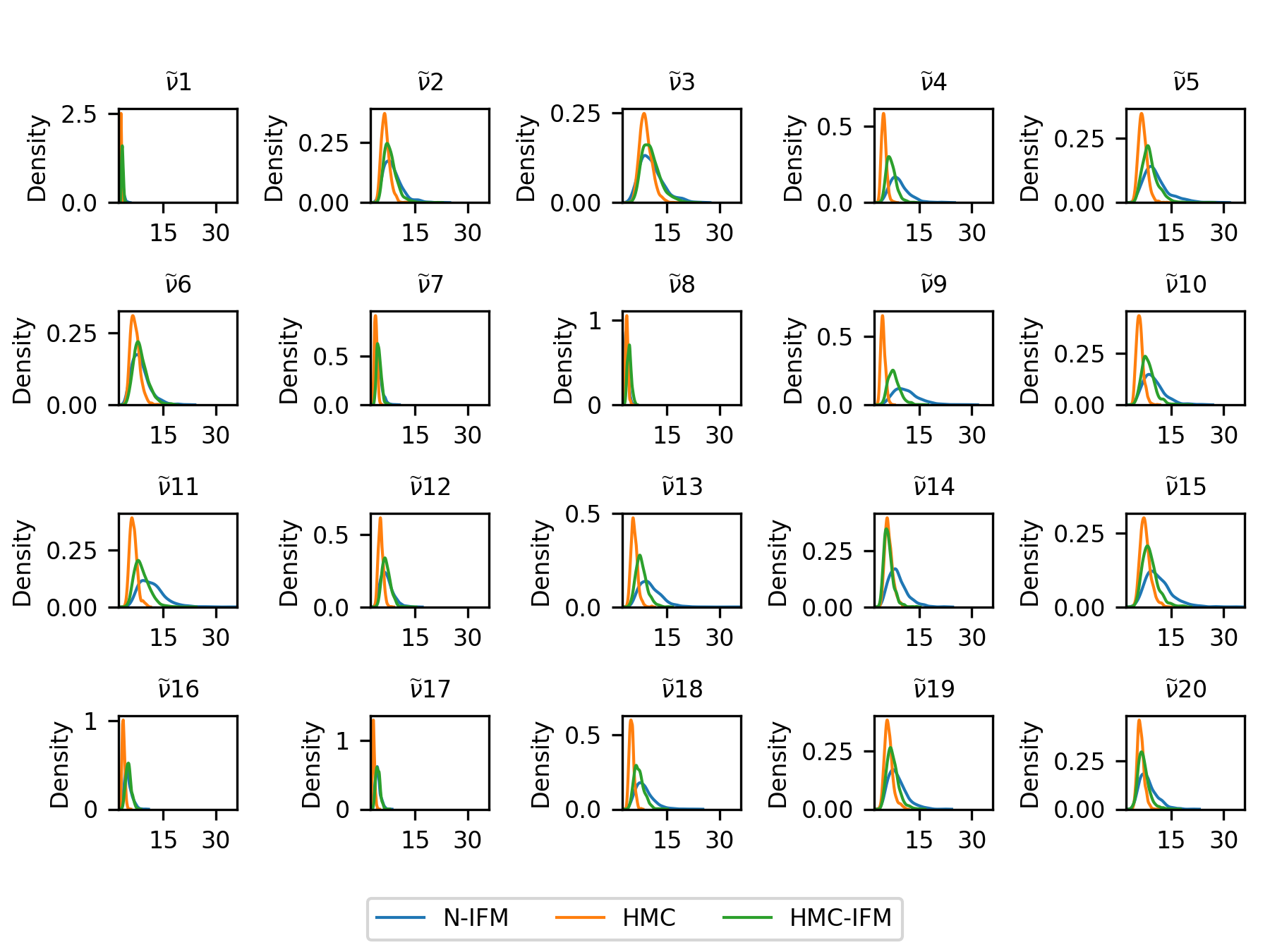}
    \caption{Posterior distributions of the GARCH$(1,1)$ parameters $\widetilde{\nu}_{d}$ for $d=1,\ldots,D$, comparing N-IFM with HMC and HMC-IFM for the industry portfolio data.}
    \label{suppfig:nu_stcop_f4}
\end{figure}

\begin{figure}[t]
    \centering
    \includegraphics[width=\linewidth]{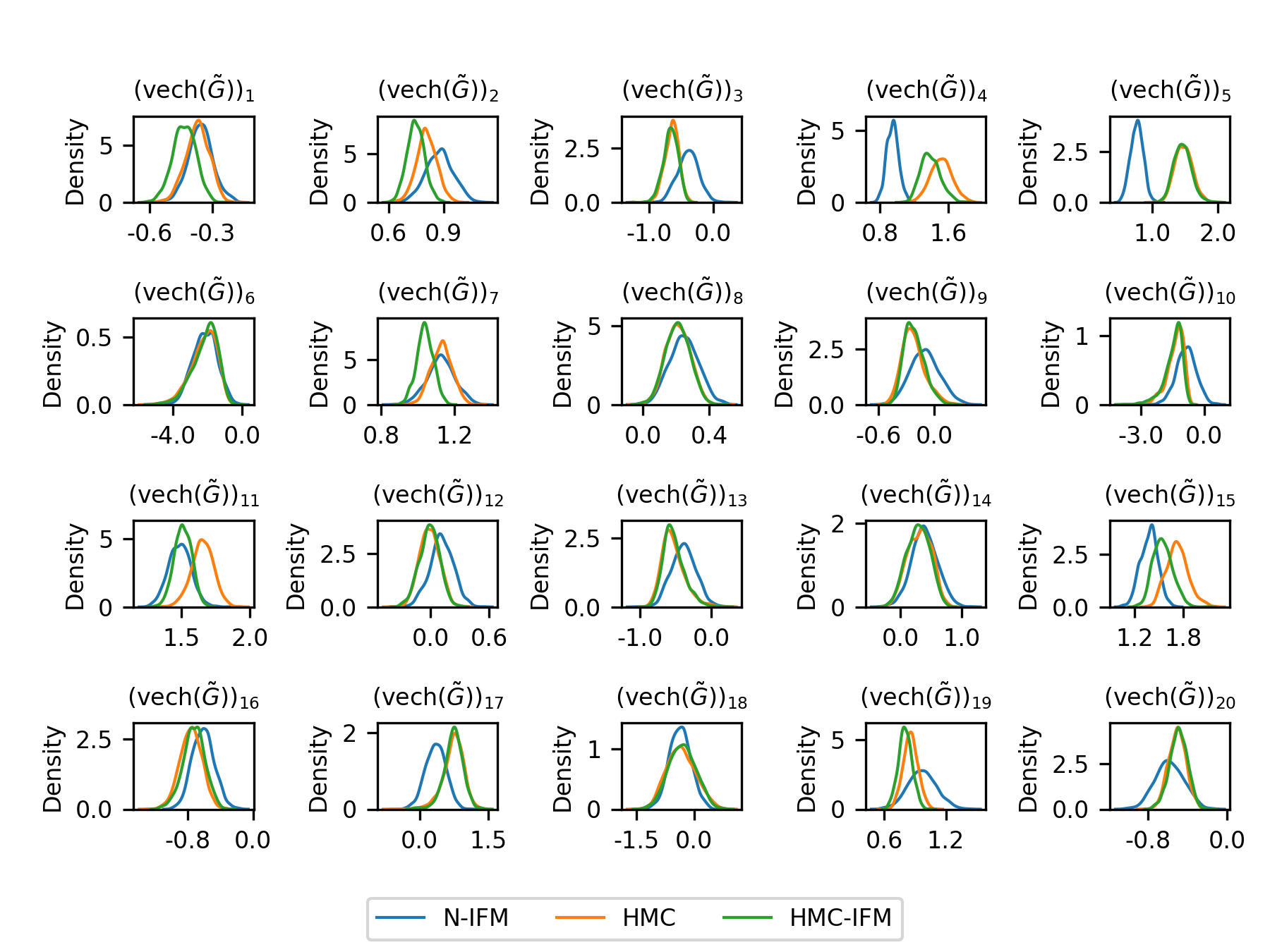}
    \caption{Posterior distributions of the $1$st factor of copula parameters $\widetilde{\boldsymbol{G}}$, comparing N-IFM with HMC and HMC-IFM for the industry portfolio data.}
    \label{suppfig:gtilde_stcop_f4_1}
\end{figure}

\begin{figure}[t]
    \centering
    \includegraphics[width=\linewidth]{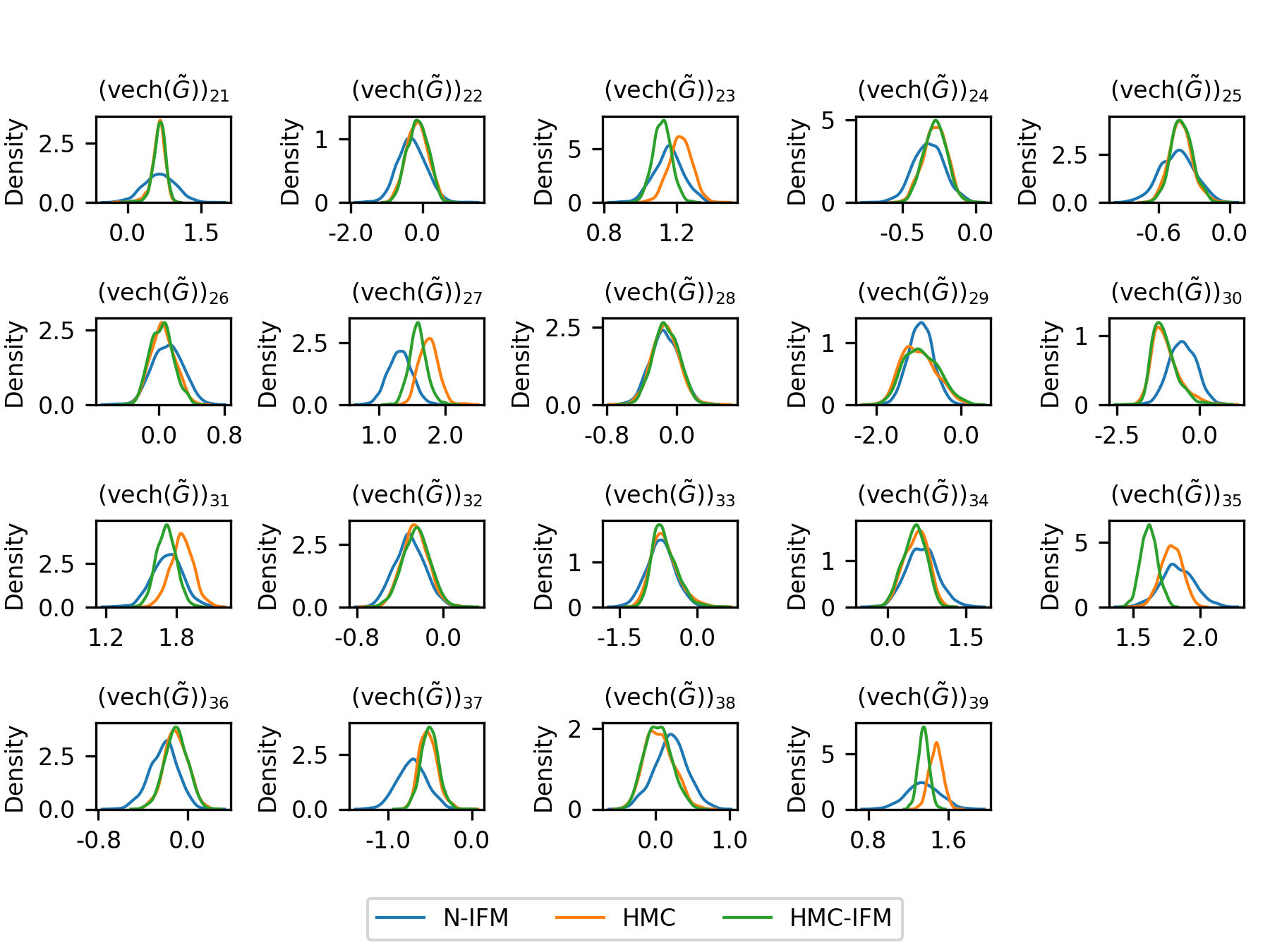}
    \caption{Posterior distributions of the $2$nd factor of copula parameters $\widetilde{\boldsymbol{G}}$, comparing N-IFM with HMC and HMC-IFM for the industry portfolio data.}
    \label{suppfig:gtilde_stcop_f4_2}
\end{figure}

\begin{figure}[t]
    \centering
    \includegraphics[width=\linewidth]{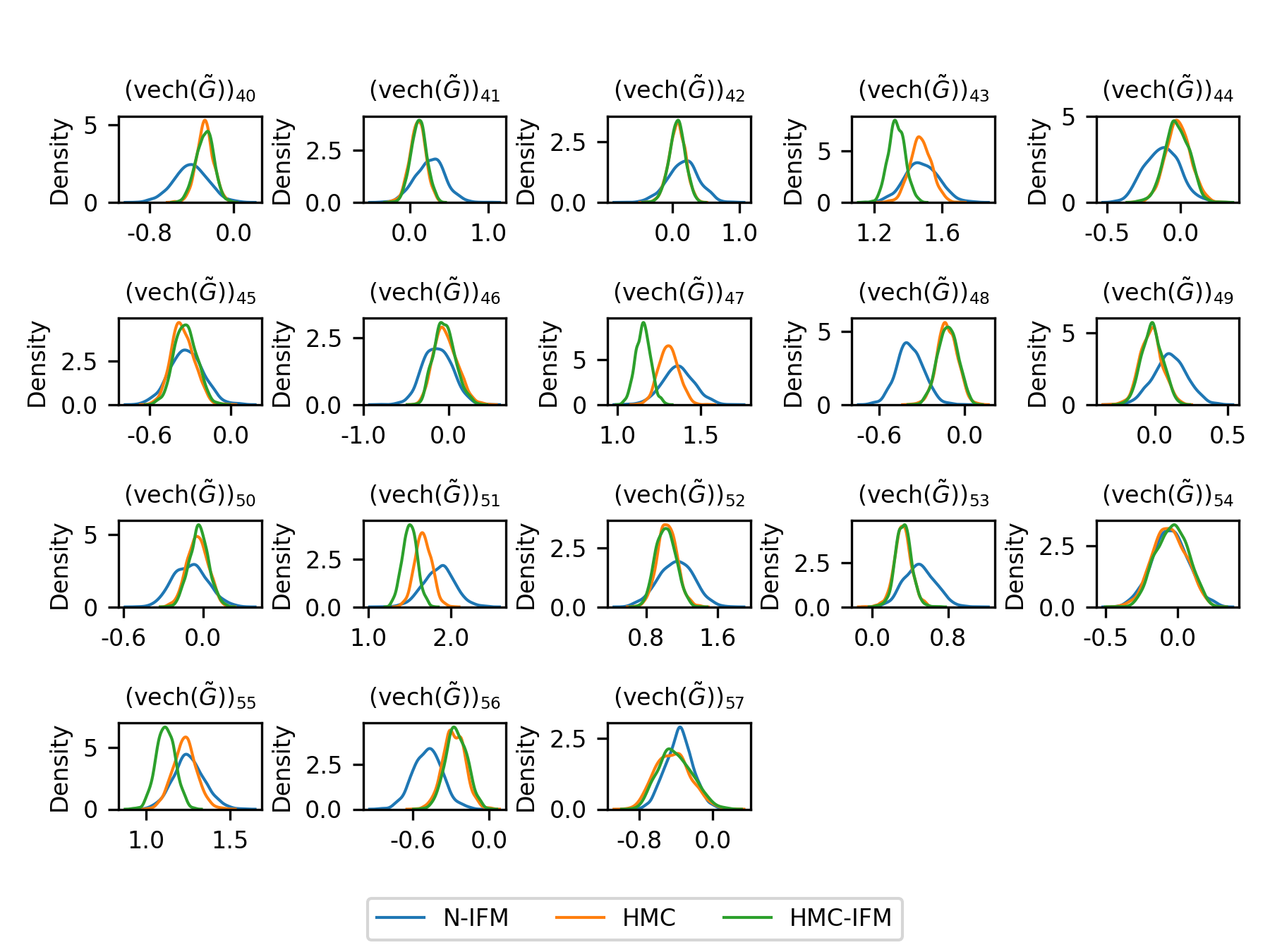}
    \caption{Posterior distributions of the $3$rd factor of copula parameters $\widetilde{\boldsymbol{G}}$, comparing N-IFM with HMC and HMC-IFM for the industry portfolio data.}
    \label{suppfig:gtilde_stcop_f4_3}
\end{figure}

\begin{figure}[t]
    \centering
    \includegraphics[width=\linewidth]{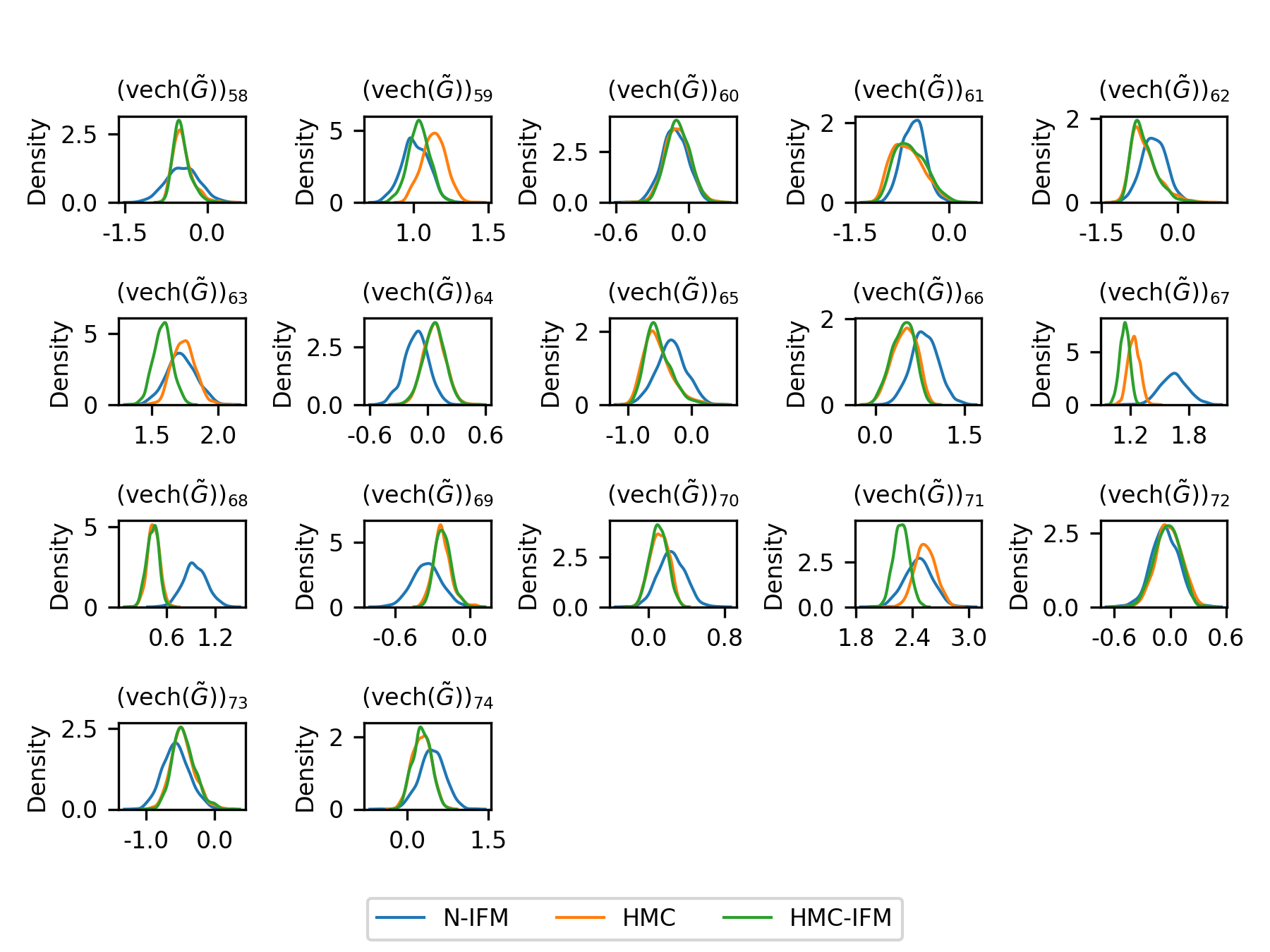}
    \caption{Posterior distributions of the $4$th factor of copula parameters $\widetilde{\boldsymbol{G}}$, comparing N-IFM with HMC and HMC-IFM for the industry portfolio data.}
    \label{suppfig:gtilde_stcop_f4}
\end{figure}

\begin{figure}[t]
    \centering
    \includegraphics[width=\linewidth]{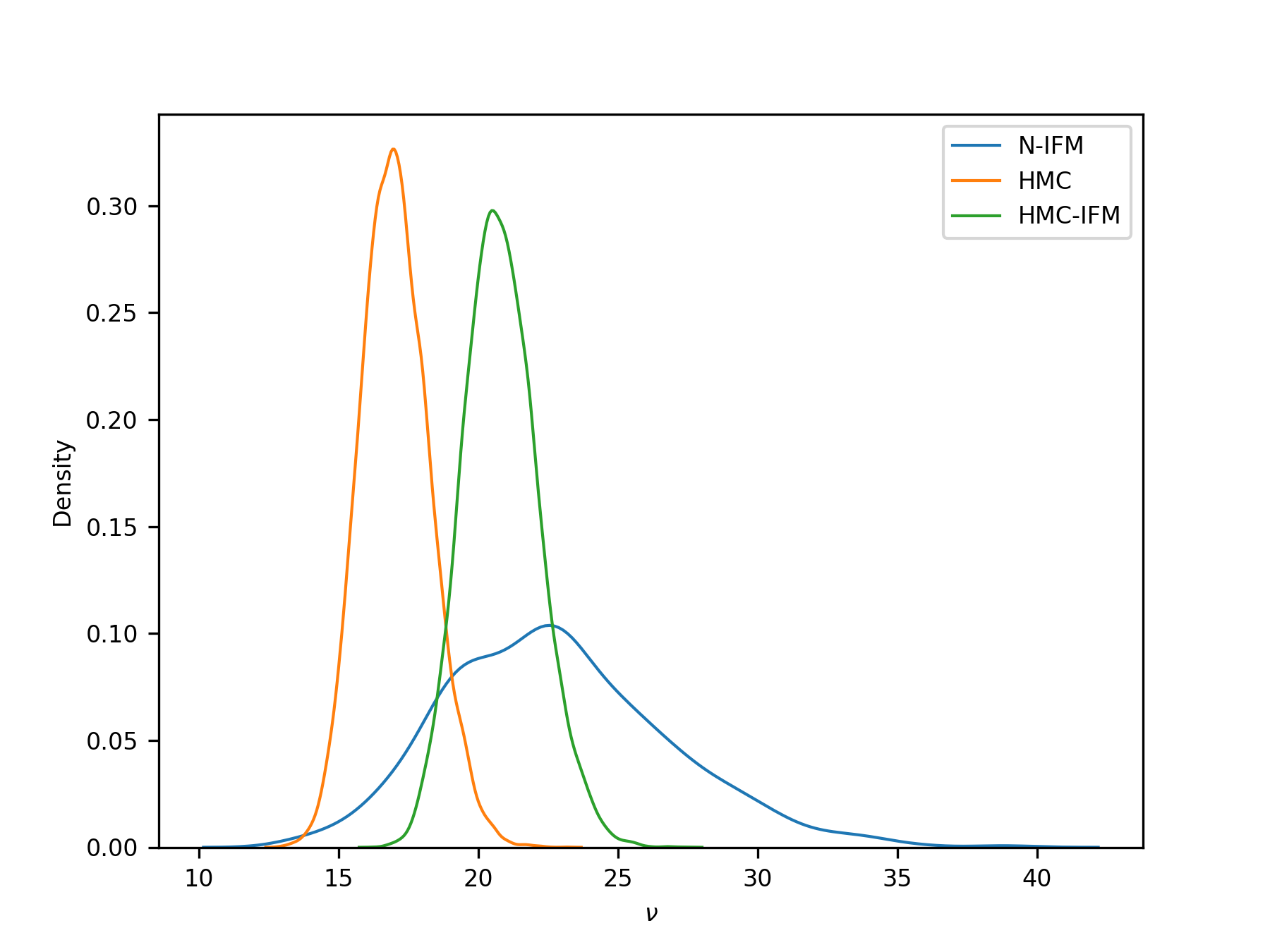}
    \caption{Posterior distributions of the $t$ copula degrees-of-freedom parameter $\nu$, comparing N-IFM with HMC and HMC-IFM for the industry portfolio data.}
    \label{suppfig:copnu_stcop_f4}
\end{figure}

\begin{table}[t]
\caption{Posterior diagnostics for model parameters estimated using HMC for the real data application, showing effective sample size (\texttt{n\_eff}) and {$\widehat{\texttt{R}}$}, for the $t$ copula model with GARCH(1,1) marginals with Student's-$t$ errors.}
\centering
{\scriptsize
\begin{tabular}{lll|lll|lll}
\hline \hline
Parameter & \texttt{n\_eff} & {$\widehat{\texttt{R}}$} & Parameter & \texttt{n\_eff} & {$\widehat{\texttt{R}}$} & Parameter & \texttt{n\_eff} & {$\widehat{\texttt{R}}$} \\
\hline
$\alpha_{1,1}$ & 34048 & 0.9999 & $\gamma_{13}$ & 31688 & 0.9999 & $(\textrm{vech}(\widetilde{G}))_{25}$ & 4809 & 1.0003 \\
$\alpha_{1,2}$ & 34450 & 0.9999 & $\gamma_{14}$ & 30362 & 1.0000 & $(\textrm{vech}(\widetilde{G}))_{26}$ & 2874 & 1.0008 \\
$\alpha_{1,3}$ & 32723 & 1.0000 & $\gamma_{15}$ & 30709 & 0.9999 & $(\textrm{vech}(\widetilde{G}))_{27}$ & 14371 & 1.0000 \\
$\alpha_{1,4}$ & 32604 & 0.9999 & $\gamma_{16}$ & 31167 & 0.9999 & $(\textrm{vech}(\widetilde{G}))_{28}$ & 6851 & 1.0001 \\
$\alpha_{1,5}$ & 30975 & 1.0001 & $\gamma_{17}$ & 27713 & 1.0000 & $(\textrm{vech}(\widetilde{G}))_{29}$ & 2929 & 1.0003 \\
$\alpha_{1,6}$ & 29648 & 0.9999 & $\gamma_{18}$ & 30422 & 1.0000 & $(\textrm{vech}(\widetilde{G}))_{30}$ & 2457 & 1.0008 \\
$\alpha_{1,7}$ & 33496 & 1.0000 & $\gamma_{19}$ & 31276 & 1.0000 & $(\textrm{vech}(\widetilde{G}))_{31}$ & 10923 & 1.0000 \\
$\alpha_{1,8}$ & 31548 & 1.0000 & $\gamma_{20}$ & 30049 & 0.9999 & $(\textrm{vech}(\widetilde{G}))_{32}$ & 5144 & 1.0000 \\
$\alpha_{1,9}$ & 30446 & 1.0000 & $\widetilde{\nu}_{1}$ & 28952 & 1.0000 & $(\textrm{vech}(\widetilde{G}))_{33}$ & 2373 & 1.0004 \\
$\alpha_{1,10}$ & 30351 & 1.0000 & $\widetilde{\nu}_{2}$ & 24742 & 1.0000 & $(\textrm{vech}(\widetilde{G}))_{34}$ & 2975 & 1.0006 \\
$\alpha_{1,11}$ & 28589 & 0.9999 & $\widetilde{\nu}_{3}$ & 28806 & 1.0000 & $(\textrm{vech}(\widetilde{G}))_{35}$ & 10524 & 1.0001 \\
$\alpha_{1,12}$ & 27543 & 1.0001 & $\widetilde{\nu}_{4}$ & 25381 & 0.9999 & $(\textrm{vech}(\widetilde{G}))_{36}$ & 5404 & 1.0001 \\
$\alpha_{1,13}$ & 29803 & 0.9999 & $\widetilde{\nu}_{5}$ & 27968 & 1.0000 & $(\textrm{vech}(\widetilde{G}))_{37}$ & 5123 & 1.0002 \\
$\alpha_{1,14}$ & 31044 & 1.0000 & $\widetilde{\nu}_{6}$ & 27345 & 0.9999 & $(\textrm{vech}(\widetilde{G}))_{38}$ & 2741 & 1.0008 \\
$\alpha_{1,15}$ & 31357 & 0.9999 & $\widetilde{\nu}_{7}$ & 25329 & 1.0000 & $(\textrm{vech}(\widetilde{G}))_{39}$ & 13931 & 1.0000 \\
$\alpha_{1,16}$ & 31255 & 0.9999 & $\widetilde{\nu}_{8}$ & 19086 & 1.0000 & $(\textrm{vech}(\widetilde{G}))_{40}$ & 6884 & 1.0000 \\
$\alpha_{1,17}$ & 31666 & 1.0000 & $\widetilde{\nu}_{9}$ & 25622 & 0.9999 & $(\textrm{vech}(\widetilde{G}))_{41}$ & 6171 & 1.0000 \\
$\alpha_{1,18}$ & 28607 & 1.0001 & $\widetilde{\nu}_{10}$ & 26673 & 1.0001 & $(\textrm{vech}(\widetilde{G}))_{42}$ & 4724 & 0.9999 \\
$\alpha_{1,19}$ & 30650 & 1.0000 & $\widetilde{\nu}_{11}$ & 25159 & 0.9999 & $(\textrm{vech}(\widetilde{G}))_{43}$ & 11999 & 0.9999 \\
$\alpha_{1,20}$ & 29667 & 1.0000 & $\widetilde{\nu}_{12}$ & 24858 & 0.9999 & $(\textrm{vech}(\widetilde{G}))_{44}$ & 5570 & 1.0002 \\
$\alpha_{2,1}$ & 34429 & 0.9999 & $\widetilde{\nu}_{13}$ & 25582 & 0.9999 & $(\textrm{vech}(\widetilde{G}))_{45}$ & 7625 & 1.0000 \\
$\alpha_{2,2}$ & 35064 & 1.0000 & $\widetilde{\nu}_{14}$ & 26667 & 0.9999 & $(\textrm{vech}(\widetilde{G}))_{46}$ & 2933 & 1.0006 \\
$\alpha_{2,3}$ & 32120 & 0.9999 & $\widetilde{\nu}_{15}$ & 29205 & 0.9999 & $(\textrm{vech}(\widetilde{G}))_{47}$ & 15201 & 0.9999 \\
$\alpha_{2,4}$ & 32834 & 1.0000 & $\widetilde{\nu}_{16}$ & 23516 & 0.9999 & $(\textrm{vech}(\widetilde{G}))_{48}$ & 6717 & 1.0000 \\
$\alpha_{2,5}$ & 27923 & 1.0000 & $\widetilde{\nu}_{17}$ & 19473 & 1.0000 & $(\textrm{vech}(\widetilde{G}))_{49}$ & 8742 & 0.9999 \\
$\alpha_{2,6}$ & 31913 & 0.9999 & $\widetilde{\nu}_{18}$ & 24210 & 1.0001 & $(\textrm{vech}(\widetilde{G}))_{50}$ & 12534 & 1.0000 \\
$\alpha_{2,7}$ & 31687 & 1.0001 & $\widetilde{\nu}_{19}$ & 26630 & 1.0000 & $(\textrm{vech}(\widetilde{G}))_{51}$ & 12375 & 1.0001 \\
$\alpha_{2,8}$ & 25229 & 1.0000 & $\widetilde{\nu}_{20}$ & 24482 & 1.0000 & $(\textrm{vech}(\widetilde{G}))_{52}$ & 9031 & 1.0000 \\
$\alpha_{2,9}$ & 29352 & 1.0000 & $(\textrm{vech}(\widetilde{G}))_{1}$ & 23746 & 0.9999 & $(\textrm{vech}(\widetilde{G}))_{53}$ & 5936 & 1.0001 \\
$\alpha_{2,10}$ & 30082 & 1.0000 & $(\textrm{vech}(\widetilde{G}))_{2}$ & 11491 & 1.0002 & $(\textrm{vech}(\widetilde{G}))_{54}$ & 3044 & 1.0003 \\
$\alpha_{2,11}$ & 29649 & 1.0000 & $(\textrm{vech}(\widetilde{G}))_{3}$ & 9311 & 1.0000 & $(\textrm{vech}(\widetilde{G}))_{55}$ & 11930 & 1.0001 \\
$\alpha_{2,12}$ & 27891 & 1.0000 & $(\textrm{vech}(\widetilde{G}))_{4}$ & 10336 & 1.0000 & $(\textrm{vech}(\widetilde{G}))_{56}$ & 6665 & 1.0001 \\
$\alpha_{2,13}$ & 29895 & 0.9999 & $(\textrm{vech}(\widetilde{G}))_{5}$ & 13356 & 1.0000 & $(\textrm{vech}(\widetilde{G}))_{57}$ & 3264 & 1.0003 \\
$\alpha_{2,14}$ & 31936 & 1.0000 & $(\textrm{vech}(\widetilde{G}))_{6}$ & 8948 & 1.0000 & $(\textrm{vech}(\widetilde{G}))_{58}$ & 2632 & 1.0007 \\
$\alpha_{2,15}$ & 30948 & 0.9999 & $(\textrm{vech}(\widetilde{G}))_{7}$ & 12234 & 1.0001 & $(\textrm{vech}(\widetilde{G}))_{59}$ & 11421 & 0.9999 \\
$\alpha_{2,16}$ & 26911 & 1.0000 & $(\textrm{vech}(\widetilde{G}))_{8}$ & 6297 & 0.9999 & $(\textrm{vech}(\widetilde{G}))_{60}$ & 7121 & 1.0001 \\
$\alpha_{2,17}$ & 24654 & 1.0000 & $(\textrm{vech}(\widetilde{G}))_{9}$ & 2531 & 1.0004 & $(\textrm{vech}(\widetilde{G}))_{61}$ & 2875 & 1.0003 \\
$\alpha_{2,18}$ & 27621 & 1.0001 & $(\textrm{vech}(\widetilde{G}))_{10}$ & 3570 & 1.0007 & $(\textrm{vech}(\widetilde{G}))_{62}$ & 2355 & 1.0006 \\
$\alpha_{2,19}$ & 33726 & 1.0000 & $(\textrm{vech}(\widetilde{G}))_{11}$ & 10238 & 1.0000 & $(\textrm{vech}(\widetilde{G}))_{63}$ & 10714 & 1.0000 \\
$\alpha_{2,20}$ & 27526 & 0.9999 & $(\textrm{vech}(\widetilde{G}))_{12}$ & 5245 & 1.0001 & $(\textrm{vech}(\widetilde{G}))_{64}$ & 5228 & 1.0000 \\
$\gamma_{1}$ & 29701 & 0.9999 & $(\textrm{vech}(\widetilde{G}))_{13}$ & 2617 & 1.0003 & $(\textrm{vech}(\widetilde{G}))_{65}$ & 2354 & 1.0006 \\
$\gamma_{2}$ & 29989 & 1.0000 & $(\textrm{vech}(\widetilde{G}))_{14}$ & 2755 & 1.0006 & $(\textrm{vech}(\widetilde{G}))_{66}$ & 2882 & 1.0005 \\
$\gamma_{3}$ & 30195 & 1.0000 & $(\textrm{vech}(\widetilde{G}))_{15}$ & 16693 & 0.9999 & $(\textrm{vech}(\widetilde{G}))_{67}$ & 11207 & 1.0001 \\
$\gamma_{4}$ & 31409 & 1.0000 & $(\textrm{vech}(\widetilde{G}))_{16}$ & 8820 & 1.0001 & $(\textrm{vech}(\widetilde{G}))_{68}$ & 6478 & 1.0000 \\
$\gamma_{5}$ & 30143 & 0.9999 & $(\textrm{vech}(\widetilde{G}))_{17}$ & 4306 & 1.0003 & $(\textrm{vech}(\widetilde{G}))_{69}$ & 3937 & 1.0004 \\
$\gamma_{6}$ & 31653 & 0.9999 & $(\textrm{vech}(\widetilde{G}))_{18}$ & 2837 & 1.0001 & $(\textrm{vech}(\widetilde{G}))_{70}$ & 2982 & 1.0004 \\
$\gamma_{7}$ & 30712 & 1.0001 & $(\textrm{vech}(\widetilde{G}))_{19}$ & 11809 & 1.0000 & $(\textrm{vech}(\widetilde{G}))_{71}$ & 11545 & 1.0001 \\
$\gamma_{8}$ & 28203 & 1.0000 & $(\textrm{vech}(\widetilde{G}))_{20}$ & 9323 & 1.0000 & $(\textrm{vech}(\widetilde{G}))_{72}$ & 5222 & 1.0001 \\
$\gamma_{9}$ & 28937 & 1.0000 & $(\textrm{vech}(\widetilde{G}))_{21}$ & 4540 & 1.0000 & $(\textrm{vech}(\widetilde{G}))_{73}$ & 3144 & 1.0003 \\
$\gamma_{10}$ & 32074 & 1.0000 & $(\textrm{vech}(\widetilde{G}))_{22}$ & 2584 & 1.0002 & $(\textrm{vech}(\widetilde{G}))_{74}$ & 3557 & 1.0008 \\
$\gamma_{11}$ & 32581 & 1.0000 & $(\textrm{vech}(\widetilde{G}))_{23}$ & 10711 & 1.0000 & $\nu$ & 25890 & 1.0002 \\
$\gamma_{12}$ & 30425 & 1.0000 & $(\textrm{vech}(\widetilde{G}))_{24}$ & 5844 & 1.0001 &  &  &  \\
\hline
\end{tabular}
}

\label{tab:hmc_stcop_realf4}

\end{table}



\end{document}